# Investigation into the foundations of the track-event theory of cell survival and the radiation action model based on nanodosimetry

**Sonwabile Arthur Ngcezu[1], Hans Rabus[2,*]**

[1] *University of the Witwatersrand, Johannesburg, 2000, South Africa*
[2] *Physikalisch-Technische Bundesanstalt (PTB), 10587 Berlin, Germany*
* email: hans.rabus@ptb.de

## Abstract

This work aims at elaborating the basic assumptions behind the "track-event theory" (TET) and its derivate "radiation action model based on nanodosimetry" (RAMN) by clearly distinguishing between effects of tracks at the cellular level and the induction of lesions in subcellular targets. It is demonstrated that the model assumptions of Poisson distribution and statistical independence of the frequency of single and clustered DNA lesions are dispensable for multi-event distributions, because they follow from the Poisson distribution of the number of tracks affecting the considered target volume. It is also shown that making these assumptions for the single-event distributions of the number of lethal and sublethal lesions within a cell would lead to an essentially exponential dose dependence of survival for practically relevant values of the absorbed dose. Furthermore, it is elucidated that the model equation used for consideration of repair within the TET is based on the assumption that DNA lesions induced by different tracks are repaired independently. Consequently, the model equation is presumably inconsistent with the model assumptions and requires an additional model parameter. Furthermore, the methodology for deriving model parameters from nanodosimetric properties of particle track structure is critically assessed. Based on data from proton track simulations it is shown that the assumption of statistically independent targets leads to the prediction of negligible frequency of clustered DNA damage. An approach is outlined how track structure could be considered in determining the model parameters, and the implications for TET and RAMN are discussed.

*Keywords:*
Nanodosimetry, track structure, track-event theory, radiation action model

## Introduction

The so-called track-event theory (TET)[1] proposed by Besserer and Schneider is a model for predicting cell survival based on the induction of DNA double-strand breaks (DSBs) by charged particle tracks (Besserer and Schneider 2015a, 2015b). The induction of pairs of DSBs within a considered target volume by a particle track is called an "event". (This is in contrast to microdosimetric terminology where "track" and "event" both refer to the statistically correlated occurrence of energy transfer points (Booz et al. 1983, Rossi and Zaider 1996, Lindborg and Waker 2017).) A low-dose approximation of the fundamental model equation was shown to be equivalent to the commonly used linear-quadratic model and to have a dose dependence that matches the experimentally observed exponential dose dependence at higher doses (Besserer and Schneider 2015a). In later work, the parameters of the model have been related to nanodosimetry (Schneider et al. 2016, 2017, 2019), and recently the TET has been developed into a radiation action model based on nanodosimetry (RAMN) that tried to resolve the shortcomings of the original TET model (Schneider et al. 2020).

In the first version of the TET (Besserer and Schneider 2015a), the basic biophysical model assumption was that a cell will be inactivated if at least two sublethal lesions in the form of DSBs are induced by direct radiation interaction with the DNA. If the two or more sublethal lesions are produced by a single track, this is called a one-track event (OTE). If a track produces exactly one sublethal lesion, then it requires at least two tracks interacting in the cell for its inactivation. This is called a two-track event (TTE). The mathematical formulation of the model further involved the assumption that OTEs and TTEs are "statistically independent events in the terminology of nanodosimetry" (Besserer and Schneider 2015a).

This statement seems paradoxical given that for a particular track and a specific target volume, an OTE and a TTE are disjoint alternatives and, hence, statistically dependent. This contradiction arises from the fact that the terms OTE and TTE were used in two different meanings (Besserer and Schneider 2015a, 2015b, Schneider et al. 2019). Namely, on the one hand, the effect of a particular track on a cell in the sense stated above, and, on the other hand, for the (multi-event) result of the irradiation on the cell. Their mathematical formulation was based on the first meaning of the terms.

In the second version of the TET (Besserer and Schneider 2015b), the model assumption was relaxed by including the possibility of DSB repair, such that cell inactivation occurs only if there are unrepaired sublethal lesions. Repair was

[1] Glossary of acronyms used in this paper:
BIV basic interaction volume
CL clustered lesion
CV cluster volume
DSB double strand break
OTE one-track event

Glossary of acronyms used in this paper (continued):
RAMN radiation action model based on nanodosimetry
ROI region of interest
SL single lesion
TET track-event theory
TTE two-track event

assumed to be of “second order”, meaning that DNA repair changes the cell survival rate only for cells with exactly two sublethal lesions. As this introduced an additional model parameter, attempts were made in further work to reduce the number of adjustable model parameters by deriving the ratio of the two model parameters (related to OTEs and TTEs) from chromatin geometry and nanodosimetric properties of ion tracks (Schneider et al. 2016, 2017).

To further reduce the number of model parameters, a first attempt was made in Schneider et al. (2019) to explicitly relate the TET model parameters and nanodosimetric parameters of track structure. This relation was derived by considering OTEs and TTEs in microscopic sites (named “lethal interaction” volumes) within which DSBs are induced in “basic interaction volumes” (BIVs). A BIV is assumed to be a sphere of 2 nm diameter that contains a DNA segment of five to ten base pairs. The size of the (spherical) sites was found to be dependent on radiation type and ranged from 5 nm diameter for carbon ions up to 35 nm for photons.

With the development of the RAMN, some methodological problems with the aforementioned first attempt to relate the TET parameters with nanodosimetry have been overcome. The radiobiological interpretation and the terminology were changed such that now clustered lesions (CLs) and single lesions (SLs) of the DNA are considered (Schneider et al. 2020). The mean frequencies of occurrences of CLs and SLs are linked to the particle fluence, while the (conditional) probability of their induction is related to nanodosimetric parameters of track structure.

This article was motivated by the following concerns of the authors regarding assumptions and methodology used in the TET and RAMN:

1. The observation of inconsistent use of terminology. Apart from already mentioned points like the terms OTE and TTE in the TET model description (Besserer and Schneider 2015a), this also applies to the RAMN model parameter σ. This parameter was initially introduced as an “intersection-cross-section” relating the fluence and frequency of lesions, whereas it was later stated that “σ contains all cell specific parameters which affect cell sterilization, as e. g. phase in cell cycle, radioresistance, repopulation and repair capability” (Schneider et al. 2020).

2. The assumption of statistical independence of lethal and sublethal (or clustered and single) lesions that seems counterintuitive given that these should be alternative outcomes of radiation interaction (Besserer and Schneider 2015a, 2015b).

3. The apparent contradiction between the concept of particle tracks as statistically correlated interactions and the assumption of statistical independence for single-event radiation effects in different sites (Schneider et al. 2020).

4. The appearance of a term in the repair model that is quadratic in the repair probability and cubic in dose (Besserer and Schneider 2015b).

5. A derivation of model parameters from nanodosimetry that considers only the case of tracks traversing the considered sites (Schneider et al. 2019, 2020). The last point has already been mentioned as one of the limitations of the RAMN in the work of Schneider et al. (2020).

This paper is intended as a critical analysis of the foundations of the TET and RAMN in terms of mathematical consistency of theory and model assumptions as well as with respect to compliance with nanodosimetric results. It is organized as follows. First, the basic TET and RAMN model formula is derived from considerations on the interaction of tracks and biological cells. Furthermore, some conceptional issues are highlighted that arise when linking the cellular-scale picture with subcellular radiation effects. Second, the inclusion of repair in the TET and RAMN is discussed. Third, the approach of Schneider et al. (2019, 2020) to link the TET and RAMN model parameters to nanodosimetric parameters of track structure is discussed with a particular focus on the range of relevant impact parameters. Finally, an outline is given how track structure could be considered in a revised TET/RAMN.

## Theoretical foundations of TET and RAMN

In this Section the fundamental model equations of the TET and RAMN are derived from an abstract perspective, with a clear distinction between the initial radiation effects at the cellular and subcellular levels and between single-event and multi-event distributions. It should be noted that this derivation is not completely aligned with the formulation of the TET by Schneider et al. (2016, 2017, 2019, 2020) but believed by the authors to be more consistent.

### *Derivation of the fundamental model equation*

A track (or event in the terminology of microdosimetry) is the set of statistically correlated loci of interactions of a primary particle and all its secondary electrons in a volume of matter. When a (single) track interacts with a biological cell, the radiation-induced damage can be classified into the three categories “lethal”, “sublethal” and “nonlethal”. A lethal event leads to cell inactivation. As this is the result of the interaction of a single track, this was called a one-track event (OTE) in the initial formulation of the TET (Besserer and Schneider 2015a). A sublethal event is not lethal on its own, but when two such events occur (i.e., two tracks interact with the cell), their combination leads to cell inactivation. This was called a two-track event (TTE) in Besserer and Schneider (2015a).

In the case of a nonlethal event by a track, the cell will only be inactivated if one of the following (not disjoint) cases occur: (1) a second track interacts with the cell and produces a lethal event; (2) at least two other tracks interact with the cell and produce sublethal events.

The (single event) probabilities of the occurrence of a nonlethal, sublethal, or lethal event will be denoted in this paper by $p_0$, $p_1$, and $p_{2+}$, respectively. The quantities $p_1$ and $p_{2+}$ are given by

$$p_1 = \frac{1}{n_t} \iint_A p_{c,1}(\boldsymbol{r}) \Phi(\boldsymbol{r}|D) \, d^2\boldsymbol{r} \tag{1}$$

$$p_{2+} = \frac{1}{n_t} \iint_A p_{c,2+}(\boldsymbol{r}) \Phi(\boldsymbol{r}|D) \, d^2\boldsymbol{r} \tag{2}$$

and $p_0 = 1 - p_1 - p_{2+}$. $p_1$ and $p_{2+}$. are the fluence averages of the conditional probabilities, $p_{c,1}(\boldsymbol{r})$ and $p_{c,2+}(\boldsymbol{r})$, that a particle trajectory produces a sublethal or a lethal event, respectively, if the primary particle trajectory passes the point given by the position vector $\boldsymbol{r}$. $\Phi(\boldsymbol{r}|D)$ is the dose-dependent area probability density (fluence) for a track passing this point.

The integrals in Eqs. 1 and 2 extend over an area $A$ that is defined by the condition that tracks passing the beam cross section within this area have a nonzero probability of producing lethal or sublethal events in the considered cell.

To avoid the notation becoming too cumbersome, we ignore in Eqs. 1 and 2 that $p_{c,1}(\boldsymbol{r})$ and $p_{c,2+}(\boldsymbol{r})$ also depend on the energy of the ionizing particle producing the track. We also do not consider explicitly that there is a dependence on the direction of motion. (In fact, the probabilities will mainly depend on the impact parameter of the track with respect to the target volume.) Furthermore, it is worth noting that Eqs. 1 and 2 work best for heavy charged particles. In the case of indirectly ionizing particles such as photons, one would have to replace the area integral by an integral over a volume in which photon interactions producing secondary electrons contribute to the induction of lesions in the considered cell.

If sublethal and lethal events are assumed to be related to the formation of DNA double-strand breaks (DSBs) and DSB clusters in subcellular target volumes that are caused by ionization clusters in the particle track (Schneider et al. 2016, 2017, 2019, 2020), the probabilities of the occurrence of these effects may be defined in an analogous way as for the cellular events. In this case, the diameter of the area $A$ may be between several hundreds of nm up to more than a µm larger than the diameter of the considered target volume (Braunroth et al. 2020). This will be further investigated in Section "Nanodosimetry in TET and RAMN".

The probabilities $p_1$ and $p_{2+}$ may be assumed to be almost independent on the absorbed dose $D$, whereas the dose dependence is included in the average number of tracks $n_t$ passing the area $A$ (Eq. 3).

$$n_t(D) = \iint_A \Phi(\boldsymbol{r}|D)\, d^2\boldsymbol{r} \tag{3}$$

It should be noted that $n_t$ is generally not an integer number; it is the expectation of the probability distribution $P_t(n)$ of the number $n$ of tracks passing area $A$ that can produce lethal or sublethal events in the considered cell. For a certain number $n$ of tracks passing $A$, the conditional probability $P_c(n_1, n_{2+}|n)$ for simultaneous induction of $n_1$ sublethal events and $n_{2+}$ lethal events is given by a multinomial distribution (Eq. 4).

$$P_c(n_1, n_{2+}|n) = \frac{n!}{n_0!\, n_1!\, n_{2+}!}\; {p_0}^{n_0} {p_1}^{n_1}\, {p_{2+}}^{n_{2+}} \tag{4}$$

where $p_0 = 1 - p_1 - p_{2+}$ and $n_0 = n - n_1 - n_{2+}$.

The (multi-event) probability $P(n_1, n_{2+})$ for $n_1$ sublethal events and $n_{2+}$ lethal events to be produced is then given by:

$$P(n_1, n_{2+}) = \sum_n P_c(n_1, n_{2+}|n)\, P_t(n) \tag{5}$$

If $P_t(n)$ is a Poisson distribution (with $n_t$ as distribution parameter), $P(n_1, n_{2+})$ is obtained as

$$P(n_1, n_{2+}) = \frac{(n_t p_1)^{n_1} (n_t p_{2+})^{n_{2+}}}{n_1!\, n_{2+}!}\, e^{-n_t(p_1 + p_{2+})} \tag{6}$$

so that the combined (multi-event) probability of $n_1$ sublethal events and $n_{2+}$ lethal events can be written as the product of the marginal distributions that are thus statistically independent and Poisson distributions. In analogy to the single-event case, cell survival occurs if $n_1 \leq 1$ and $n_{2+} = 0$, i.e.,

$$S = (1 + n_t p_1) e^{-n_t(p_1 + p_{2+})}. \tag{7}$$

Defining the parameters $p$ and $q$ as

$$p = \frac{n_t(D) \times p_{2+}}{D} \qquad q = \frac{n_t(D) \times p_1}{D} \tag{8}$$

transforms Eq. 7 into

$$S = (1 + qD) e^{-(p+q)D}\,. \tag{9}$$

Eq. 9 has the functional form of the basic TET model formula (Besserer and Schneider 2015a). It should be noted, however, that the parameters $p$ and $q$ in Eq. 9 are the expected mean numbers of lethal and sublethal events per dose, not the number of subcellular DNA lesions.

The derivation of Eq. 9 did not require presuming the (multi-event) distributions of lethal events and sublethal events to be statistically independent and to be Poisson distributed as was done in previous work (Besserer and Schneider 2015a, 2015b, Schneider et al. 2016, 2017, 2019, 2020). Both properties follow from the assumption of the Poisson distribution of the number of primary tracks interacting with the cell. Therefore, it seems that these two model assumptions are dispensable, at least when considering events at the cellular level.

### *Comparison with the original TET and the RAMN*

The original formulation of the TET (Besserer and Schneider 2015a) suffered from a somewhat unclear terminology. Examples are the confusing use of the term “event” for radiation effects in subcellular targets or the use of the term “TTE” for a track inducing a single sublethal lesion as well as for the occurrence of two tracks inducing sublethal lesions that form a lethal lesion. Furthermore, a TTE in the first sense was identified with a DSB and an OTE with the occurrence of “two lethal DSBs on the same or different chromosomes” (Besserer and Schneider 2015a). Thus, it was unclear whether, for example, three DSBs produced by a single track would be considered as the simultaneous occurrence of an OTE and a TTE or whether this would also count as an OTE.

The mathematical formulation of the model in Besserer and Schneider (2015a) suggests that the case of more than two sublethal lesions was implicitly subsumed when talking about two sublethal lesions. On the other hand, the illustration of the basic interactions considered in the model shown in Fig. 1 of Besserer and Schneider (2015a) suggests that the possibility of more than one track affecting the target volume is considered. At the same time, cases such as a track inducing exactly one or more than two sublethal lesions do not seem to be included.

The conceptional and terminology problems of the original TET seem to have been overcome with the RAMN. In the RAMN, the fundamental model equation relates survival of a cell to the average frequency of occurrence of single or clustered DNA lesions (Schneider et al. 2020). The latter are related to the particle fluence and single-event probabilities of the induction of clustered DSBs within subcellular targets. These subcellular targets were called “lethal interaction volumes” in preliminary attempts to derive the ratio of the TET model parameters $p$ and $q$ (Schneider et al. 2016, 2017) or the absolute parameter values (Schneider et al. 2019) from nanodosimetric parameters of track structure.

Within the RAMN, these (spherical) volumes are called cluster volumes (Schneider et al. 2020). These cluster volumes (CVs) contain an integer number of basic interaction volumes (BIVs). The BIVs have a diameter of 2.5 nm such as to represent a DNA segment of ten base pairs. It is assumed that a DNA lesion in the form of a DSB is induced if at least two ionizations occur within the BIV.

The formalism used in Subsection "Derivation of the fundamental model equation" can also be applied for determining the multi-event frequency distribution of single-track interactions that induce clustered or single DNA lesions in a single nanometric volume. For the case of a (single) subcellular target this approach has also been used in Schneider et al. (2017) and implicitly also in Schneider et al. (2020).

However, there is a (potentially large) number of such subcellular targets. For example, the diameter of the spherical cluster volume best fitting experimental relative biological effectiveness (RBE) data reported in Schneider et al. (2020) for soft X-ray photons was 7.5 nm. Thus, such a volume covers only a small fraction of the volume of the cell nucleus on the order of $2\times10^{-9}$. Of course, one has to consider that DNA accounts for only a small fraction of the mass content in the nucleus and that, in addition, chromatin organization may play a role such that certain regions of the chromosome may be more prone to radiation damage (Schneider et al. 2016). However, even if there were only as few as 50 such sites per chromosome, the total number of CVs in a cell nucleus would be on the order of $10^3$.

Therefore, the question arises how the occurrence of DNA lesions in this large number of subcellular targets relates to the induction of lethal and sublethal events at the level of a cell. In the first publication of the TET, “two lethal DSBs on the same or different chromosomes” was the definition of an OTE, i.e., a lethal event at the cellular level (Besserer and Schneider 2015a). In the RAMN, this was replaced with the occurrence of a cluster of DNA lesions (CL) within a nanometric CV. The procedure used in Schneider et al. (2020) for determining the probability of this happening suggests that only a single CV is considered.

In the RAMN it is explicitly assumed that different CVs have the same probabilities of receiving a CL or a SL and that these probabilities are statistically independent (i.e., the probability of obtaining for example a CL in the second CV does not depend on whether there is a CL in the first CV or not).

If different CVs are assumed to be statistically independent, then the convolution of the Poisson distributions of the (multi-event) frequencies of CLs and SLs in all CVs leads to statistically independent Poisson distributions of the number of CLs and SLs per cell. The assumption that “a cell will survive irradiation if no CL [and] at most one SL occurs” then leads to the model equation (1) in Schneider et al. (2020). This has the same form as our Eq. 7 but slightly modified as follows:

$$S = (1 + N\, n_t\, p_{SL}) e^{-N\, n_t (p_{SL} + p_{CL})} \tag{10}$$

where the parameters $p_{SL}$ and $p_{CL}$ are the probabilities of the induction of an SL and CL, respectively, in a CV when a single track interacts with the cell. $N$ is the number of CVs in the cell and $n_t$ is the dose-dependent number of tracks interacting with the cell.

If the meaning of the parameter σ used by Schneider et al. (2020) is that of a geometrical cross section, Eq. 10 is the same as their Eq. 1.[2]

By adapting the definition of the model parameters in Eq. 8, Eq. 10 transforms again into the fundamental model equation (Eq. 9). The problem is then that the values obtained by Besserer and Schneider (2015a) for the model parameters by fitting to measured survival curves does not corroborate the identification of a sublethal lesion with a single DSB and a lethal lesion with a cluster of DSBs. The values for parameter $q$ shown in Table 1 of Besserer and Schneider (2015a) suggest that around one DSB is induced per Gy of absorbed dose, whereas evidence in radiobiological literature indicates that there are generally on the order of several tens per Gy (Ward 1990).

A potential solution to this dilemma may be to consider only severe lesions in the form of complex DSBs. However, such a distinction of DSBs with respect to their complexity has not been considered in the RAMN (Schneider et al. 2020). A second option could be that only a subset of all possible CVs is relevant for radiation-induced cell killing (Schneider et al. 2016). Then, one could hypothesize that a cell survives irradiation if all critical CVs receive at most one SL (and no CL). However, then a cell will survive with a probability $S$

$$S = (1 + n_t\, p_{SL})^N e^{-N n_t (p_{SL} + p_{CL})} \tag{11}$$

where all parameters have the same meaning as in Eq.10. For a large value of $N$, the first factor on the right-hand side of Eq. 11 can be approximated by Eq. 12.

$$(1 + n_t\, p_{SL})^N \approx e^{N(n_t\, p_{SL} - (n_t\, p_{SL})^2/2)} \tag{12}$$

Similar to Besserer and Schneider (2015a), a second-order Taylor expansion of the logarithm is used here. With this, the survival probability becomes

$$S = e^{-pD - (qD)^2/(2N)} \tag{13}$$

where the notation of Eq. 9 was re-used with the numerators in Eq. 8 replaced by $N\, n_t\, p_{CL}$ and $N\, n_t\, p_{SL}$, respectively. If SLs are identified with single DSBs then the quadratic term is negligible for all practically relevant values of dose. The reason is that the average number of DSBs per Gray in a cell is on the order of a few tens (Ward 1990). If $N$ is the number of possible CVs, i.e., on the order of $5\times10^8$, and if 40 DSBs are produced per Gy, then the quadratic term would be unity for a dose on the order of 500 Gy.

If $N$ is the number of critical CVs as considered in Schneider et al. (2016), the probability that in a cell a DSB is induced in such a CV is reduced by a factor equal to the ratio of $N$ and the number of such possible CVs. Hence, the quadratic term would be smaller by the same factor, as the numerator scales quadratically with this factor. Thus, the quadratic term would be significant only for even higher doses than 500 Gy. Therefore, in the practically relevant dose range up to 80 Gy, the survival curve would be approximately a pure exponential function as for radiation qualities of high linear energy transfer (Goodhead et al. 1993).

[2] The parameter is introduced by Schneider et al. ( 2020) as “intersection-cross-section”, and it is stated that its product with the particle fluence is “probability that a particle track intersects any BIV in the cell nucleus”, which suggests a geometrical interpretation. But a few lines below this quote, it is said that “σ contains all cell specific parameters which affect cell sterilization, as e. g. phase in cell cycle, radioresistance, repopulation and repair capability”, which suggests a completely different meaning.

Therefore, it seems that the assumption of statistical independence of the probabilities of the induction of SLs and CLs in different CVs does not lead to a model function compatible with radiobiological evidence. Furthermore, it should be noted that for single event distributions, the assumption of statistical independence of CLs and SLs in different targets contradicts the definition of a track as a set of statistically correlated energy transfer points. This will be further investigated in Section "Nanodosimetry in TET and RAMN".

# Repair

DNA damage repair has not been explicitly addressed in the previous Section. Similar to the original TET in Besserer and Schneider (2015a), however, the notion of sublethal events implicates that the associated damage is repaired. Repair was explicitly introduced in the TET in a second paper by Besserer and Schneider (2015b). The model assumptions with respect to repair were that

a) if exactly one DSB is induced by the irradiation of the cell, this DSB is always repaired.
b) if exactly two DNA lesions are induced either by one OTE or two TTEs, they are both repaired with a probability $R$.

In the respective model equation derived as Eq. 7 in Besserer and Schneider (2015b), the factor in front of the exponential in Eq. 9 is replaced by a third-order polynomial in the absorbed dose.

Within the framework of (multiple) tracks interacting with a cell that was adopted in Subsection "Derivation of the fundamental model equation", the above model assumptions would translate into assuming that radiation-induced damage is a') always repaired if only one of the tracks interacting with the cell produces a sublethal event while all others are nonlethal events.
b') repaired with a probability $R$ if one track produces a lethal event and all others are nonlethal events or if two tracks are sublethal events and all other tracks are nonlethal events.

A cell survives if the radiation-induced damage is repaired. Using the probabilities $P$ from Eq. 6, the probability $S$ for survival is thus given by

$$S = P(0,0) + P(1,0) + R[P(0,1) + P(2,0)] \tag{14}$$

Using Eqs. 6 and 8 this transforms into

$$S = \left(1 + qD + R\left[pD + \frac{(qD)^2}{2}\right]\right) e^{-(p+q)D}. \tag{15}$$

Equation 15 differs from the model equations used in the TET (Besserer and Schneider 2015b, Schneider et al. 2017, 2019) by the absence of mixed terms (containing $p \times q$) and the absence of a term that is quadratic in the repair probability and cubic in dose.

*Critical observations on the TET model with repair*

The reason why the approach of Besserer and Schneider (2015b) leads to the additional terms that are not appearing in Eq. 15 is that they seem to have implicitly assumed that the frequency distributions of unrepaired DSBs produced by OTEs and TTEs would also be statistically independent if the frequency distributions of OTEs and TTEs are statistically independent.

This assumption is not plausible, however, as the probability of repair should depend on the total number of DSBs produced in the cell and not how they are produced, as long as they are produced by tracks arriving with a time delay much smaller than the time needed for DSB repair. The latter is on the order of tens of minutes (Metzger and Iliakis 1991), so that for therapeutic beams, the DSBs produced by different tracks can be assumed to occur simultaneously.

Therefore, outcomes of the irradiation with the same number of DSBs in the cell should be treated in the same way. From Eqs. 4 and 5 in Besserer and Schneider (2015b), the mixed term (containing the product of $p$ and $q$) corresponds to the case of survival after two tracks interacted with the target volume; one track produces one DSB which is repaired with probability 1 and the other track two DSBs that are both repaired with probability $R$. The term quadratic in $R$ would correspond to three tracks, of which one produces a pair of DSBs that are both repaired with probability $R$ while the other two tracks each produce a single DSB and the two DSBs coming from these two tracks are also repaired with a probability $R$.

From the point of view of DNA damage repair, there are two equivalent situations to the first case (mixed terms), namely one track that produces three DSBs or three tracks that each produce one. Similarly, the quadratic term involves four DSBs which would also be obtained by (a) one track producing four DSBs, (b) one track producing three DSBs and a second track producing one DSB, (c) two tracks producing two DSBs or (d) four tracks each producing one DSB. Hence, all these cases would have to be considered as well. However, this would require the respective probabilities to be used as further parameters of the model.

*Consistent DSB-based repair model*

To avoid a "Ptolemaic" model with too many parameters, the pragmatic approach taken by Besserer and Schneider (2015b) to assume that up to two DSBs can be repaired and to use only one model parameter for the repair of exactly two DSBs seems advisable. However, the correct functional form of the model curve for such an assumption is different from Eq. 7 in Besserer and Schneider (2015b) and from Eq. 15 above.

The reason for this is that there is implicitly another model assumption involved regarding the relation between the conditions for lethality of events (i.e., tracks interacting with a cell) and the number of DSBs produced by such tracks. In the work of Besserer and Schneider (2015b), the fate of a cell in which a single track produces more than two DNA lesions has not been explicitly addressed. From their Fig. 1 one may infer that if a track induces four or more DSBs, the cell is killed.[3] However, if a cell is killed when a track induces four or more DSBs, this implies that one has to consider four categories of events in the repair model (see Fig. 1): nonlethal, sublethal, potentially lethal (i.e. lethal if not repaired), and definitely lethal events.[4]

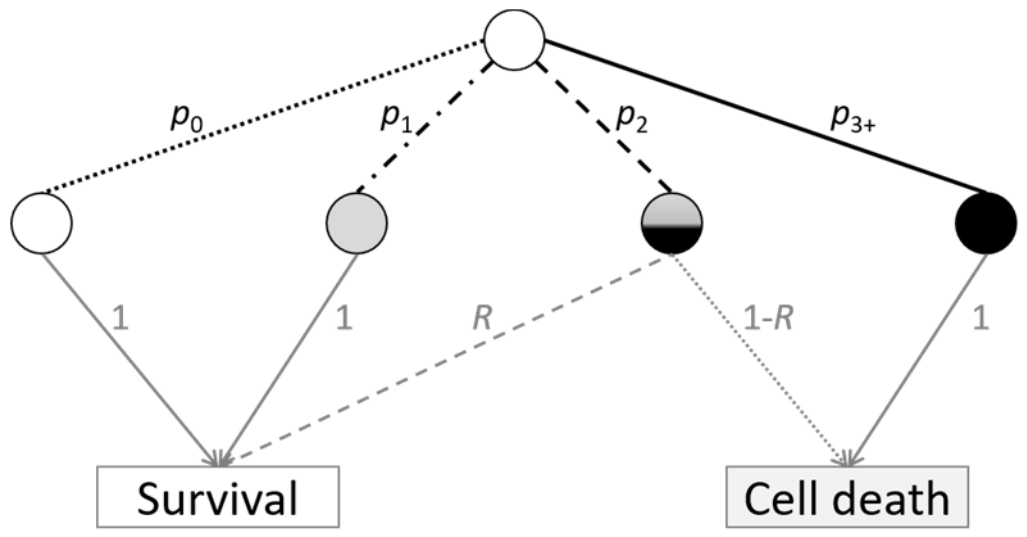

**Fig. 1** Illustration of the fate of a cell interacting with a single track. The upper open circle symbolizes the cell prior to the radiation interaction. The interaction with the track may be a nonlethal event (dotted line), a sublethal event (dot-dashed line), a potentially lethal event (dashed line), or a definitely lethal event (solid line). In the first case, the cell remains in an essentially unaltered state (open circle) and survives. The second case leads to a cell with a sublethal damage (gray circle) that is repaired with 100% probability (solid gray line). A cell with potentially lethal damage (circle filled half with gray and half with black) has a probability of surviving if the radiation damage is repaired (dashed gray line) and otherwise dies (dotted gray line). A cell with damage from a definitely lethal event dies at 100% probability.

If the induction of potentially and definitely lethal events occurs with average probabilities $p_2$ and $p_{3+}$, respectively, the conditional probability $P_c(n_1, n_2, n_{3+}|n)$ for simultaneous occurrence of $n_1$, $n_2$, and $n_{3+}$ tracks inducing sublethal, potentially lethal, and definitely lethal events in the considered cell is given by:

$$P_c(n_1, n_2, n_{3+}|n) = \frac{n!\ {p_0}^{n_0}{p_1}^{n_1}\ {p_2}^{n_2}\ {p_{3+}}^{n_{3+}}}{n_0!\,n_1!\,n_2!\,n_{3+}!} \qquad (16)$$

Weighting with the Poisson distribution of the number of tracks leads to the probability distribution $P'(n_1, n_2, n_{3+})$ as given in Eq. 17.

$$P'(n_1,n_2,n_{3+}) = \frac{(n_t p_1)^{n_1}(n_t p_2)^{n_2}(n_t p_{3+})^{n_{3+}}}{n_1!\,n_2!\ n_{3+}!}e^{-n_t(p_1+p_2+p_{3+})}\,. \qquad (17)$$

If potentially lethal events are repaired with probability $R$, a cell survives with probability $S$ given by

$$S = P'(0,0,0) + P'(1,0,0) + R[P'(0,1,0) + P'(2,0,0)] \quad (18)$$

Using Eq. 8 this transforms into

$$S = \left(1 + qD + R\left[p'D + \frac{(qD)^2}{2}\right]\right)e^{-(p+q)D} \qquad (19)$$

where $p'$ is a fourth model parameter which is related to the probability that a track produces a potentially lethal event:

$$p' = \frac{n_t(D) \times p_2}{D}\,. \qquad (20)$$

The respective cell fate for the case of exactly two tracks is schematically illustrated in Fig. 2. The third row of circles shows the possible results of the interactions of the two tracks in the cell before repair. The possible results are a cell with nonlethal (open circle), sublethal (gray circles), potentially lethal (half gray and half black circles) and definitely lethal (black circles) damage. The solid gray lines indicate 100% repair probability, the dashed gray lines indicate repair with probability $R$, and the dotted lines repair failure with probability (1-$R$).

Figure 2 can also be seen as an illustration for more than two tracks interacting with the cell, if the second row of symbols is interpreted as the cell damage produced by all previous tracks where equivalent cases have been combined.

Furthermore, Figs. 1 and 2 can also be used as illustrations of the repair model that derives from the model assumptions made by Besserer and Schneider (2015b), if one distinguishes between tracks producing exactly two DSBs and those that produce three or more DSBs and assumes that the latter case is a definitely lethal event. Alternatively, one may assume that a definitely lethal event requires four or more DSBs induced by a track. Then the probability $p_2$ would refer to two or three DSBs produced and $p_{3+}$ to four or more DSBs. In both cases, however, the correct model equation is Eq. 19 and not Eq. 7 given by Besserer and Schneider (2015b).

Only if, in contrast to the illustration in Fig. 1 of Besserer and Schneider (2015b), one excludes that a single track can induce a definitely lethal event are the parameters $p'$ and $p$ identical and the model has only three parameters. Furthermore, the term quadratic in $R$ in Eq. 7 of Besserer and Schneider (2015b) would only appear if damage from different tracks was repaired independently. As repair occurs at a much longer timescale than the production of the damage by the different tracks, there will not be quadratic terms in $R$. In summary, the considerations in this Subsection mean that the model equations for the second version of the TET used in Besserer and Schneider (2015b) and Schneider et al. (2017, 2019) are incompatible with the model assumptions.

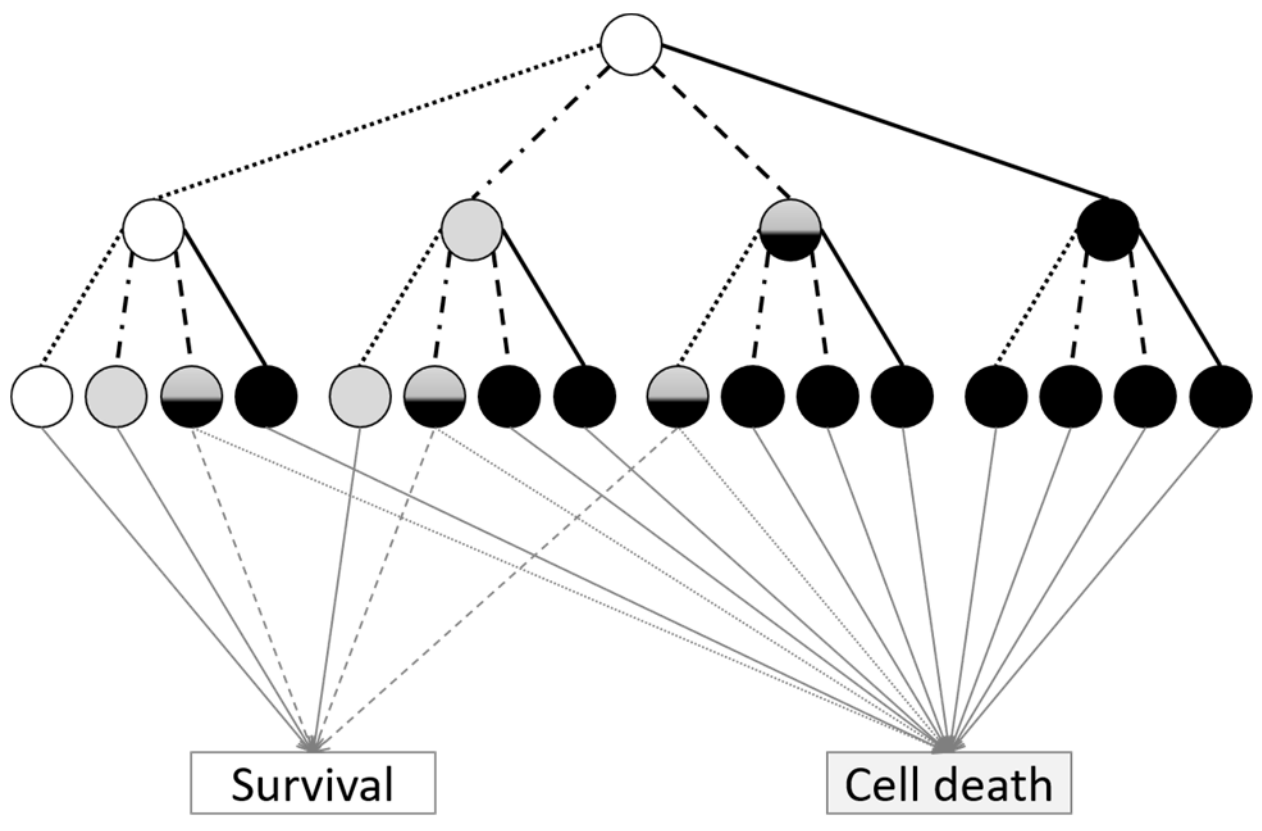

**Fig. 2** Illustration of the outcome when two tracks interact with a cell. The meanings of the symbols and lines are the same as in Fig. 1.

[3] The case of three DSBs produced cannot be inferred and may have been assumed to equal the simultaneous occurrence of a DSB pair that is repaired with probability R and a single DSB that is repaired with probability 1.

[4] Assuming a repair probability implies that the considered events are no longer lethal but only potentially lethal. As there may be radiation-induced damage of different complexity, the possibility of definitely lethal events that cannot be repaired appears plausible.

*Repair model used in the RAMN*

The treatment of repair in the RAMN appears a bit confusing at first glance. The double definition of the model parameter $\sigma$ suggests that the number of CLs and SLs appearing in model Eq. 1 of Schneider et al. (2020) are the number of lesions after repair. This is further suggested by the use of a "persistence parameter" appearing in the expression for the number of SLs that is determined in the appendix of that paper as the ratio of the frequencies of unrepaired SLs and CLs. On the other hand, in the investigation of the dose-rate dependence of cell survival, a repair factor $R$ was introduced that affects the probability of SL formation (Eq. 11 in Schneider et al. (2020)).

In any case, the fundamental model equation of the RAMN appears to be based on the implicit assumptions that

a) a cell survives if there is no unrepaired CL and at most one unrepaired SL.
b) SLs and CLs are repaired independently with constant probabilities $R_1$ and $R_{2+}$, respectively.

The ratio of the two complementary probabilities, $(1-R_1)$ and $(1-R_{2+})$, is the "persistence parameter" in the terminology used by Schneider et al. (2020). The wording of assumption a) above differs from Schneider et al. (2020) in that the condition is not referring to the occurring CLs and SLs, but to the persistent CLs and SLs after repair.

If assumption b) applies and if $P(n_1, n_{2+})$ is the (multi-event) probability of the induction of $n_1$ SLs and $n_{2+}$ CLs, then the distribution $P^*(k_1, k_{2+})$ of the numbers $k_1$ and $k_{2+}$ of unrepaired SLs and CLs, respectively, is given by

$$P^*(k_1, k_{2+}) = (1-R_1)^{k_1}(1-R_{2+})^{k_{2+}} \times \sum_{n_1=k_1}^{\infty}\sum_{n_2=k_2}^{\infty}\binom{n_1}{k_1}R_1^{\,n_1-k_1}\binom{n_2}{k_2}R_{2+}^{\,n_{2+}-k_{2+}}\;P(n_1,n_{2+}). \quad (21)$$

From Eq.21, it is evident that if the distributions of induced SLs and CLs are statistically independent, i.e., $P(n_1, n_{2+}) = P_1(n_1)\,P_{2+}(n_{2+})$, then the same is also true for the distributions of persistent SLs and CLs, whether the marginal distributions $P_1(n_1)$ and $P_{2+}(n_{2+})$ are Poisson distributed or not. If they are Poisson distributed, this is also the case for the distributions of $k_1$ and $k_{2+}$.

However, it is important to note that the statistical independence and Poisson distributions for lesions in cells or subcellular targets found in Subsection "Derivation of the fundamental model equation" does not warrant that the distributions of CLs and SLs in a cell also have these properties. The reason is that there is more than one target volume involved and that the statistical independence between different target volumes cannot be inferred from the statistical independence of the tracks interacting with a cell. To assess the relation between distributions of CLs and SLs and those of tracks requires the single event distributions of CLs and SLs to be considered, which brings nanodosimetry into play (cf. Section "Nanodosimetry in TET and RAMN").

*An alternative repair model*

It is plausible that the repair capacity of a cell is limited so that for a large number of DSBs the average probability of an individual DSB to be repaired will decrease. However, it seems rather implausible that this should already be the case for three (or four) DSBs in a cell. In radiobiological assays, often a large number of DSB repair foci are observed (MacPhail et al. 2003, Ponomarev and Cucinotta 2006, Ponomarev et al. 2008, Martin et al. 2013). Hence, it might have been more appropriate to rather assume in the model a constant probability of the repair of an individual DSB. Deriving a respective model equation becomes very intricate, however, as an analytical treatment of this case would require knowledge of all probabilities $p_k$ for induction of $k$ DSBs by a single track.

As this would make the model rather complex, an alternative simple repair model would be to assume that repair with probability $R$ occurs whenever there is more than one DSB. Then the probability of cell survival $S'$ would be given by Eq. 22.

$$S' = (1+qD)e^{-(p+q)D} + R\left[1-(1+qD)e^{-(p+q)D}\right] \quad (22)$$

The trivial reason is that the first term of the sum is the probability that at maximum one DSB is produced so that the term in the square brackets is the probability of more than one DSB. If the other two model parameters can be determined from nanodosimetry, this model Eq. 22 has only one free parameter.

## Nanodosimetry in TET and RAMN

In further work, Schneider et al. elaborated an approach to derive the ratio of model parameters $p$ and $q$ (Schneider et al. 2016, 2017) or the absolute parameter values (Schneider et al. 2019, 2020) from nanodosimetric parameters of track structure. To determine the absolute values of the parameters, they added the following model assumptions:

a) Existence of subcellular target volumes of identical size within which the induction of two or more (unrepaired) DSBs leads to cell death.
b) Such a target volume contains a number of "basic interaction volumes" (BIVs) in which a DSB is produced with a probability equal to the nanodosimetric parameter $F_2$, i.e., the probability of two or more ionizations within that BIV.

The BIVs are assumed to be spheres enclosing a short strand of DNA of five to ten base pairs (Schneider et al. 2019). The sphere diameter was assumed to be 2.0 nm (Schneider et al. 2019) or 2.5 nm (Schneider et al. 2020). The nanometer-sized spherical volumes from assumption a) were named "lethal interaction volume" in Schneider et al. (2019) and "cluster volume" (CV) in Schneider et al. (2020). The size of the CV was assumed to depend on radiation quality (Schneider et al. 2019).

Based on the two aforementioned assumptions, the probabilities of OTEs and TTEs (within the TET) and of CLs and SLs (in the RAMN) were then derived by binomial statistics. These probabilities were finally used to obtain an expression for RBE (Schneider et al. 2019, 2020).

*Issues with the TET's and RAMN's link to nanodosimetry*

The preliminary attempt to link the track-event theory with nanodosimetry presented in Schneider et al. (2019) had its deficiencies that have been healed in the RAMN where a similar approach as presented in Subsection "Derivation of the fundamental model equation" was used in which the probabilities of the occurrence of CLs and SLs are given by multiplications of three factors. One is the fluence $\phi$, while another one is the respective conditional probabilities, $P_{CL}$ and $P_{SL}$, for the induction of these lesions in a CV (Schneider et al. 2020). The third factor is the model parameter $\sigma$, which is defined ambiguously, but appears to be meant as the product of the geometrical cross section of a BIV and the probability of a CL not being repaired. Thus, Eqs. 2 and 3 of Schneider et al. (2020) could be rewritten as

$$\overline{CL} = (1 - R_{2+}) \times \phi \times \sigma \times P_{CL} \tag{23}$$

$$\overline{SL} = (1 - R_1) \times \phi \times \sigma \times P_{SL} \,. \tag{24}$$

Equations 23 and 24 are expressions of a form that would also be obtained by inserting Eq. 6 in Eq. 21 and then calculating the mean numbers of persistent SLs and CLs. The difference would be that the number of contributing tracks would relate to a cross-sectional area that is potentially much larger than the cross section of a BIV (cf. Subsection "Probability of inducing an IC in a BIV by proton tracks"). Even if only tracks passing the target region mattered, $\sigma$ would be the cross section of the CV and not of the BIV.

A second issue with the approach used by Schneider et al. (2019, 2020) to derive the model parameters from nanodosimetry is the assumed one-to-one correspondence between DSBs and the formation of ionization clusters.[5] While this has also been hypothesized in other work (Grosswendt 2005, 2006), comparisons with dedicated radiobiological experiments in work by Garty et al. showed the relation between ionization clusters and DSBs to require the use of a (one-parameter) combinatorial model (Garty et al. 2006, 2010). This was later demonstrated to imply the one-to-one correspondence between the probability of two or more ionizations to apply only approximately and only for low-LET radiation (Nettelbeck and Rabus 2011, Rabus and Nettelbeck 2011). Conte et al. (2017, 2018) and Selva et al. (2019) demonstrated that a link between nanodosimetry and cell survival can be based on cumulative probabilities of ionization clusters, if in addition to $F_2$ also the probability of clusters with three or more ionizations, $F_3$, is included in the model.

However, even if the assumption holds that a DSB in a BIV occurs with the same probability $F_2$ as an ionization cluster is formed by a passing track in this BIV, a further issue arises: the derivation of the parameters $P_{CL}$ and $P_{SL}$ in Schneider et al. (2019, 2020) ignores the fact that $F_2$, $P_{CL}$ and $P_{SL}$ are all conditional probabilities. They all relate to the occurrence of the respective radiation effect if a track interacts with the considered target.

If the cross section of the CV and of the BIV is taken as the area used in Eq. 3, the respective mean number of tracks, $n_t$, interacting with the CV or BIV is very small compared to unity and can be interpreted as the probability of a track interacting with the target. The probability of the formation of a DSB in any BIV within the CV is then $n_t \times F_2$. The total probability of an SL and a CL is then given by the right-hand sides of Eqs. 4 and 5 in Schneider et al. (2020) but with $F_2$ replaced by $n_t \times F_2$. The conditional probabilities are then obtained by dividing with $n_t$, so that the correct expressions for $P_{CL}$ and $P_{SL}$ are as follows:

$$P_{SL} = F_2 \times n \times (1 - n_t F_2)^{n-1} \tag{25}$$

$$P_{CL} = \frac{1 - (1 - n_t F_2)^n}{n_t} - F_2 \times n \times (1 - n_t F_2)^{n-1} \tag{26}$$

where $n$ is the number of BIVs traversed by a track intersecting the CV. As $n_t$ is small compared to unity, one can use an expansion of the binomials and discard terms quadratic in $n_t$:

$$P_{SL} \approx F_2 \times n \tag{27}$$

$$P_{CL} \approx n_t \times n \times (n - 1) \times {F_2}^2. \tag{28}$$

Therefore, the magnitude of $P_{CL}$ derived in this way depends on both the number of BIVs per CV (or per mean chord length through the CV) and the cross-sectional area considered in determination of $n_t$. This leads to a further potential issue which is related to the determination of the nanodosimetric parameter $F_2$ from track structure simulations, where the illustrations in Fig. 1 of Schneider et al. (2019) and Fig. 1 of Schneider et al. (2020) suggest that only a central passage of the primary particle through a BIV is considered. This conjecture is corroborated by the number of BIVs in a CV used in the binomial, namely the ratio of the mean chord length in the CV and the BIV diameter.

In the work of Schneider et al. (2020), the simulations were performed for secondary electrons from photon irradiation taking into account the spectral fluence of the electrons. The electron fluence can be expected to be isotropic, so that normal incidence to the BIV surface can be assumed. For determining the probability of CLs, however, it would be better to perform the simulations with the electrons impinging on the surface of a sphere (of diameter equal to a CV) and to score ionizations in all BIVs within this sphere, not only those aligned along the initial direction of motion.

If heavy charged particles (protons, ions) are considered, as was the case in Schneider et al. (2019), one has to take into account that a significant proportion of ionization clusters are produced at radial distances of several tens to several hundreds of nm from the primary particle trajectory (Braunroth et al. 2020, Rabus et al. 2020). For determining the fluence-averaged probabilities of CLs and SLs in a CV, a better assumption would thus be that all BIVs in a CV have the same probability of receiving an ionization cluster. The importance of heavy charged particle tracks with large impact parameters is demonstrated in the following Subsections.

*Probability of inducing an IC in a BIV by proton tracks*

In this Subsection results are presented for single-event and multi-event averages of the nanodosimetric parameter $F_2$ for induction of an ionization cluster (IC) in a BIV by passing proton tracks. The methodology used is described in detail in Supplement 1.

In brief, it is assumed that the probabilities of IC formation in different sites are statistically independent and that the dependence of the probability of the formation of an IC in a site, $F_2(r)$, on the impact parameter $r$ of the primary particle trajectory with respect to the center of the site is known.[6] Spherical sites are considered that are located within a spherical region of interest (ROI) with radius $R_L$. The primary particle trajectory is assumed to pass the ROI within an annulus (see Supplementary Fig. 1) whose inner and outer radii are successive integer multiples of $R_L$.

[5] In the terminology of nanodosimetry, the ionization cluster size is the number of ionizations in a considered target volume, which may also take the values zero or one. As only two or more ionizations constitute a cluster of ionizations, term ionization cluster is used here for this case only.

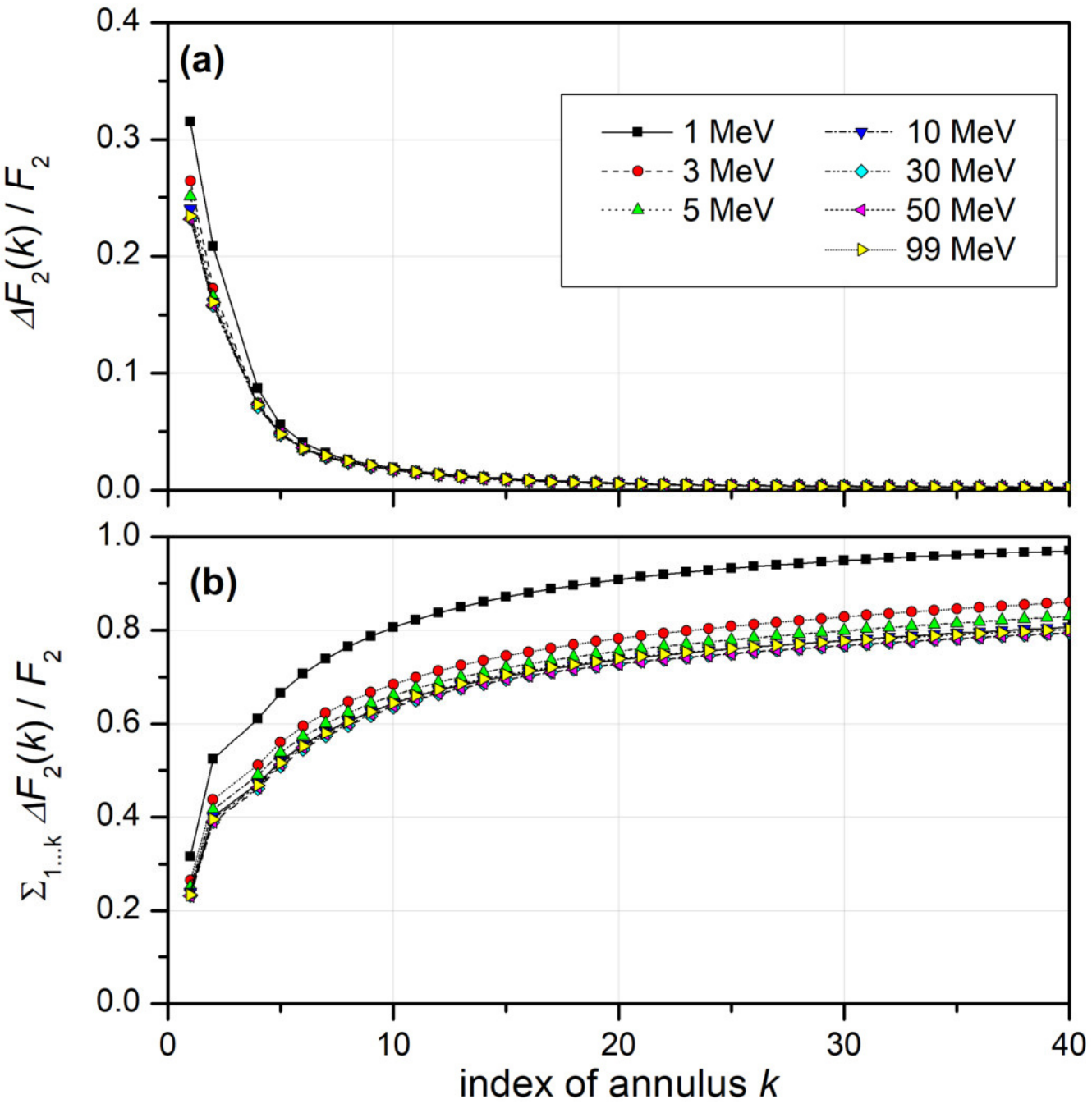

**Fig. 3**: (a) Relative contribution to the total probability $F_2$ of obtaining an ionization cluster in a BIV of 3 nm from a proton of the energies given in the legend that passes through the $k$-th annulus around the BIV or through the BIV ($k$ = 1). (b) Relative contribution from protons passing the first $k$ annuli around a BIV to the total probability of obtaining an ionization cluster. (BIV - basic interaction volume; $F_2$ - probability for induction of an ionization cluster. For details see text.)

For determining the probability of induction of an IC in a BIV by proton tracks, the ROI was chosen identical to the site and the site diameter was chosen as 3 nm to have a volume identical to the cylindrical targets used in the analysis of simulated proton tracks by Braunroth et al. (2020). The results for the contributions of the different annuli to the total probability $F_2$ are shown in Fig. 3(a) for a number of proton energies, and Fig. 3(b) shows the respective cumulative contributions. While protons traversing the BIV have the highest contribution to the total probability of inducing an IC in the BIV, about 70% to 75% of the probability $F_2$ is due to protons passing the BIV for the considered BIV size of 3 nm diameter. It is to be expected that for smaller BIVs this contribution is even higher.

With the exception of the lowest considered energy of 1 MeV, the contributions of the different annuli are almost independent of energy, and convergence of the cumulative distribution is relatively slow. For an energy of 3 MeV or higher the relative cumulative contribution is seen in Fig. 3(b) to be below 80% up to the maximum annulus index of 40, which corresponds to an outer radius 60 nm in this case.

Thus, determining the value of $F_2$ from simulations where the primary particle traverses the BIV is problematic in two respects. One is that considering only traversing tracks leads to a significant underestimation of the actual value that would be obtained in a real broad-beam irradiation. The other is that the values obtained from such simulations are only conditional probabilities and need to be corrected for the probability of such a primary particle traversal to occur.

For a fluence value estimated from the ratio of an absorbed dose of 2 Gy and the mass stopping power of protons[7], the total probability of the formation of an IC in a particular BIV is between $1.5\times10^{-6}$ and $1.4\times10^{-5}$ (depending on proton energy). These values suggest that the probability of simultaneous occurrence of several BIVs within a CV should be negligibly small.

*Frequency of BIVs inside a CV receiving an IC by protons*

To determine the mean number of sites within a CV that receive an IC from protons passing an annulus around the CV, a ROI diameter of 18.0 nm was chosen that contains the same number of BIVs as the CVs reported by Schneider et al. (2019) for protons. The results are shown in Fig. 4(a) as a function of the annulus index $k$ (ratio of outer radius and $R_L$). The values shown apply to a single event, i.e., one proton passing the cross section of a spherical cell nucleus of 6 µm diameter. It can be seen that the expected number of BIVs with ICs produced by a single event in the considered CV decrease with increasing proton energy and also with increasing annulus index. For the 1 MeV data the decrease with the annulus index is much more pronounced. This can be explained by the smaller energy transfer to the secondary electrons. It should be noted that the maximum annulus index shown corresponds to a maximum impact parameter of the proton track of 180 nm.

Fig. 4(b) shows the respective multi-event values of the probability $p_1$ that exactly one site within the CV receives an IC for a proton fluence corresponding to an absorbed dose of 2 Gy, i.e., for a typical treatment fraction in radiation therapy. The maximal values are in the $10^{-4}$ range so that they approximate well the mean number of sites with ICs for Poisson and binomial distributions. The probability $p_{2+}$ of two or more sites in the CV receiving an IC, i.e., that a CL is produced, is shown in Fig. 4(c). These probabilities are on the order of $10^{-8}$ or lower and have been calculated assuming Poisson statistics, but using a binomial distribution would give practically the same values.

The dependence on proton energy is less pronounced in Figs. 4(b) and 4(c) as compared to Fig. 4(a), because the fluence corresponding to a value of dose increases with increasing proton energy (at least for energies above the Bragg peak energy of around 80 keV). The relative dependence on the annulus index is naturally the same as in Fig. 4(a) for the probability $p_1$, whereas a much stronger decrease with increasing annulus index is observed for probability $p_{2+}$. This is expected as ICs formed in different BIVs are assumed to be statistically independent, so that the probability of two or more ICs should be approximately equal to the square of the probability of a single IC if the latter probability is small, as is seen in Fig. 4(b).

[6] This impact parameter is equal to the magnitude of the position vector $\boldsymbol{r}$ in Eqs. 1 to 3 if the center of the target volume is chosen as the origin of the coordinate system and lies on a plane perpendicular to the primary particle trajectory that contains area $A$.

[7] The resulting fluence values are between $4.8\times10^{-7}$ nm$^{-2}$ and $1.7\times10^{-5}$ nm$^{-2}$ for the proton energies considered.

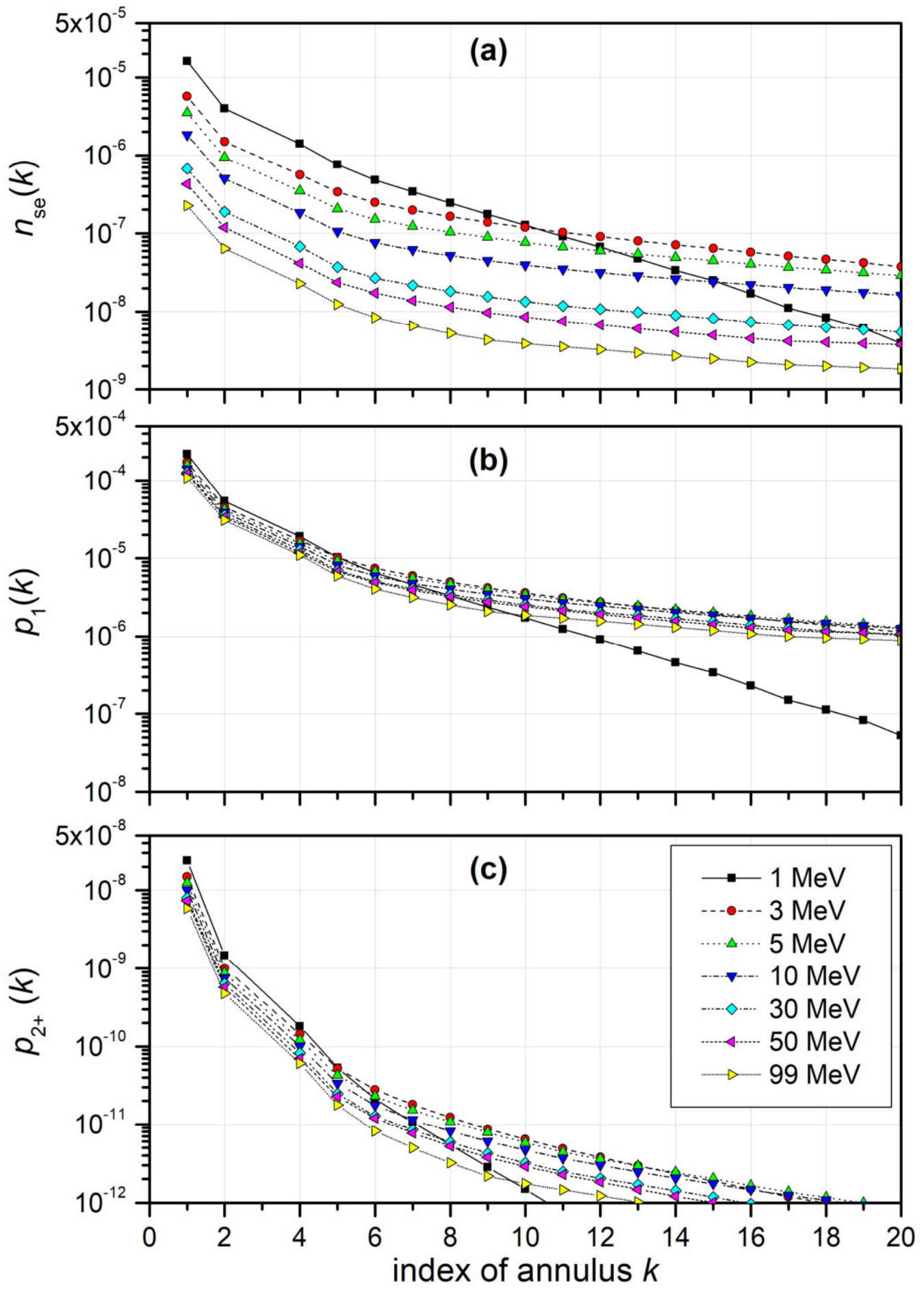

**Fig. 4**: (a) Mean number of BIVs of 3.0 nm diameter inside a CV of 18 nm diameter that receive an ionization cluster when a proton of the energies given in the legend passes through the $k$-th annulus around the CV or through the CV ($k$ = 1). The data correspond to a single event, i.e., a fluence of one proton per cross section of a cell nucleus (assumed to have 6 µm diameter). (b) Probabilities that exactly one site in the CV receives an ionization cluster when a proton passes through the $k$-th annulus for an absorbed dose of 2 Gy. (c) Corresponding probabilities of two or more sites in the CV receiving an ionization cluster. cluster. (BIV - basic interaction volume; CV - cluster volume; for details see text.)

The pronounced decrease with the annulus index seen in all panels of Fig. 4 implies that the cumulative probabilities converge fast with increasing annulus index (see Supplementary Fig. S4). Therefore, it seems that despite the large proportion of ICs formed at large radial distances seen in Supplementary Fig. S3(c), the probability of the formation of two or more ICs in BIVs within a CV (of the sizes used in the present analysis) is mostly determined by proton tracks passing through the CV with a small minor additional contribution from the first real annulus (with outer radius of twice the CV radius). These two regions of impact parameters also account for more than about 80% of the probability of a single IC within the CV. This suggests that, depending on the accuracy aspired, it may be sufficient to consider tracks with impact parameters up to a few times the CV radius when determining the numbers of single and multiple ICs in a CV (SLs and CLs).

It is important to note, however, that there is several orders of magnitude difference between the values of $p_1$ and $p_{2+}$ seen in Figs. 4(b) and 4(c). This is at variance with the results obtained in the approach of Schneider et al. (2019, 2020) and it also does not seem to be compatible with the values reported earlier for the TET model parameters (Besserer and Schneider 2015a). This indicates that the assumption of statistical independence of the probabilities of IC formation in different targets is not only conceptionally at variance with the definition of tracks and in contradiction to recent experimental evidence for correlated IC formation in adjacent sites (Pietrzak et al. 2018, Hilgers and Rabus 2020), but also leads to a severe underestimation of the probabilities of clusters of ICs (i.e., CLs).

## Outline of a tentative approach to consider track structure in the TET and RAMN

The small absolute values of the probabilities found in Section "Nanodosimetry in TET and RAMN" are due to the fact that fluence averaging has been performed for a single site, where the geometrical relation with the track is generally not known. On the other hand, a track traversing a cell will also traverse or closely pass by some of the sites in the cell nucleus. These close encounters correspond to a locally high value of fluence which, in turn, results in much higher probabilities of the induction of single or multiple ICs within the affected CVs.

Capturing this stochastic process requires a paradigm shift for nanodosimetry that was first proposed by Selva et al. (2018). The further elaboration of these ideas by Braunroth et al. (2020), Rabus et al. (2020), and Rabus (2020) that was used in Section "Nanodosimetry in TET and RAMN" essentially considered amorphous tracks. This Section gives an outline how this paradigm shift for nanodosimetry could be used for the purposes of the TET and RAMN.

### *Nanodosimetry of track structure at the micrometer level*

For this purpose, the track structure simulation data from Rabus et al. (2020) were analyzed using a development of the methods used by Braunroth et al. (2020) for scoring ICs in the penumbra. In this new approach, a full segmentation of three-dimensional space was performed using the Wigner-Seitz cells of a face-centered cubic Bravais lattice for scoring. A face-centered cubic lattice has a coordination number of 12; its Wigner-Seitz cell is a rhombic dodecahedron which may be considered a reasonable approximation of a sphere.

The scoring approach was used twice. In the first pass, the number of ionizations in the Wigner-Seitz cells were scored. The Bravais lattice constant was chosen such that the volume of the Wigner-Seitz cells was the same as of a sphere of either 2.0 nm, 2.5 nm, or 3.0 nm diameter. The first two dimensions correspond to the BIV sizes assumed in the publications of Schneider et al. (2019, 2020). The third one is the sphere diameter used in Subsection "Probability of inducing an IC in a BIV by proton tracks", i.e., of the same volume as the cylindrical targets used by Rabus et al. (2020) and Braunroth et al. (2020).

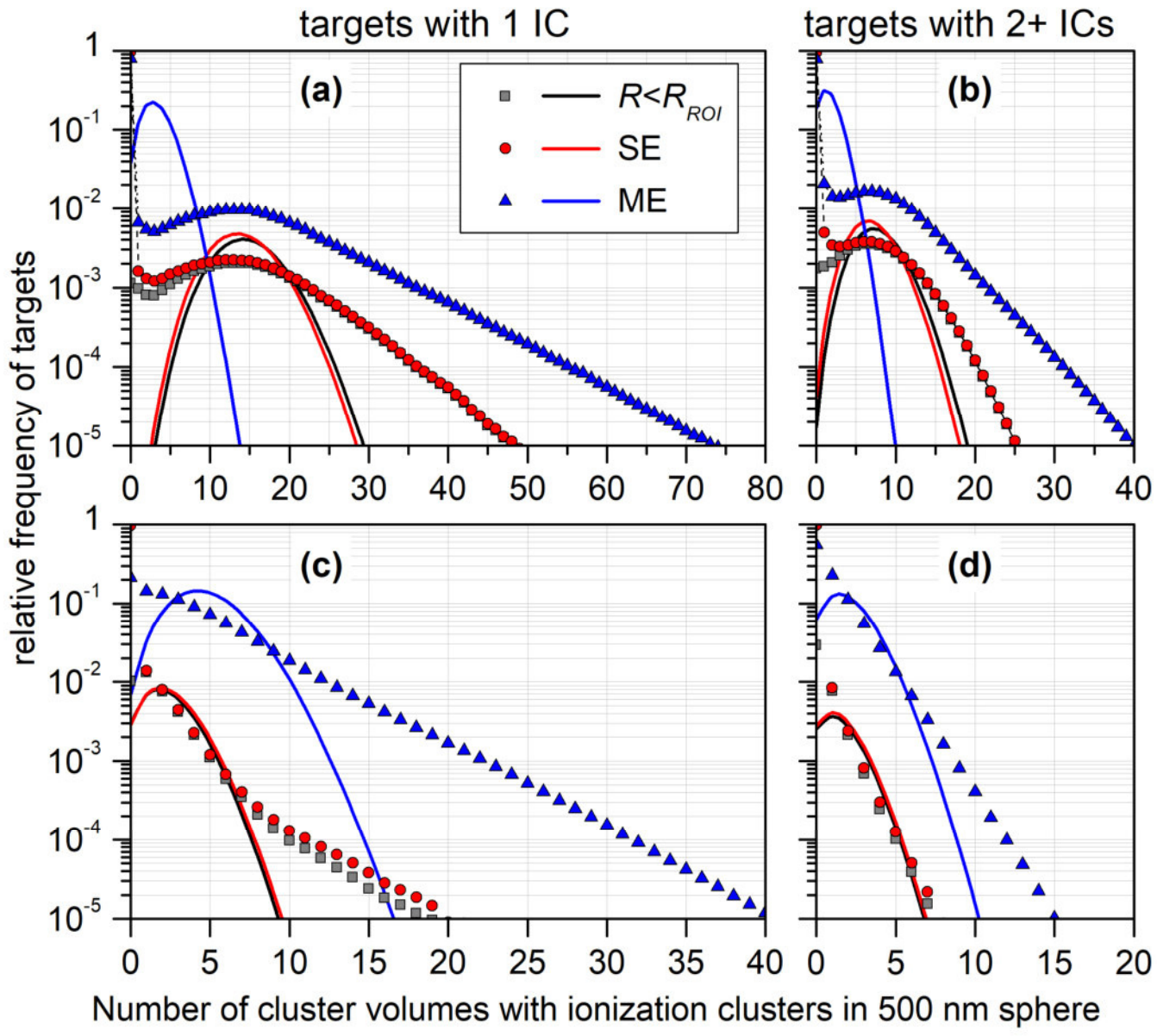

**Fig. 5**: Single-event (SE) and multi-event (ME) distributions of the number of cluster volumes inside a spherical region of interest (ROI) with radius $R_{ROI}$ = 250 nm that receive a single ionization cluster (IC) (left column) or two or more ionization clusters (right column) from proton tracks. In (a) and (b), the proton energy is 3 MeV and in (c) and (d) 50 MeV. The spherical cluster volume has 12 nm diameter and the spherical sites 2 nm. The squares indicate the contribution to the SE distribution from tracks intersecting the ROI. The solid lines represent Poisson distributions of the same mean value as the corresponding data marked by symbols when the data point at 0 is omitted. For details see text.

When an IC was found within a Wigner-Seitz cell, the center of gravity of the ionization points in that cell was taken as the position of the IC. In the second pass, the number of ICs was scored within larger cells which had the same volume as spheres of either 12 nm, 7.5 nm, or 18 nm diameter. The first two values correspond to the CV diameters used by Schneider et al. (2019, 2020). The last value is the one used in Subsection "Frequency of BIVs inside a CV receiving an IC by protons".

The outcome of this scoring were the relative positions with respect to the proton trajectory of CVs in which either a single or multiple ICs were found. In the next step, ROIs in the form of large spheres were placed at different radial distances from the primary particle trajectory and the numbers of CVs with single and multiple ICs inside the ROIs were scored.

The positions of the ROIs with respect to the primary particle trajectories were the centers of cylinder shell sectors around the primary particle trajectory similar to those used by Braunroth et al. (2020). Thus, a segmentation of the ROI's cross section is obtained that allows the integrals in Eqs. 1 and 2 to be calculated by deterministic sampling. To also account for contributions from primary particle trajectories passing the ROI without intersection, radial distances up to five times the radius of the ROI cross section were included.

Single-event distributions of CVs with single and multiple ICs were determined for spherical ROIs of 500 nm diameter. The restriction in ROI size was imposed by the fact that the simulated proton tracks covered a path length of only 650 nm (Braunroth et al. 2020, Rabus et al. 2020). (The first 100 nm and the distal 50 nm of the track were not used in the analysis.)

Multi-event distributions were obtained by calculating the weighted sum of $n$-fold convolutions of the single-event distributions using the probability of $n$ tracks interacting with the ROI as weights. This probability was calculated from Poisson statistics using a primary particle fluence corresponding to a dose of 2 Gy. Results are shown in Fig. 5 as well as in Supplementary Figs. S5 to S8. In Fig. 5, results are shown for a BIV of 2 nm and a CV of 12 nm diameter. The top and bottom panels correspond to proton energies of 3 MeV and 50 MeV, respectively. The panels on the left-hand and right-hand sides show the frequencies of cluster volumes with exactly one and more than one IC, respectively. The red circles correspond to the single-event distributions and the blue triangles to the multi-event distributions. The gray squares show the contribution to the single-event frequency coming from proton tracks traversing the ROI. The solid lines represent Poisson distributions with a distribution parameter equal to the mean number of targets obtained for the corresponding data set.

Figures 5(a) and 5(c) show that the frequency distribution of CVs with a single IC has a shape that does not resemble the Poisson distributions obtained using the mean values as Poisson parameter (solid lines). In contrast, the single-event distribution of CVs with more than one IC has some similarity with the respective Poisson distribution, but for the multi-event distributions a non-Poisson shape is observed again. These findings are corroborated by Supplementary Figs. S5 and S6, which show comparisons of the results obtained for 3 MeV and 50 MeV protons, respectively, with the three choices of BIV and CV dimensions. As can further be seen in Supplementary Figs. 7 and 8, also for single tracks with a defined impact parameter, the distributions of CVs with exactly one IC are not well described by Poisson distributions. For single tracks traversing the ROI, a Poisson distribution is an approximation for the distribution of CVs with multiple ICs, but with a tail at the right-hand side of the peak that seems to become more pronounced with increasing impact parameter.

To further investigate whether the distributions of CVs with single or multiple ICs are statistically independent, the bivariate distributions of the frequencies of CVs with one and more than one IC have also been sampled. Results for the cases of 3 MeV and 50 MeV proton energy are shown in Fig. 6. The $z$-axis is the ratio of the frequency for simultaneous occurrence of a certain number of CVs with one ($x$-axis) and with more than one IC ($y$-axis) to the product of the marginal probabilities of observing the respective number of CVs, i.e., the data shown in Fig. 5. Statistical independence of the induction of CVs with exactly one or with two or more ICs would be confirmed if this ratio plotted on the *z*-axis has values around unity. However, this is not observed in Fig. 6.

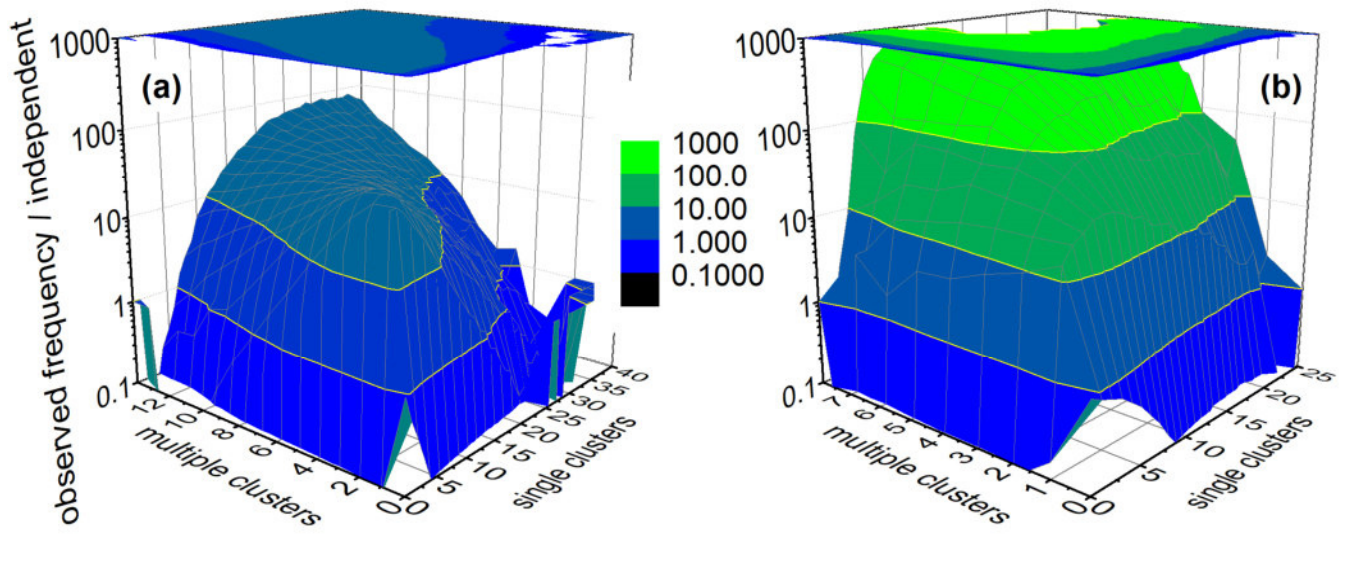

**Fig. 6**: Ratio of the observed frequencies for pairs of numbers of CVs with single and with multiple ICs to the expected frequency for the case that the two marginal distributions are statistically independent. The data have been obtained in a spherical region of interest (ROI) with radius $R_{ROI}$ = 250 nm and single events of protons of energy 3 MeV (left) and 50 MeV (right).

In contrast, values between 10 and 100 are found for most elements of the bivariate distribution for the case of 3 MeV protons. In the case of 50 MeV protons, the values are even an order of magnitude higher, which is presumably due to the fact that the decrease of both marginal frequencies is much faster than for the 3 MeV data. This is to some extent expected as secondary electrons produce ICs at their track ends that may be more important for the sparser ionizing 50 MeV protons.

With respect to the formation of ionization clusters in spherical sites within a ROI, the message of Fig. 5 and Supplementary Figs. S5 to S8 is that the respective frequency distributions are not Poisson distributed. And Fig. 6 shows that the frequency distributions of the spherical sites with exactly one or with two or more ICs are not statistically independent. This is essentially reflecting the statistical correlation of the energy transfer points that is at the basis of the definition of events in microdosimetry.

*Track structure at the micrometer level and DSBs*

Closer inspection of Fig. 5 and Supplementary Figs. S5 and S6 reveals that the mean number of targets receiving single or multiple ICs is far too high for a 500 nm diameter ROI as compared to the expected number (which is on the order of 30 to 40) of DSBs in a cell nucleus of ten times larger diameter and, hence, thousand times larger volume. The reason is that not all CV-sized spherical volumes in a cell nucleus contain DNA and thus can be considered to be a target of radiation effects.

The effect of the spatial filtering induced by the sparsity of potential targets has been estimated in this work by assuming that the potential targets have a uniform spatial density within the cell nucleus. If this assumption holds, each CV containing ICs has the same probability $p_d$ for being a “true” target in which ICs lead to DSBs. The conditional probability $P(k_1, k_{2+}|n_1, n_{2+})$ that $n_1$ CVs with one IC and $n_{2+}$ CVs with more than one IC result in $k_1$ CVs with one DSB and $k_{2+}$ CVs with two or more DSBs is then given by the product of two binomial probabilities:

$$P(k_1, k_{2+}|n_1, n_{2+}) = B(k_1|n_1, p_d)\times B(k_{2+}|n_{2+}, p_d) \qquad (29)$$

where

$$B(k|n,p) = \binom{n}{k} p^k (1-p)^{n-k} \, . \qquad (30)$$

Inferring the resulting distribution of the number of CVs with single and multiple DSBs from the data obtained for the 500 nm ROIs in Subsection 0 was then done by first determining the distributions of CVs with ICs within a cell nucleus by repeated convolution of the data shown in Fig. 5 and Fig. 6. However, this implied the assumption that the ROIs are statistically independent, which may introduce a bias in the results and make them unsuitable for assessing the statistical independence of CVs with single and multiple DSBs.

Therefore, this part of the investigation has been based on simulation data obtained in the frame of the BioQuaRT project (Palmans et al. 2015). The number of tracks was comparatively small compared to the 50,000 used by Braunroth et al. (2020): only 50 for 3 MeV protons and 250 for 50 MeV. However, the tracks covered a path of 10 µm (Alexander et al. 2015). Hence, despite the low statistical power, it was possible to study the frequency distribution of CVs with ICs and DSBs for ROIs in the size of a cell nucleus. Here, a ROI diameter of 6 µm has been used and a beam diameter of 9.9 µm. The scoring has been done similar to Subsection "Nanodosimetry of track structure at the micrometer level". The resulting frequency distributions of CVs with single and multiple ICs are shown in Supplementary Fig. S9 for the BIV and CV dimensions used in Schneider et al. (2019). Similar to what can be seen in Fig. 5, these distributions are also evidently not Poisson distributions.

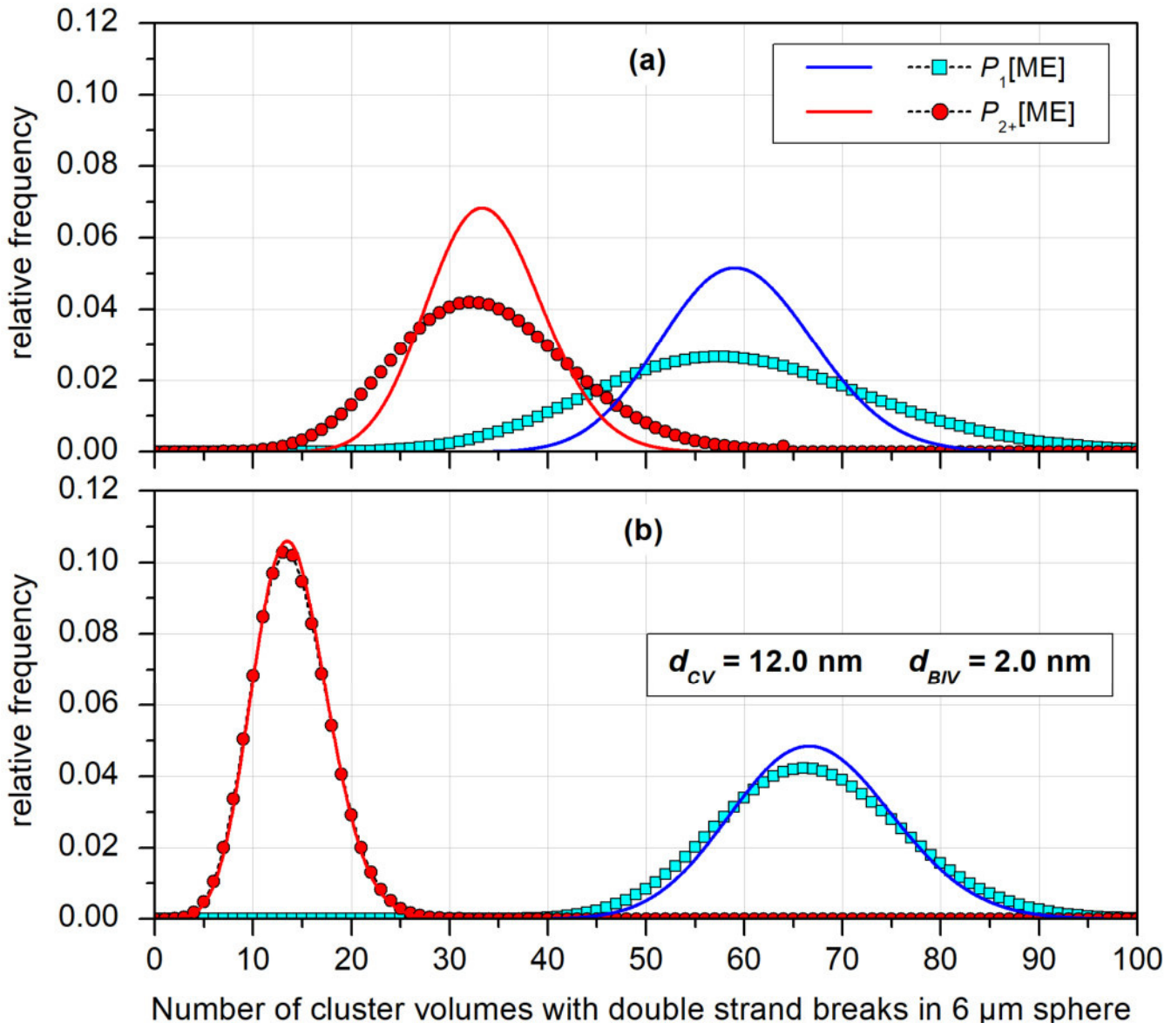

**Fig. 7**: Multi-event (ME) distributions of cluster volumes inside a spherical region of interest (ROI) with radius $R_{ROI}$ = 6 µm that receive a single DSB (squares) or two or more DSBs (circles) from protons of (a) 3 MeV and (b) 50 MeV energy. The data apply to a particle fluence corresponding to an absorbed dose of 2 Gy and a constant probability of 0.01 for an ionization cluster (IC) to be converted to a DSB. The solid lines are Poisson distributions with the same expectation as the data shown by symbols. (BIV - basic interaction volume; CV - cluster volume. For details see text.)

Figure 7**Fig. 7** shows the distributions of CVs with single (squares) or multiple DSBs (circles) obtained with a value of 0.01 for probability $p_d$. The solid lines indicate Poisson distributions with the same mean value as the data marked by symbols. Contrary to what can be seen in Fig. 5, the distribution of CVs with multiple DSBs for the case of 50 MeV protons in Fig. 7(b) is seen to be relatively well fitted by a Poisson distribution, whereas the other distributions are overdispersed compared to the related Poisson distributions. This overdispersion is more pronounced for the 3 MeV data and may be related to this radiation quality being more densely ionizing than 50 MeV protons.

The bivariate distributions of CVs with single and multiple DSBs for the two proton energies are shown in Fig. 8, overlaid by a contour plot of the ratio between bivariate frequency and the product of the marginal frequencies. The bivariate distribution for 3 MeV protons in Fig. 8 (a) is tilted with respect to the coordinate axes, which suggests that there is a correlation between the occurrence of CVs with single and multiple DSBs. This suggestion is further corroborated by the observation that near the maximum of the distribution, the ratio of the bivariate frequency to the product of the marginal frequencies is between 1.2 and 1.3 and that values of this ratio as high as 6 are found for bivariate frequencies within the top 95% of observed values (see Supplementary Fig. S10(a)).

In contrast, the bivariate distribution shown in Fig. 8(b) is aligned with the coordinate axes and in this case the ratio of bivariate frequency to the product of the marginal frequencies is close to unity near the maximum of the distribution and between 0.6 and 1.4 for bivariate frequencies higher than 5% of the maximum (see Fig. 8(b)). Thus, for this case the distributions of CVs with single and multiple DSBs seem to be statistically independent. Furthermore, the marginal distribution of CVs with multiple DSBs is well described by a Poisson distribution and the distribution of CVs with single DSBs is at least well approximated.

These observations seem surprising, given the large discrepancy between the distributions of CVs with single and multiple ICs and the respective Poisson distributions of the same mean value (see Supplementary Fig. S9). And they are not explained by the fact that a very small value has been used for the probability $p_d$, so that the binomials appearing in Eq. 29 can be well approximated by Poisson distributions (Schneider et al. 2017). In contrast, the single event distributions of CVs with one and multiple DSBs also show significant discrepancies from the respective Poisson distributions of the same mean value (cf. Fig. 9). However, the deviations from the Poisson distributions are more pronounced for the more densely ionizing 3 MeV protons.

For the 50 MeV protons the average number of tracks corresponding to a dose of 2 Gy and the considered beam diameter of 9.9 µm is about 800. If the bivariate single-event distribution is convoluted 800 times with itself, the result converges to a curve resembling a Gaussian. As can be seen in Supplementary Figs. S11 and S12, however, the overdispersion with respect to the respective Poisson distribution seems to be independent of dose. This suggests that the assumption of statistical independence may be justified for sparsely ionizing radiation, and that, in this case, the frequency distributions of CVs containing single and multiple DSBs may be assumed to have a Poisson-like shape.

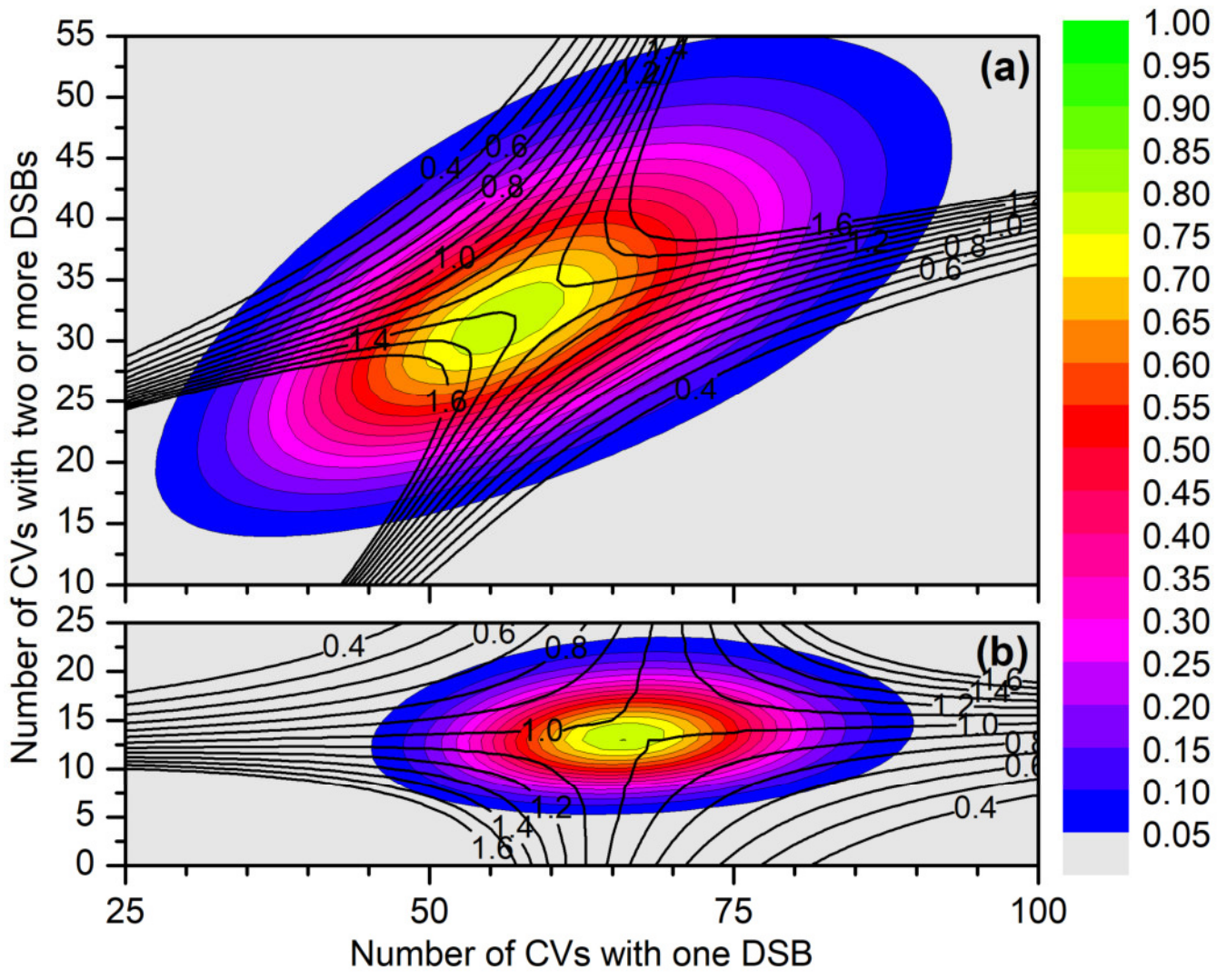

**Fig. 8**: Bivariate frequency distributions of simultaneous occurrence of a number of cluster volumes (CVs) with one DSB (shown on the *x*-axis) and a number of CVs with two or more DSBs (*y*-axis) from protons of (a) 3 MeV and (b) 50 MeV energy. The data apply to a particle fluence corresponding to an absorbed dose of 2 Gy and a constant probability of 0.01 for an ionization cluster (IC) to be converted to a DSB. The colored areas indicate the distribution in increments of 5% of the maximum value. The thick contour lines refer to the ratio of the bivariate distribution to the product of the marginals distributions. For details see text.

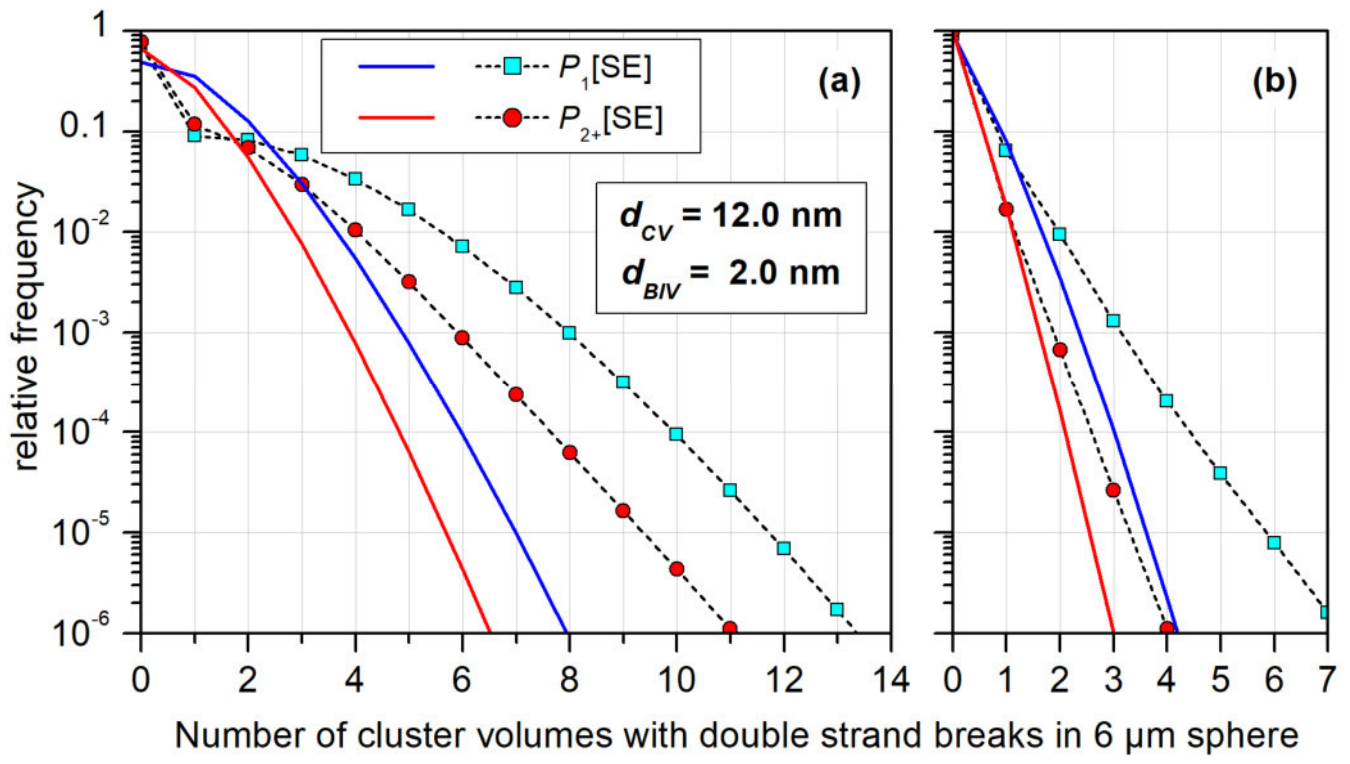

**Fig. 9**: Single-event frequency distributions of cluster volumes (CVs) with one DSB (squares) and CVs with two or more DSBs (circles) from protons of (a) 3 MeV and (b) 50 MeV energy. The data apply to a particle fluence corresponding to an absorbed dose of 2 Gy and a constant probability of 0.01 for an ionization cluster (IC) to be converted to a DSB. The solid lines are Poisson distributions of the same average as the data represented by symbols. For details see text.

## Discussion

The results shown in Fig. 7 are the predicted number of CVs in which single or multiple DSBs are produced. Therefore, the next methodological step along the lines of the TET/RAMN would be to include the repair of DSBs to derive the respective distributions of unrepaired single DSBs and DSB clusters. In principle, this can be done in the same way as in Subsection "Track structure at the micrometer level and DSBs". Following the approach of Schneider et al. (2020), one has to consider different repair probabilities of SLs and CLs. This would then be equivalent to using two different compound probabilities of the production and non-repair of single and clustered DSBs.

To separate physical and biological radiation effects, as proposed by the BioQuaRT project (Palmans et al. 2015), it would be more consistent to maintain separate parameters for the spatial density of target volumes and for the repair of single and multiple DSBs. Similar to the approach of Schneider et al. (2019), a large number of cell irradiation experiments could be analyzed with a model that considers two cell-line-specific parameters for repair and three cell-line-independent parameters: the parameter $p_d$ for the target density and the physical parameters used for scoring ICs and clusters of ICs, namely the diameters of the BIVs and CVs, $d_{BIV}$ and $d_{CV}$. In the work of Schneider et al. (2019, 2020), the value of $d_{BIV}$ was always set by a model assumption, but it would be more convincing if the parameter value (or its likelihood distribution) could be inferred from radiobiological data rather than arbitrarily chosen in the range of possible values compatible with existing evidence.

The essential model assumption would be that $d_{BIV}$ is independent of the biological system and the radiation quality since it is related to the properties of the DNA molecule. It is very likely that also $d_{CV}$ could be assumed to be independent of both radiation quality and cell type. (The latter would be accounted for by the repair parameters.)

The elaboration of a revised RAMN based on a comprehensive characterization of particle track structure is a major endeavor and, hence, beyond the scope of this work, which focusses for the rest of the article on a few methodological aspects.

*Connection between ICs and DSBs*

The approach presented in Subsection "Track structure at the micrometer level and DSBs" has similarities with the combinatorial model of Garty et al. (2006, 2010), where the parameter used in the binomial was the conditional probability of an ionization to result in a DNA (single) strand break. In a second step, they considered a random distribution of the strand breaks over the DNA double helix to derive the probability of the formation of a DSB. The analysis in Subsection "Track structure at the micrometer level and DSBs" was, however, based on identifying a BIV with an IC with a DSB. As discussed in Subsection "Issues with the TETs and RAMNs link to nanodosimetry", this is at variance with evidence for IC complexity (number of ionizations) to play a role (Nettelbeck and Rabus 2011, Rabus and Nettelbeck 2011, Conte et al. 2017, 2018, Selva et al. 2019).

Following the line of arguments of Garty et al. (2006, 2010), a better approach would be to use the following hypothesis: an IC in a short segment of the DNA double helix (represented by the BIV) leads to a DSB if ionizations occur on both strands of the DNA. If only the number of ionizations in the BIV are known, it is straightforward to assume that each ionization has a probability of 0.5 to occur on one strand or the other. Then the conditional probability of the formation of a DSB in a site on the DNA where an IC occurs is given by Eq. 31.

$$P(DSB|IC) = \frac{1}{F_2} \sum_{k=2}^{\infty} \frac{1}{2^{(k-1)}} F_k \ . \tag{31}$$

*Relevant length scales and model parameters*

As has already been discussed by Schneider et al. (2020), a RAMN needs to consider interactions of distant (single) lesions as is done in some other approaches to connect microscopic radiation effects and cellular outcome, such as the BIANCA model (Ballarini et al. 2014). The relevance of radiation action on both the micrometric and nanometric scales has been the hypothesis underlying the generic multi-scale model of the BioQuaRT project (Palmans et al. 2015) and demonstrated later by radiobiological evidence (Friedrich et al. 2018).

The extension of the approach outlined in Section "Outline of a tentative approach to consider track structure in the TET and RAMN" toward also including frequency distributions in subcellular volumes, for example of CVs with single ICs, is straightforward. The disadvantage is that further model parameters are introduced. However, the evidence presented by Friedrich et al. (2018) and the success of the BIANCA model (Carante et al. 2018) make such a future extension of the RAMN probably a necessity. The approach presented in Section "Outline of a tentative approach to consider track structure in the TET and RAMN" can be easily extended to include clustering at different spatial scales. And the number of extra parameters coming into play can be handled by assuming them as independent of cell type and radiation quality and taking a big-data approach, as already done to some extent by Schneider et al. (2019).

*Limitations*

The approach outlined in Section "Outline of a tentative approach to consider track structure in the TET and RAMN" overcomes two of the limitations of the RAMN discussed by Schneider et al. (2020), since it neither considers only straight segments of tracks in the BIV or CV nor ignores the extension of tracks. Like the RAMN, the present approach also relies on the CVs being homogeneously filled with DNA such that ionization clusters within the CV can be interpreted as DSBs. It does not consider the actual spatial arrangement of DNA in the cell nucleus that may have a role in the formation of DSBs (Kellerer and Rossi 1978, Schneider et al. 2016) and the contribution of radiation damage due to free radicals from water radiolysis.

Furthermore, the limitations discussed by Schneider et al. (2020) regarding the role of different biological endpoints and interference of pathways leading to them as well as interactions between complex and simple DNA lesions also apply. It should be noted, however, that the mean values of CVs with single and multiple DSBs (cf. Fig. 7) are compatible with the rule of thumb that 30 to 40 DSBs are induced per Gy, if an average DNA content in the cell nucleus on the order of 1% is assumed (Goodhead and Brenner 1983). (The multi-event distributions have been calculated for a dose of 2 Gy.)

In addition, there are three potential limitations inherent to the scoring procedure used. First, in the calculation of the multi-event distributions, the possibility of several tracks interacting in the same CV has been ignored. However, this is justified, since the probability of this occurring has been shown in Subsection "Frequency of BIVs inside a CV receiving an IC by protons" to be negligibly small. Second, the spherical target volumes (for the formation of DSBs as well as DSB clusters) are approximated by polyhedrons of the same volume. As it has been shown by Grosswendt (2002) that the geometric shape of the target volume has only a minor influence on the IC distributions, this should not be a major issue.

The third limitation is that a regular array of such target volumes is used. This may potentially introduce bias toward a smaller probability of IC formation and toward smaller clusters of ICs within a CV. As the face-centered cubic Bravais lattice has octahedral symmetry, this shortcoming of the scoring geometry could be overcome to a major extent by considering different orientations of the track with respect to the lattice within its small fundamental domain and additionally considering different impact parameters.

However, it should be noted that the procedure outlined in Section " Outline of a tentative approach to consider track structure in the TET and RAMN" is not reliant on the particular scoring method, which may be substituted in the future with more sophisticated techniques from database analysis (Francis et al. 2011, Bueno et al. 2015).

## Conclusions

The track event model was developed as an alternative model for the dose dependence of cell survival that takes into account the radiation quality by including properties of particle track structure in the form of nanodosimetric probabilities of ionization cluster formation. The radiation action model based on nanodosimetry of Schneider et al. (2020) has been a development that tried to overcome some of the deficiencies of the TET by rebuilding the link to radiobiology.

The original version of the TET produced a model equation (cf. Eq. 9) which offered the advantage of a functional shape that is equivalent to the linear-quadratic model in the dose range in which the latter describes the trend of experimental data well and is superior to it at higher doses where a pure exponential dose dependence is observed experimentally (Besserer and Schneider 2015a).

In this article, it has been shown that some of the assumptions in the original model are dispensable: the Poisson statistics of the frequency distributions of OTEs and TTEs or SLs and CLs and their statistical independence can be derived from an assumed Poisson distribution of the number of tracks contributing to the numbers of OTEs and TTEs or SLs and CLs formed in the considered target volume.

On the other hand, it has also been found that the formula used within the extension of the TET for repair (Besserer and Schneider 2015b) is not consistent with the underlying model assumptions. The correct model equation has been derived in this work and includes a further model parameter, namely the probability that repairable lesions are produced. This parameter can only be ignored if one assumes that there are no single-event lesions that are unrepairable. But even in this case, the model equation is different from the one used in the TET. Unfortunately, this fault in the mathematical model makes the comparison of the TET model with experimental data and the assessment of its performance with respect to the linear-quadratic model questionable. This deficiency of the repair model became obsolete when the TET was replaced with the RAMN.

It has further been demonstrated that the implicit assumption of independent subcellular targets leads to a survival model that is almost purely exponential for relevant dose ranges. This was further corroborated by an evaluation of the probabilities of single and multiple ICs from nanodosimetric simulations for protons, which showed that the probability of multiple ICs would be negligibly small for amorphous tracks and CV dimensions analogous to those used by Schneider et al. (2019, 2020). This evaluation further revealed that with proton tracks more than 50% of the probability of the formation of an IC in a spherical BIV of 3 nm diameter is due to tracks passing the BIV at impact parameters larger than ten times the BIV radius. For proton energies of 3 MeV and higher, more than 25% of the total IC probability comes from impact parameters larger than fifty times the BIV radius.

On the other hand, only tracks passing at impact parameters up to three times the CV radius contribute to the probability of the formation of several ICs within a spherical CV of 18 nm diameter. For single ICs in the CV, impact parameters up to about ten times the CV radius contribute. This suggests that for amorphous tracks and independent BIVs the central passage used in the simulations of Schneider et al. (2019, 2020) to determine model parameters from nanodosimetry should be replaced by simulations where impact parameters up to about 100 nm are considered.

The probabilities of CVs with multiple ICs were found to be negligibly small when IC formation in different targets was assumed to be statistically independent. This confirms that statistical correlations of interactions within particle tracks must be taken into account to obtain reasonably large probabilities of CLs. This has been shown in Section “Outline of a tentative approach to consider track structure in the TET and RAMN” where a paradigm shift was applied for nanodosimetry: instead of considering IC formation in defined targets, the spatial distribution of targets with ICs was used to obtain frequency distributions in micrometric volumes of CVs with single and multiple BIVs with an IC.

Assuming a constant probability that the ICs in the CVs occur within DNA, the frequency distributions of CVs with single and multiple DSBs can be obtained. Using this approach for proton tracks revealed large deviations of the frequency distributions for CVs with ICs from Poisson distributions and a strong correlation between the frequencies of CVs with single and multiple ICs. Despite this, the resulting distributions for CVs with single and multiple DSBs for sparsely ionizing radiation were found to be almost statistically independent and to have shapes that can be roughly approximated by Poisson distributions. For densely ionizing radiation, the frequency distributions of CVs with single and multiple DSBs were found to remain correlated and strongly departing from Poisson shape. For both sparsely and densely ionizing protons, the deviation between actual distribution and the corresponding Poisson distribution was found to be invariant with dose.

In summary, the analysis presented here has revealed some inconsistencies and weaknesses of the TET and RAMN, but also determined that some of the precarious assumptions made in their development, such as the statistical independence of relevant targets, only seem to contradict the concept of particle tracks, but are at least approximately true for sparsely ionizing radiation. The decisive ingredient of a revised TET/RAMN appears to be a consistent description of the relation between tracks interacting with cells and radiation action in subcellular targets. This requires a paradigm shift from the single-target perspective of nanodosimetry to a track-oriented view. The first steps toward this goal have been outlined in Section “Outline of a tentative approach to consider track structure in the TET and RAMN” The results seem very promising and warrant further endeavor in this direction.

## Acknowledgments

This work was in part supported by the German Federal Ministry for Economic Cooperation and Development (BMZ) in the frame of the Technical Cooperation Project “Upgrading of quality infrastructure in Africa”. The National Metrology Institute of South Africa (NMISA) and the PTB Guest Researcher Program are acknowledged for sponsoring guest researcher stays of S.A.N. at PTB. Carmen Villagrasa is credited for performing the track simulations in the frame of the BioQuaRT project. The BioQuaRT project was funded within the European Metrology Research Program (EMRP). The EMRP was jointly funded by the European Union and the EMRP-participating countries.

# Supplemental material to "Investigation into the foundations of the track-event theory of cell survival and its link to nanodosimetry" in Radiation and Environmental Biophysics

**Sonwabile Arthur Ngcezu[1], Hans Rabus[2,*]**

[1] *University of the Witwatersrand, Johannesburg, 2000, South Africa*
[2] *Physikalisch-Technische Bundesanstalt (PTB), 10587 Berlin, Germany*
* email: hans.rabus@ptb.de

## Supplementary Figures for Section "Nanodosimetry and RAMN" and Methodology used for producing the data

In this Section, an approach is presented to determine single-event and multi-event averages of the nanodosimetric parameter $F_2$ for induction of an ionization cluster (IC) in a basic interaction volume (BIV) as well as for the number of BIVs in a cluster volume (CV) that receive an ionization cluster (IC). The approach assumes that the probabilities of IC formation in different sites are statistically independent, and it requires knowledge of the dependence of the probability of the formation of an IC in a site, $F_2(r)$, on the impact parameter $r$ of the primary particle trajectory with respect to the center of the site.

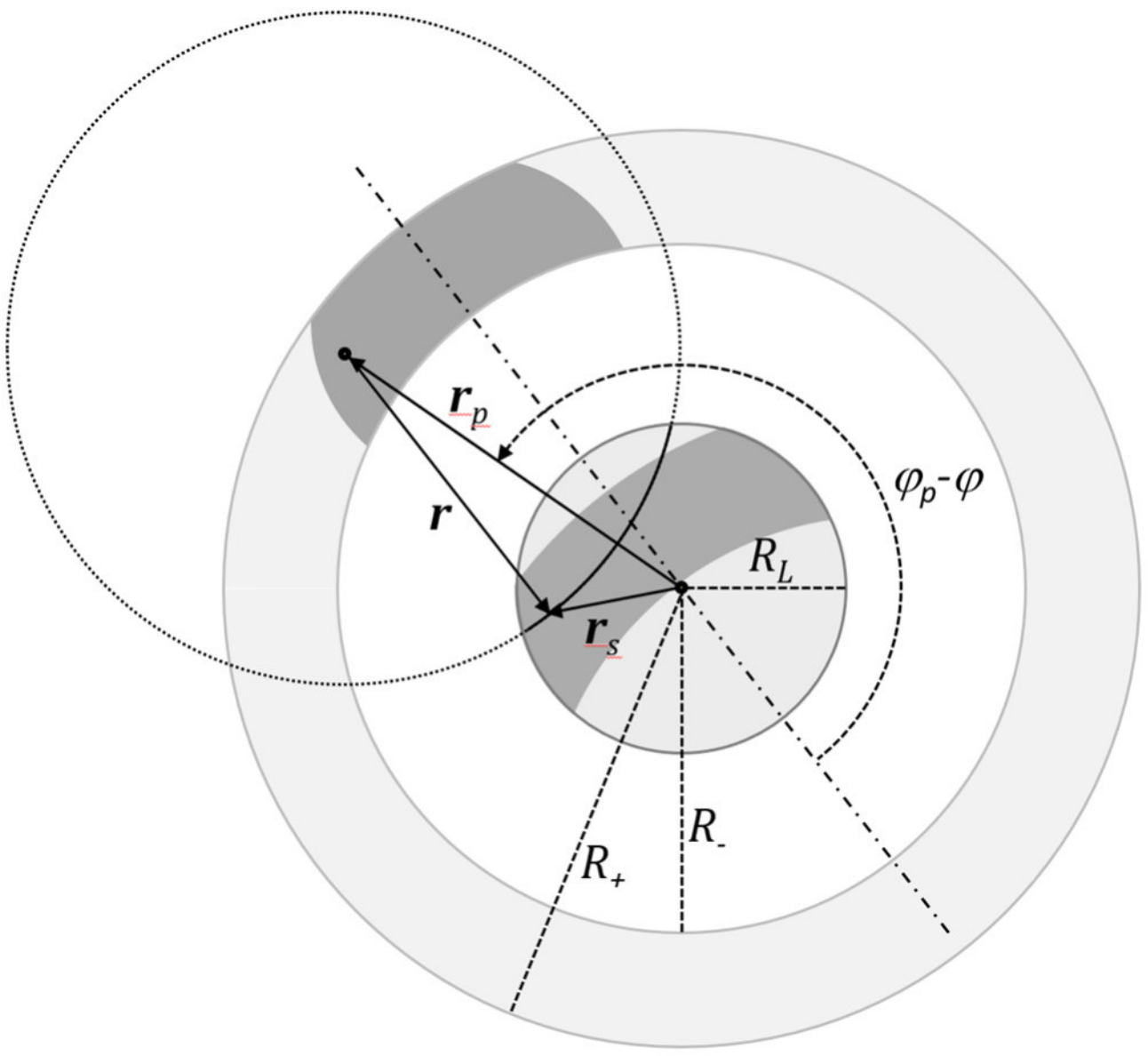

**Fig. S1**: The shaded circle represents a central cross section of the spherical CV (radius $R_L$). The small filled black circle (dot) indicates a primary particle trajectory (perpendicular to the drawing plane) that passes the CV at the point described by vector $\boldsymbol{r_p}$ within an annulus of inner radius $R_-$ and outer radius $R_+$. The solid circle segment inside the CV cross section indicates the loci of sites within the CV that have a radial distance $r$ from this primary particle trajectory. The dark-shaded part of the annulus represents all primary trajectory positions for which the endpoint of a radial vector $\boldsymbol{r}$ that is parallel to the dot-dashed line is within the CV cross-section. The loci of these endpoints cover the dark-shaded area within the CV cross section.

For this purpose, spherical sites are considered that are located within a spherical region of interest (ROI) with radius $R_L$. Furthermore, the primary particle trajectory is assumed to pass the ROI within an annulus of inner radius $R_-$ and outer radius $R_+$ in a plane perpendicular to the trajectory that passes through the ROI center (see Fig. S1). The expected number of sites $\bar{n}$ within the spherical ROI that receive an IC when the primary particle passes the annulus is then given by Eq. (S1).

$$\bar{n} = \oint \int_0^\infty \frac{F_2(r)}{V_s} \iint_{A_p} L(r_p, r, \Delta\varphi_p)\Phi_p r_p dr_p d\varphi_p r dr d\varphi. \quad \text{(S1)}$$

In Eq. (S1), the first integral extends over the full planar angle, the second integral extends over all radial distances from the primary particle trajectory, and the double integral extends over the area of the annulus, $A_p$ (see Fig. S1). $r$ and $\varphi$ are polar coordinates of a point in a plane perpendicular to the primary particle trajectory relative to the point of its intersection with this plane. $r_p$ and $\varphi_p$ are the polar coordinates of this intersection point with respect to the center of the CV. $F_2(r)$ is the complementary cumulative probability of an IC that depends on the radial distance $r$ of a site from the primary particle trajectory. $V_s$ is the volume of the site.

$L(r,r_p,\Delta\varphi_p)$ is the length of a chord through the CV that is parallel to the primary particle trajectory and passes the point described by vector $\boldsymbol{r_s}$. The length of this chord depends on the distance $r_s$ of this point from the CV center, which is a function of $r$, $r_p$ and the difference $\Delta\varphi_p$ of the azimuth angles $\varphi_p$ and $\varphi$. $L$ has nonzero values only if point $\boldsymbol{r_s}$ is within the cross section of the CV in the plane. (That is, for the points on the arc shown as a solid black line in Fig. S1.)

Finally, $\Phi_p$ is the fluence of primary particles given by

$$\Phi_p = \frac{n_t(D)}{A} \quad \text{(S2)}$$

where $n_t$ and $A$ are the mean number of primary particle tracks and the area $A$ considered in Eqs. (1) and (2). (For a single event, $n_t$ has to be replaced by unity.)

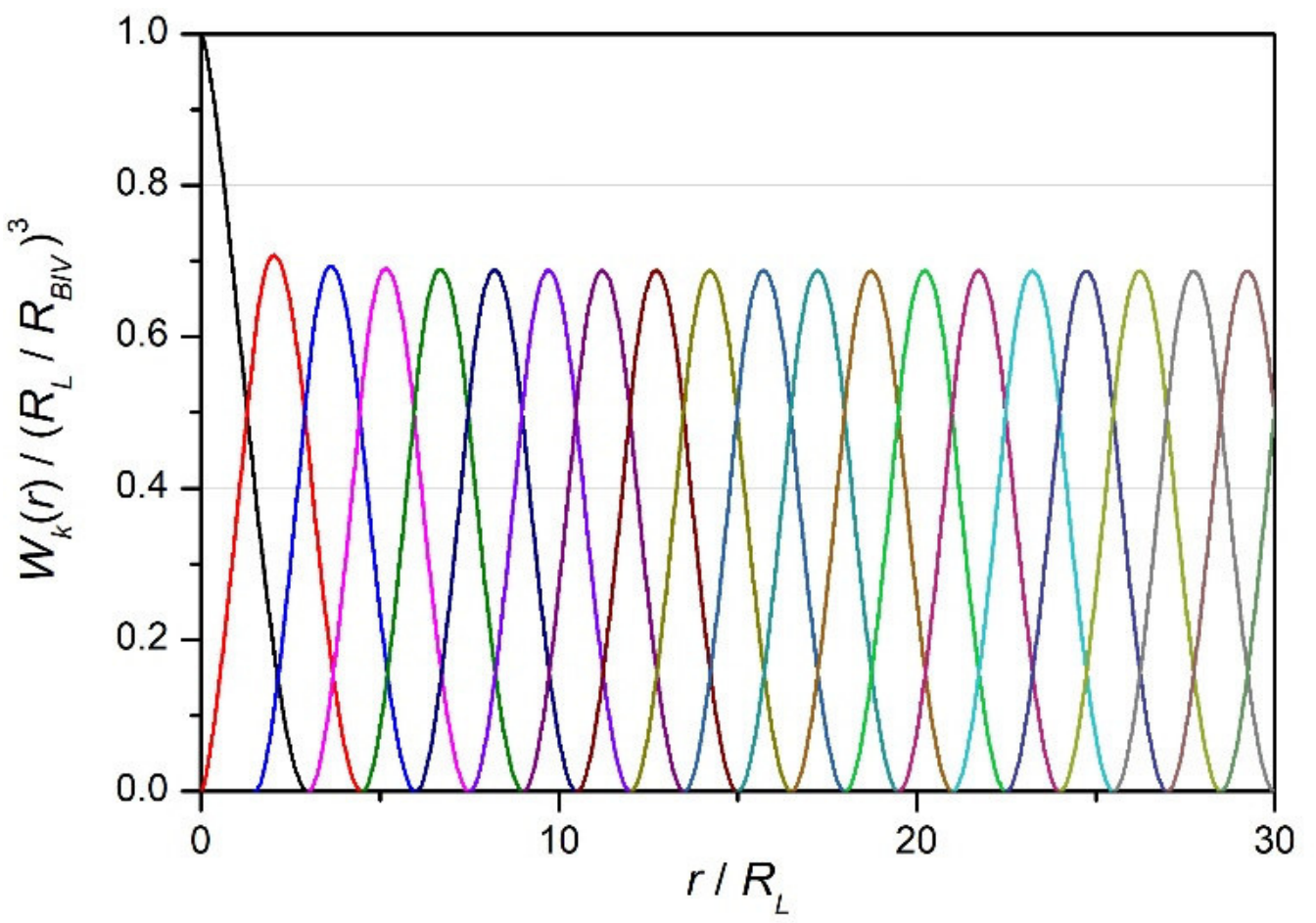

**Fig. S2**: Weighting function for the radial distribution of ionization clusters around a proton track used in Eq. (14) in the paper. $R_{BIV}$ is the radius of the (spherical) basic interaction volume in which ionization clusters are scored. $R_L$ is the radius of the spherical region of interest. If the region of interest is the BIV, then the outcome of Eq. (S3) is the probability of the formation of an ionization cluster in the BIV when proton tracks traverse the respective annulus around the BIV cross section. If the region of interest is a cluster volume (CV) containing several BIVs, then the result of Eq. (S3) is the expected number of BIVs within the LIV that receive an ionization cluster when a proton track passes the respective annulus.

For a fixed value of $\varphi$, only points within the dark-shaded part of the annulus are linked to nonzero chord lengths where the endpoints of the vector ***r*** cover the dark-shaded area within the CV cross section in Fig. S1. Therefore, it is convenient to rewrite Eq. (S1) as follows:

$$\bar{n} = \Phi_p \int_0^\infty F_2(r) \times W(r|R_L, R_-, R_+)\, 2\pi r dr \qquad \text{(S3)}$$

where $W(r|R_L, R_-, R_+)$ is a geometrical weighting function that is given by Eq. (S4).

$$W(r|R_L, R_-, R_+) = \frac{1}{V_S} \iint_{A_p} L(r_p, r, \Delta\varphi_p) r_p dr_p d\varphi_p. \qquad \text{(S4)}$$

In this work, the weighting functions were evaluated for the cases that the radii $R_-$ and $R_+$ are successive integer multiples of the ROI radius $R_L$.[1] The respective weighting functions are denoted by the upper integer as subscript, i.e., $W_k(r)$, where $k = 1$ corresponds to a passage of the primary particle trajectory through the CV. $W_1(0)$ is the ratio of the volumes of the ROI and the BIV, i.e., unity if the ROI is identical to the BIV. If the ROI is the CV of Schneider et al. (2020), $W_1(0)$ is the number of BIVs per CV. (Whereas Schneider et al. (2019, 2020) used the ratio of mean chord length through the CV and BIV diameter for the number of BIV per CV that potentially receive an ionization cluster.)

The resulting weighting functions $W_k$ are dimensionless and have domains and co-domains that scale with $R_L$ and the third power of $R_L/R_{BIV}$, respectively, and are shown for the first 20 annuli in Fig. S2. Except for the case of ROI traversal, the weighting functions of all annuli have the same maxima and functional shape and are only shifted with respect to each other.

Proton tracks simulated in previous work (Rabus et al. 2020) are used as data for a showcase example. In the simulations, protons were started in water with start energies between 1 MeV and 99 MeV at the surface of a slab of water of 650 nm thickness. Positions and energy transfers were recorded for each interaction of the proton or its secondary electrons within this slab. Secondary electrons were tracked until their energy dropped below the ionization threshold of water (about 11 eV). To obtain reasonable statistics, $5\times10^4$ individual tracks were simulated for each proton energy.

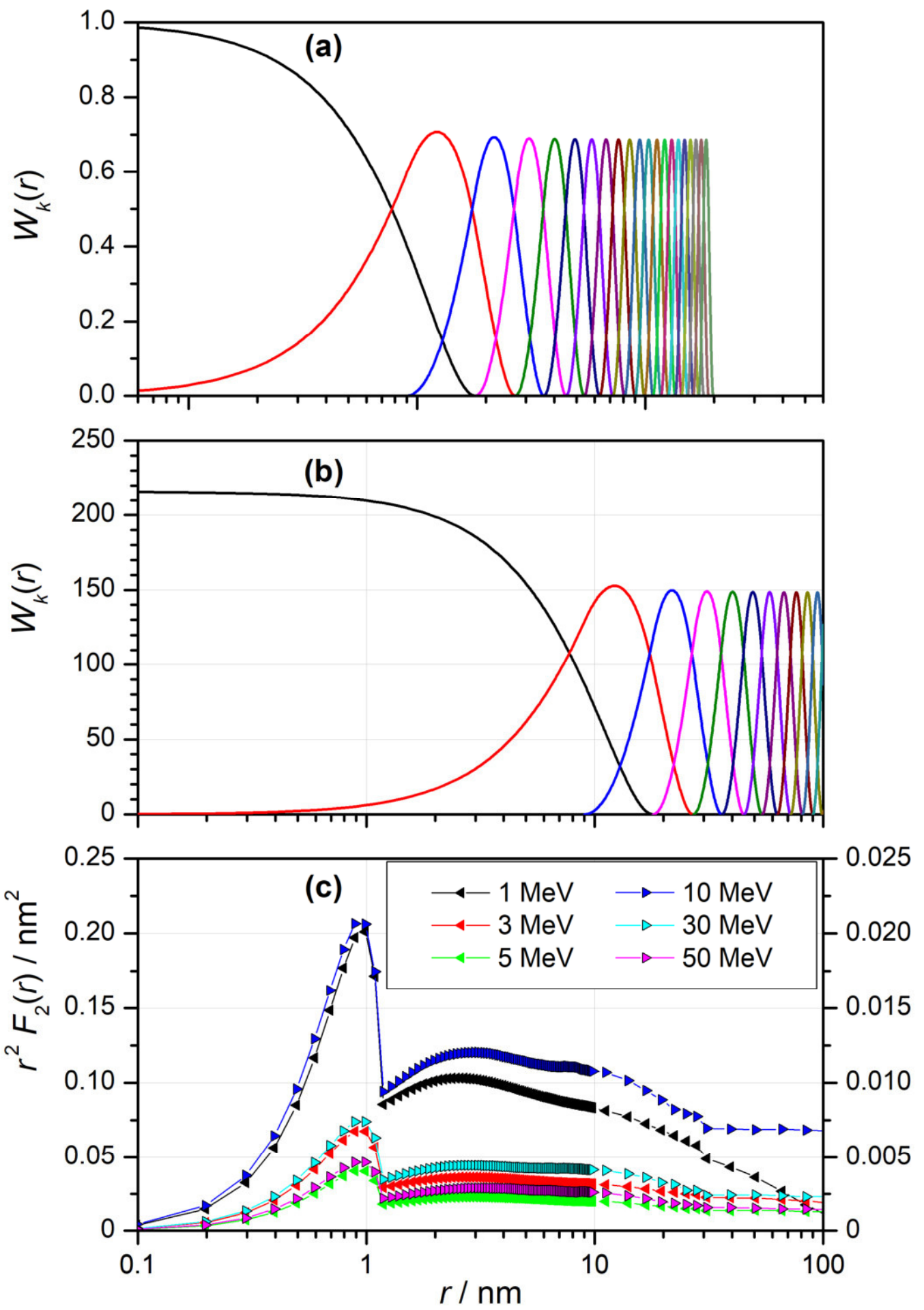

**Fig. S3**: (a) Weighting function for the radial distribution of ionization clusters considering the contribution of protons passing through one of the first 20 annuli around the cross section of a spherical BIV of 3 nm diameter. (b) Weighting function for the radial distribution of ionization clusters considering the contribution of protons passing through one of the first ten annuli around the cross section of a spherical CV of 18 nm. The weighting functions apply to the condition that the proton trajectory traverses the BIV or CV (black line) or through the *k*-th annulus (inner radius (*k*-1) times the BIV or CV radius, outer radius *k* times this radius). (c) Radial distribution of the probability of inducing a true ionization cluster in a cylindrical target of the same volume as the BIV located around trajectories of protons of different energies (Braunroth et al. 2020). (The right-hand side *y*-axis applies to the data points marked with right-pointing triangles.)

[1] The respective code is listed in Subsection "FORTRAN source code of program Radial_Weight" of Supplement 2.

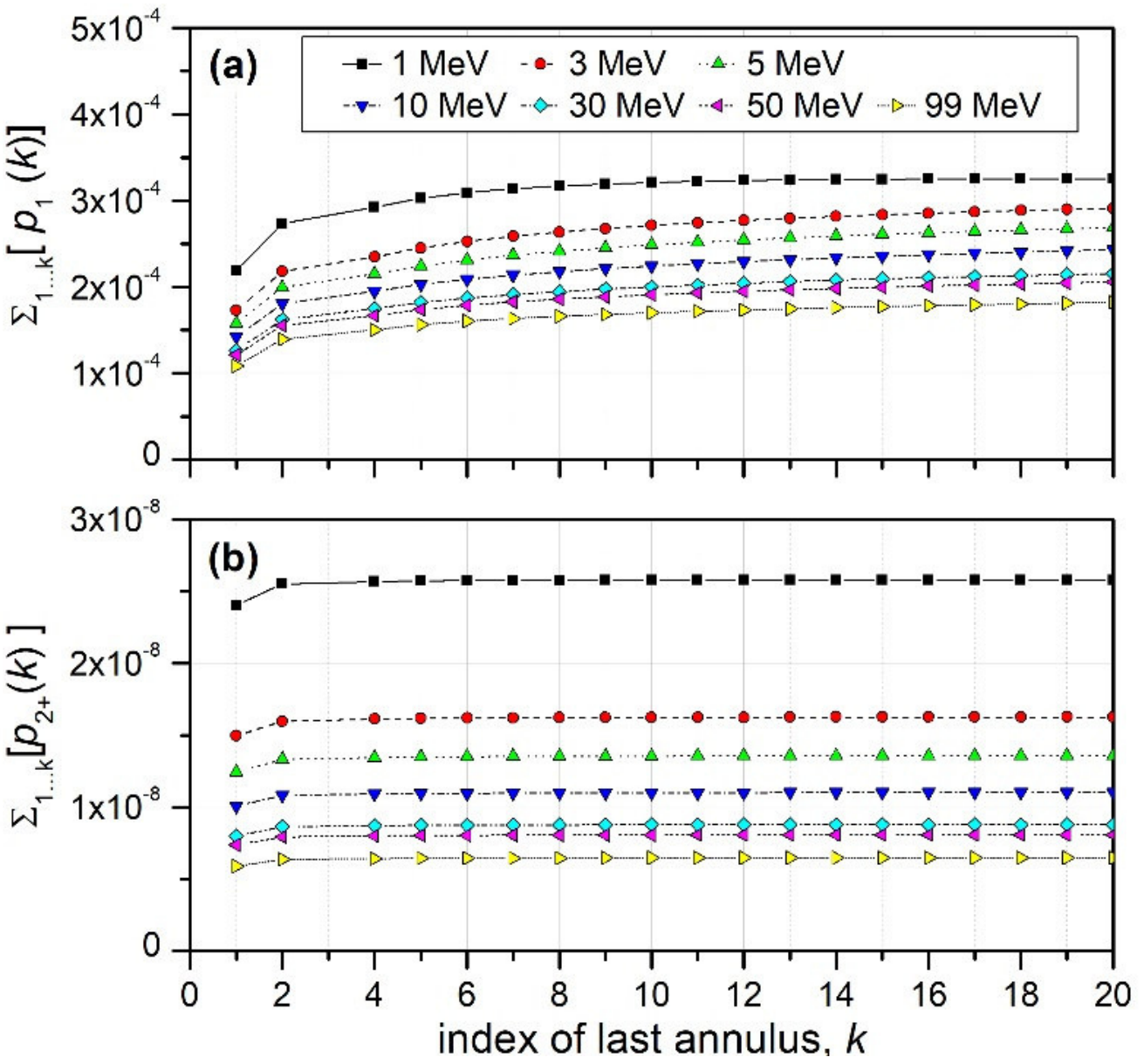

**Fig. S4**: Probabilities of (a) exactly one and (b) two or more sites in a CV receiving an ionization cluster when a proton passes through one of the first *k* annuli for an absorbed dose of 2 Gy. The data apply to spherical BIVs of 3.0 nm diameter inside a CV of 18 nm diameter.

The proton tracks were analyzed *a posterior* to determine the radial distributions of the frequency of nanometric targets with a certain ionization cluster size, i.e., number of ionizations (Braunroth et al. 2020). The targets were either cylindrical with a diameter of 2.3 nm and a height of 3.4 nm or cylinder shell sectors of Equal volume to the aforementioned cylinders. The cylinder axis was perpendicular to the shortest radial vector from a point on the proton trajectory to the center of the cylinder. Therefore, the frequency of targets with a certain number of ionizations Equals the probability of the formation of such an ionization cluster size when a proton track passes at an impact parameter Equal to the shortest radial distance of the target from the primary particle trajectory.

In the following, it is assumed that the probability of the formation of an ionization cluster in the aforementioned targets is the same as within a spherical target of the same volume (that has a diameter of about 3.0 nm). Ignoring potential changes in the values of $F_2$ is justified, as the purpose of the discussion given here is to demonstrate the order of magnitude of the effects to be considered.

The weighting functions applying to the case that the ROI is identical to the BIV are shown in Fig. S3(a) for a BIV of 3.0 nm diameter, i.e., of the same volume as the cylindrical targets used by Braunroth et al. (2020). Fig. S3(b) shows the respective weighting functions for the same BIV and a ROI of 18.0 nm diameter that contains the same number of BIVs as the “lethal interaction volumes” reported by Schneider et al. ( 2019) for protons. The black lines in Fig. S3(a) and Fig. S3(b) refer to a proton traversing the ROI, the red lines to a proton traversing the first annulus, and so forth.

Fig. S3(c) shows the results from Braunroth et al. ( 2020) for the radial distribution of targets receiving an IC in the tracks of protons of different energies. Triangles pointing left refer to the *y*-axis on the left-hand side and those pointing right to the right-hand side *y*-axis. Owing to the logarithmic *x*-axis, the values were multiplied by $r^2$ so that the integral under the plot curve is proportional to the contribution of the respective radial interval to the total radial integral. It should be noted that even though there is a pronounced peak for proton tracks passing the target cylinders, a significant proportion of sites with ICs lie at radial distances up to 100 nm and beyond (not shown).

The data presented Fig. S3(a) and Fig. S3(c) have been used to determine the relative contributions to the total probability of induction of an IC in a BIV from protons that pass the BIV within an annulus around the BIV cross section assuming a uniform fluence profile.[2] The respective results are presented in Fig. 3 of the paper.

Using the data shown in Fig. S3(b) and Fig. S3(c) allows the determination of the mean number of sites within a CV that receive an IC from protons passing an annulus around the CV cross section. The results are presented in Fig. 4 of the paper. The respective cumulative contributions of the first *k* annuli are shown in Fig. S4.

[2] The respective code is listed in Subsection “Excel VBA source code of routine convol” of Supplement 2.

## Supplementary Figures for Section "Outline of a tentative approach to consider track structure in the TET and RAMN"

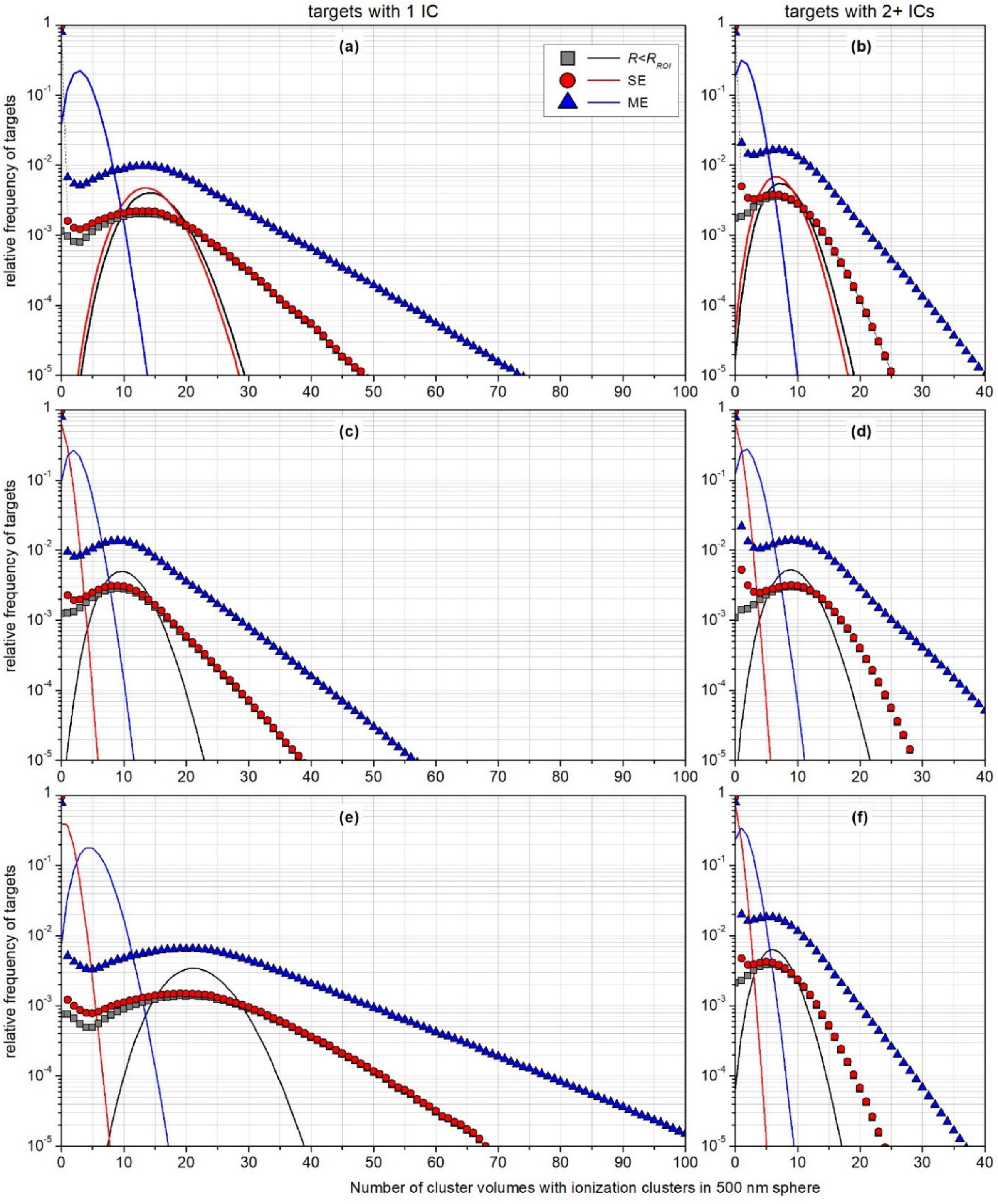

**Fig. S5:** Frequency distribution of the number of cluster volumes (targets) inside a spherical region of interest (ROI) with radius $R_{ROI}$ = 250 nm that receive a single ionization cluster (IC) (left column), or two or more ionization clusters (right column) induced by protons of 3 MeV energy. Red circles: Single event (SE) distribution, i.e., for a proton track passing the region of interest with a maximum impact parameter of $5R_{ROI}$. Blue triangles: Multi-event (ME) distribution for a dose of 2 Gy. Gray squares: Contribution to the SE distribution coming from proton tracks intersecting the ROI. The ICs are scored in spherical targets of (a) and (b) 2 nm, (c) and (d) 3 nm, and (e) and (f) 2.5 nm diameter. The cluster volumes are spheres of targets of (a) and (b) 12 nm, (c) and (d) 18 nm, and (e) and (f) 7.5 nm diameter. The solid lines indicate Poisson distributions that have the same mean value as the respective corresponding distributions indicated by symbols.

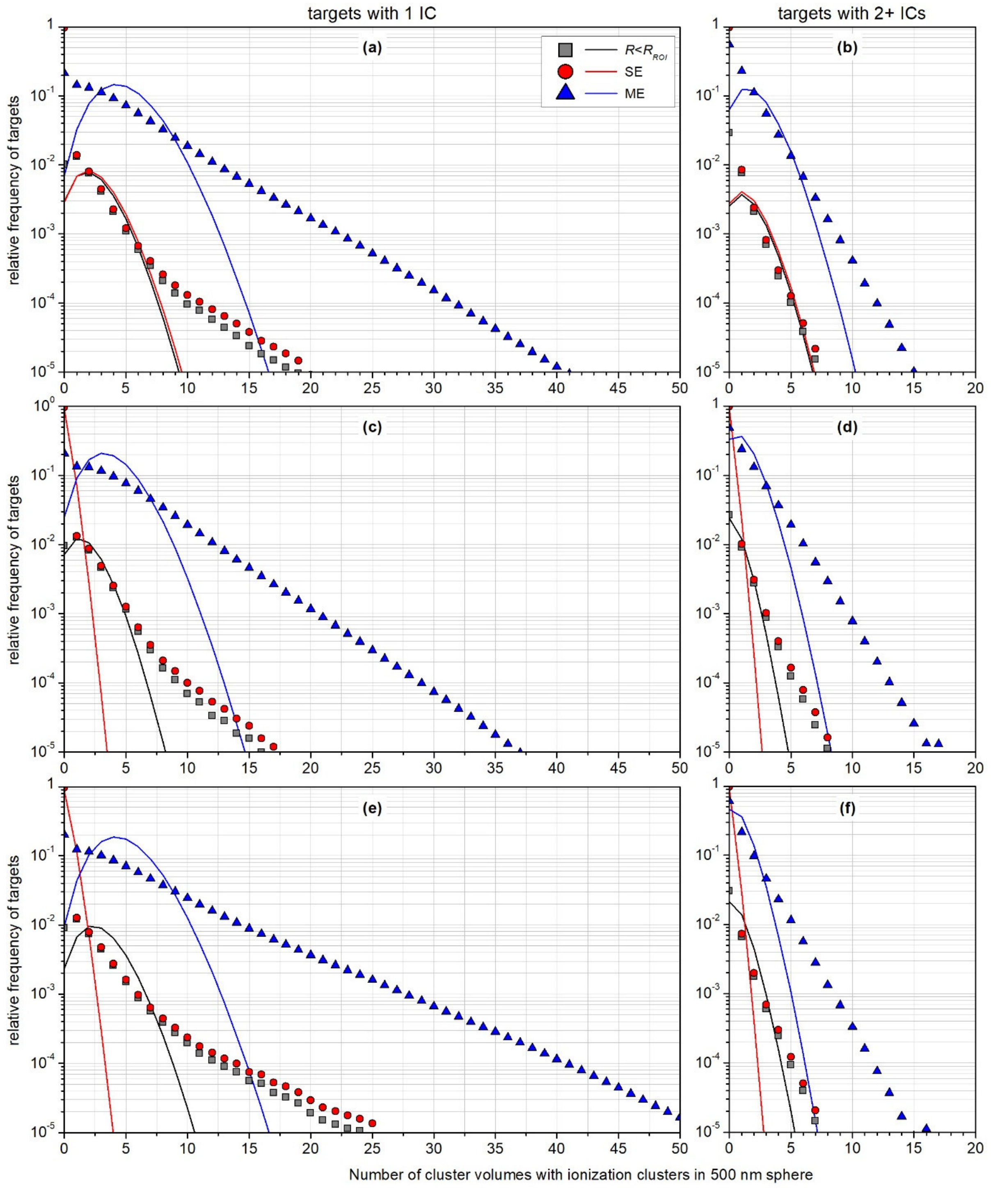

**Fig. S6:** Frequency distribution of the number of cluster volumes (targets) inside a spherical region of interest (ROI) with radius $R_{ROI}$ = 250 nm that receive a single ionization cluster (IC) (left column), or two or more ionization clusters (right column) induced by protons of 50 MeV energy. Red circles: Single event (SE) distribution, i.e., for a proton track passing the region of interest with a maximum impact parameter of $5R_{ROI}$. Blue triangles: Multi-event (ME) distribution for a dose of 2 Gy. Gray squares: Contribution to the SE distribution coming from proton tracks intersecting the ROI. The ICs are scored in spherical targets of (a) and (b) 2 nm, (c) and (d) 3 nm, and (e) and (f) 2.5 nm diameter. The cluster volumes are spheres of targets of (a) and (b) 12 nm, (c) and (d) 18 nm, and (e) and (f) 7.5 nm diameter. The solid lines indicate Poisson distributions that have the same mean value as the respective corresponding distributions indicated by symbols.

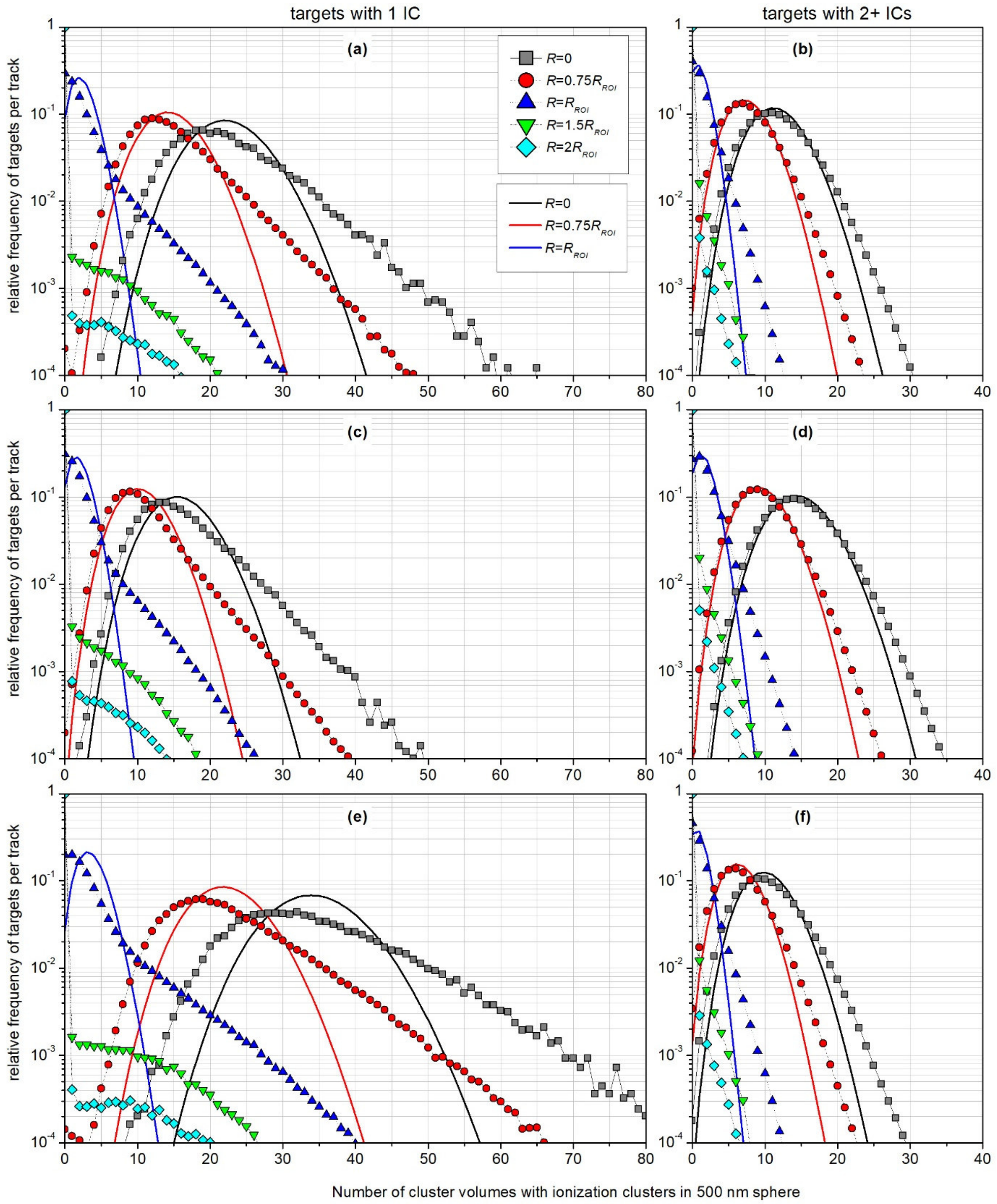

**Fig. S7:** The symbols indicate the frequency distributions of the number of cluster volumes (targets) inside a spherical region of interest (ROI) with radius $R_{ROI}$ = 250 nm that receive a single ionization cluster (IC) (left column) or two or more ionization clusters (right column) when a proton track (3 MeV proton energy) passes the ROI with an impact parameter $R$ as given in the legend. The ICs are scored in spherical targets of (a) and (b) 2 nm, (c) and (d) 3 nm, and (e) and (f) 2.5 nm diameter. The cluster volumes are spheres of targets of (a) and (b) 12 nm, (c) and (d) 18 nm, and (e) and (f) 7.5 nm diameter. The solid lines indicate Poisson distributions that have the same mean value as the respective corresponding distributions indicated by symbols.

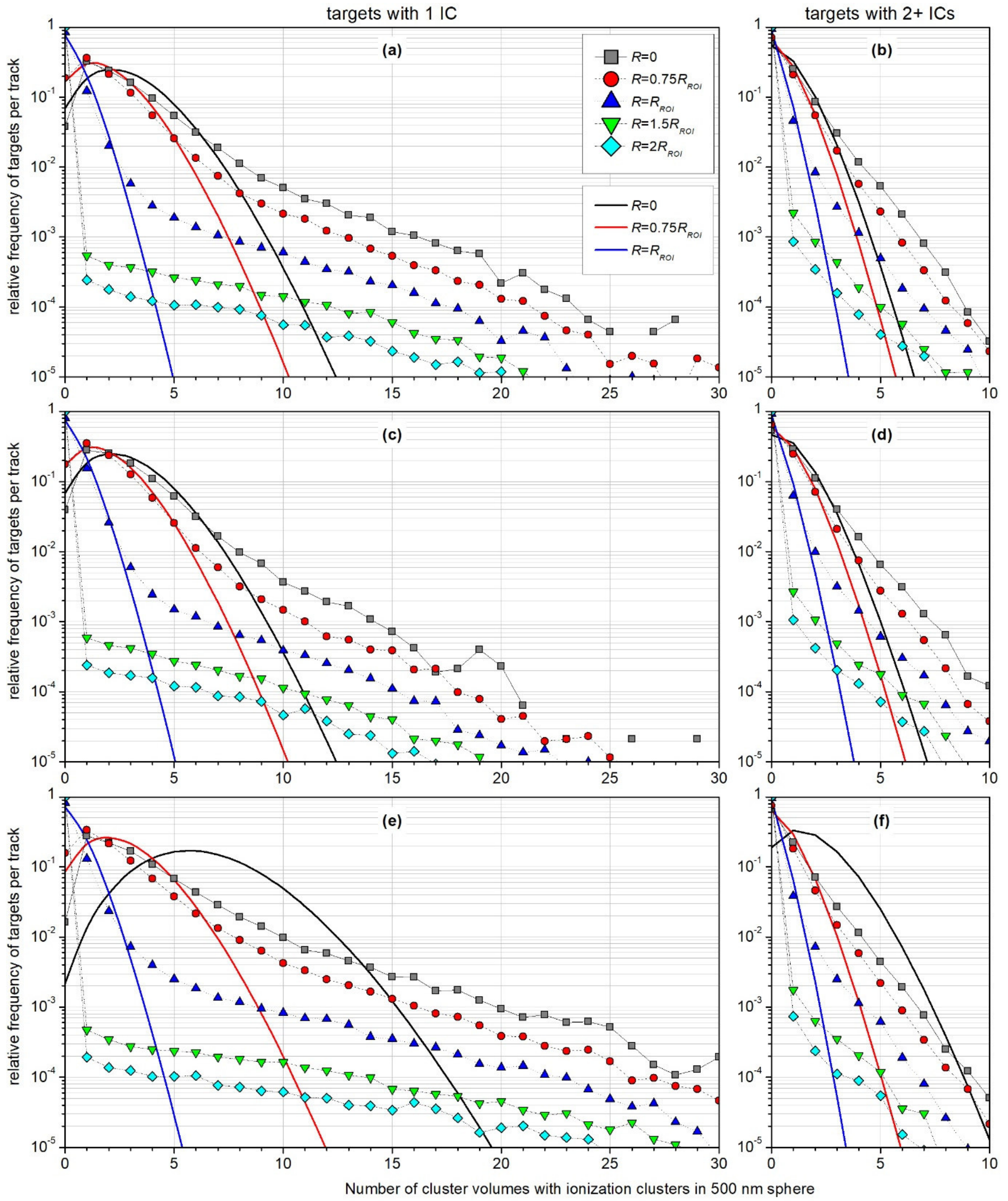

**Fig. S8:**The symbols indicate the frequency distributions of the number of cluster volumes (targets) inside a spherical region of interest (ROI) with radius $R_{ROI}$ = 250 nm that receive a single ionization cluster (IC) (left column) or two or more ionization clusters (right column) when a proton track (50 MeV proton energy) passes the ROI with an impact parameter $R$ as given in the legend. The ICs are scored in spherical targets of (a) and (b) 2 nm, (c) and (d) 3 nm, and (e) and (f) 2.5 nm diameter. The cluster volumes are spheres of targets of (a) and (b) 12 nm, (c) and (d) 18 nm, and (e) and (f) 7.5 nm diameter. The solid lines indicate Poisson distributions that have the same mean value as the respective corresponding distributions indicated by symbols.

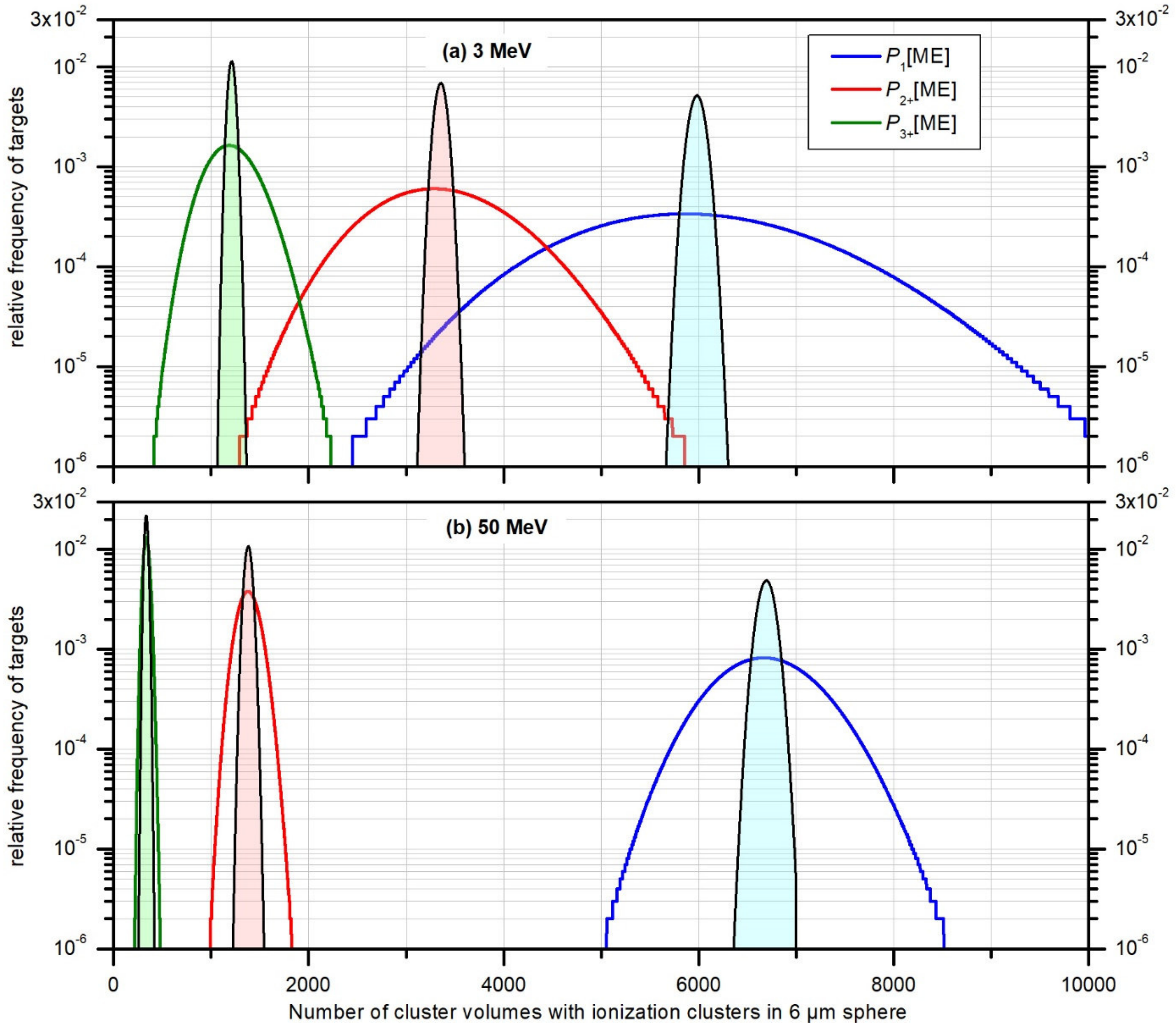

**Fig. S9:** Frequency distributions of the number of cluster volumes (targets) inside a spherical region of interest (ROI) with radius $R_{ROI}$ = 6 µm that receive a single ionization cluster (blue line), two or more ionization clusters (red line), or three and more ionization clusters (green line) for irradiation with a proton beam of 9.9 µm diameter at a fluence corresponding to an absorbed dose of 2 Gy. The proton energy is (a) 3 MeV and (b) 50 MeV, the ionization clusters are scored in spherical targets of 2 nm diameter, and the considered cluster volumes are spheres of 12 nm diameter as taken from Schneider et al. (2019). The track data used for evaluation have been taken from the work of Alexander et al. (2015).

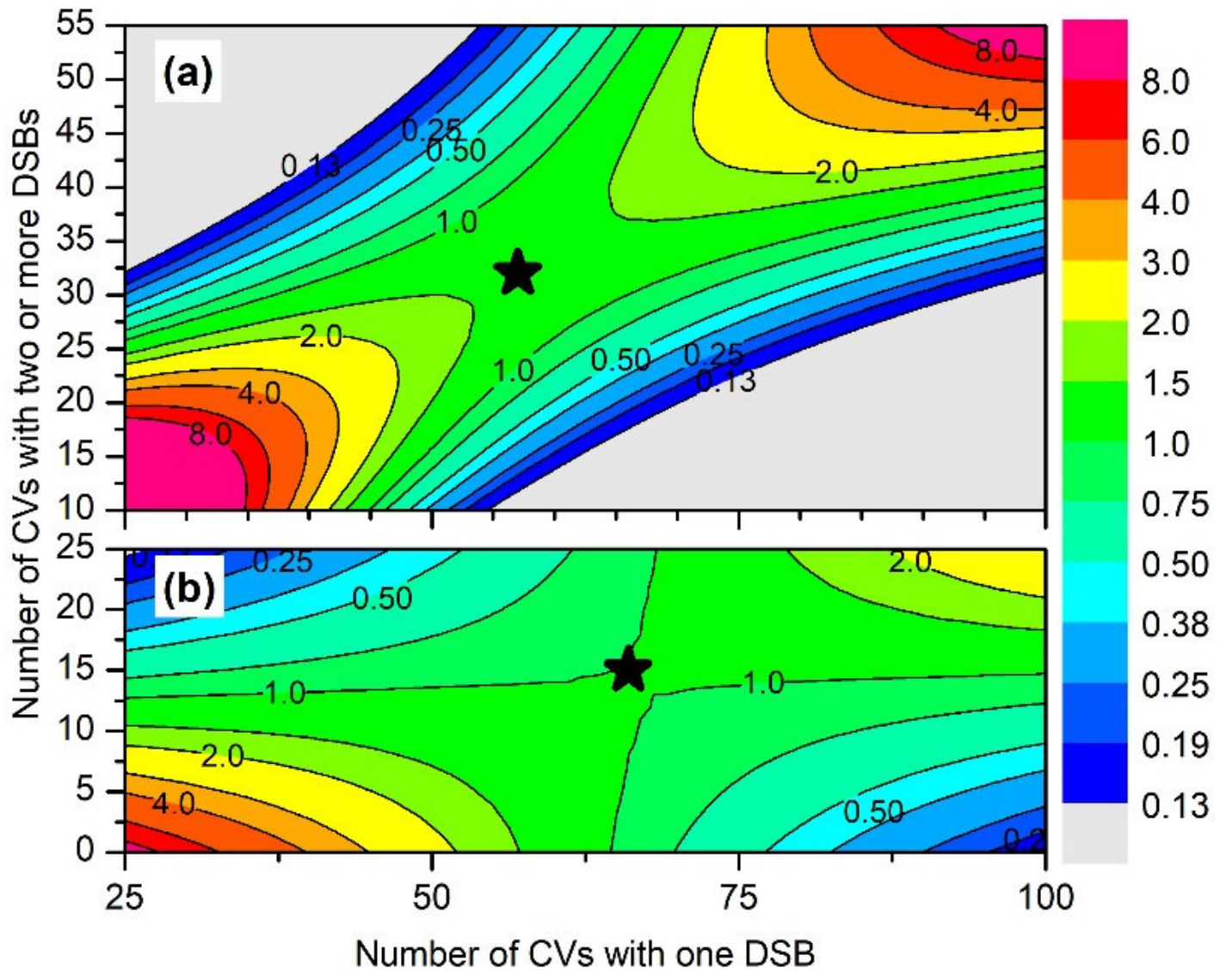

**Fig. S10:** Bivariate frequency of simultaneous occurrence of a number of cluster volumes (CVs) with one DSB (shown on the *x*-axis) and a number of CVs with two or more DSBs (*y*-axis) normalized to the product of the marginal frequencies. The asterisks mark the location of the modal values of the marginal distributions. The data apply to protons of (a) 3 MeV and (b) 50 MeV energy, an absorbed dose of 2 Gy, and a constant probability of 0.01 for an ionization cluster to be converted to a DSB. The basic interaction volume diameter was 2.0 nm and the cluster volume diameter 12.0 nm as in (Schneider et al. 2019). The track data used for evaluation have been taken from the work of Alexander et al. (2015).

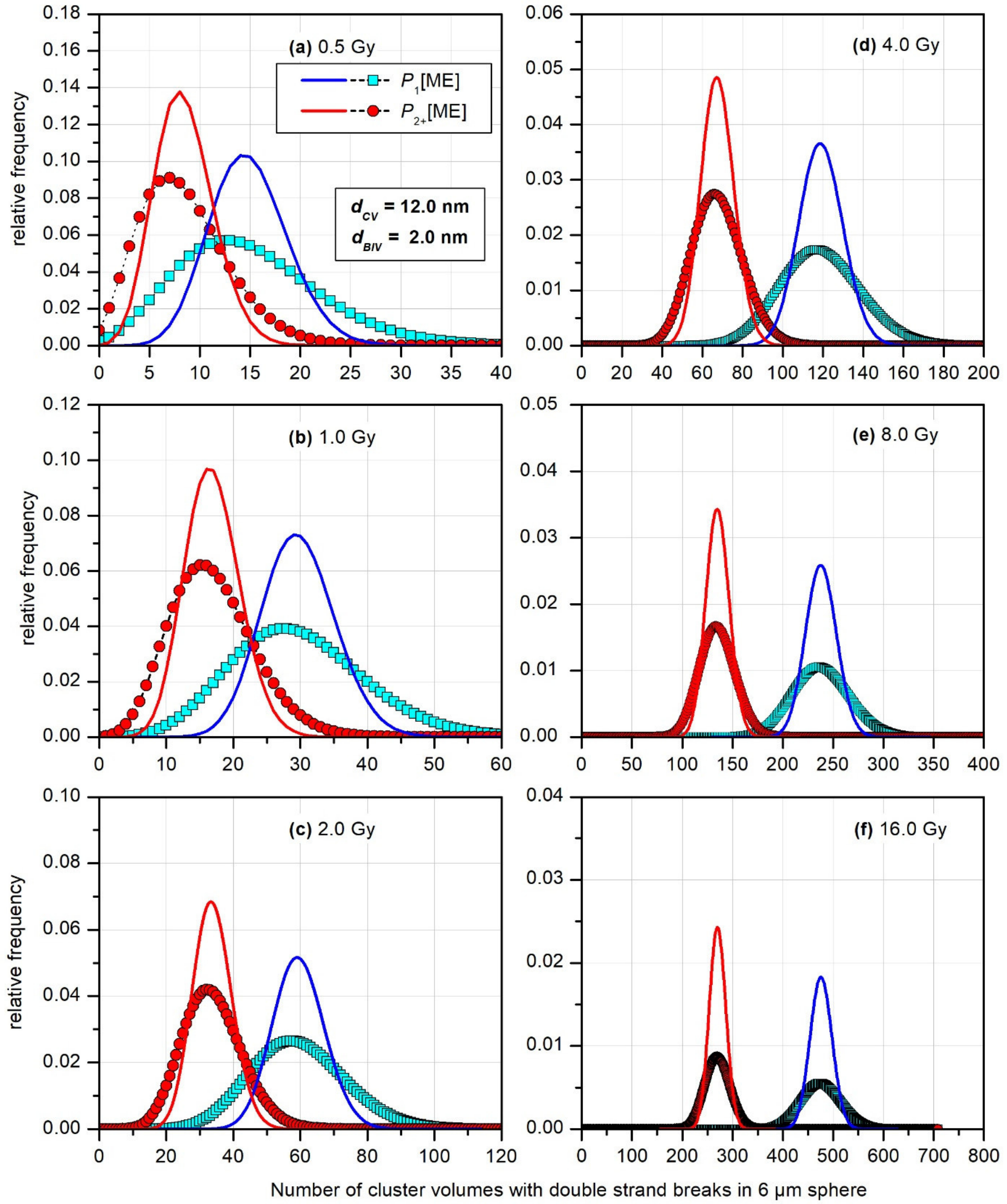

**Fig. S11:** Frequency distributions of the number of cluster volumes (CVs) inside a spherical region of interest (ROI) with radius $R_{ROI}$ = 6 µm that receive a single ionization cluster (IC) (blue squares) or two or more ICs (red circles) for irradiation with a proton beam of 9.9 µm diameter and 3 MeV energy at different values of absorbed dose: (a) 0.5 Gy, (b) 1 Gy, (c) 2 Gy, (d) 4 Gy, (e) 8 Gy, and (f) 16 Gy. The blue and red solid lines are Poisson distributions of the same mean value as the corresponding data represented by symbols. The results correspond to ICs scored in spherical targets of 2 nm diameter, spherical CVs of 12 nm diameter as taken from (Schneider et al. 2019), and an assumed uniform probability of 0.1% that a CV is filled with DNA. The track data used for evaluation have been taken from the work of Alexander et al. (2015).

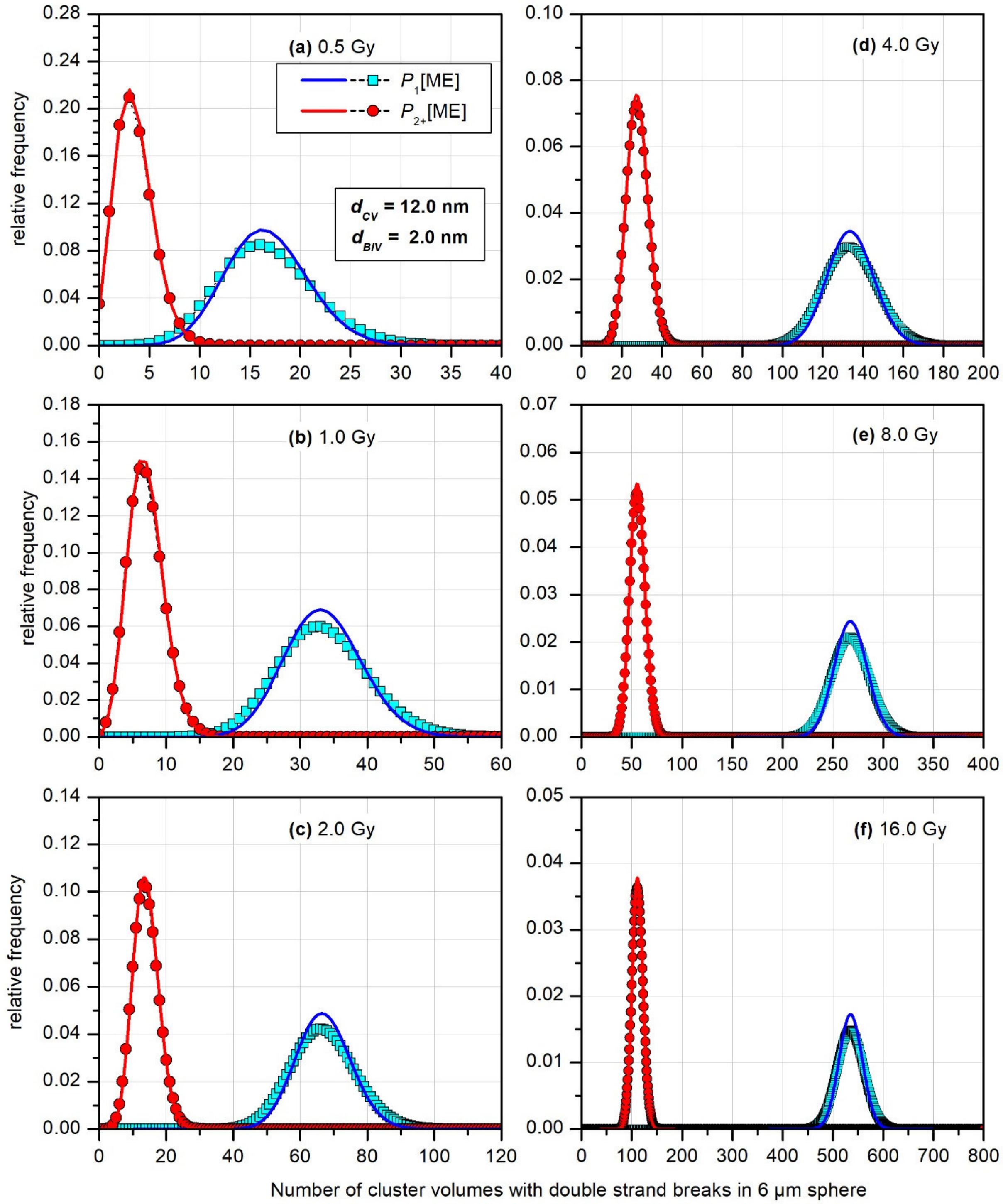

**Fig. S12:** Frequency distributions of the number of cluster volumes (CVs) inside a spherical region of interest (ROI) with radius $R_{ROI}$ = 6 µm that receive a single ionization cluster (IC) (blue squares) or two or more ICs (red circles) for irradiation with a proton beam of 9.9 µm diameter and 50 MeV energy at different values of absorbed dose: (a) 0.5 Gy, (b) 1 Gy, (c) 2 Gy, (d) 4 Gy, (e) 8 Gy, and (f) 16 Gy. The blue and red solid lines are Poisson distributions of the same mean value as the corresponding data represented by symbols. The results correspond to ICs scored in spherical targets of 2 nm diameter, spherical CVs of 12 nm diameter as taken from (Schneider et al. 2019) and an assumed uniform probability of 0.1% that a CV is filled with DNA. The track data used for evaluation have been taken from the work of Alexander et al. (2015).

# Supplemental material to “Investigation into the foundations of the track-event theory of cell survival and the radiation action model based on nanodosimetry” in Radiation and Environmental Biophysics

**Sonwabile Arthur Ngcezu[1], Hans Rabus[2,*]**

[1] *University of the Witwatersrand, Johannesburg, 2000, South Africa*
[2] *Physikalisch-Technische Bundesanstalt (PTB), 10587 Berlin, Germany*
* email: hans.rabus@ptb.de

## Source codes of the programs used in this study

*FORTRAN source code of program Radial_Weight*

This program calculates the weighting factors used in Subsection "Probability of inducing an IC in a BIV by proton tracks" to get from the cumulative nanodosimetric ionization cluster frequencies $F_k$ as a function of radial distance $r$ from the primary particle trajectory the expected (mean) number of targets within a sphere for the cases that the primary particle trajectory intersects the sphere or that it is within cylinder shells whose inner and outer radii are subsequent integer multiples of the sphere radius.

----- FORTRAN source code of program Radial_Weight -----

```
      PROGRAM Radial_Weight
C!>>> Declarations >>>>>>>>>>>>>>>>>>>>>>>>>>>>>>>>>>>>>>>>>>>>>>>>>>>>
      IMPLICIT NONE
C!    ----- Parameters: Version Date
      CHARACTER VDATE*11
      PARAMETER(VDATE='12-MAR-2021')
C
C!    PURPOSE: Calculates the weighting factors that are needed to get
C!    from the cumulative nanodosimetric ionization cluster frequencies
C!    F_k as a function of radial distance r from the primary particle
C!    trajectory the expected (mean) number of targets within a sphere
C!    for the cases that the primary particle trajectory intersects the
C!    the sphere or that it is within cylinder shells whose inner and
C!    outer radii are subsequent integer multiples of the sphere radius.
C
C!    ----- Parameters: General purpose constants
      INTEGER*4 LUN
      REAL*8 ONE, PI, ZERO
      PARAMETER(LUN=11, ONE=1.0, PI=3.1415926, ZERO=0.0)
C!    ----- Parameters for #####
      INTEGER*4 MCYLSH, MRSTEP, MXRVAL, MXYCOL
      PARAMETER(MCYLSH=201, MRSTEP=121, MXRVAL=3*MRSTEP+1,
     &          MXYCOL=2*MCYLSH)
C!    ----- Parameters for #####
      REAL*8 DCYLNM, HCYLNM
      PARAMETER(DCYLNM=2.3E0, HCYLNM=3.4E0)
C!    ----- Functions
      REAL*8 RANDY
      EXTERNAL RANDY
C!    ----- Input parameters for geometry
      INTEGER*4 NBATCH, NCYLSH, NRSTEP, NSAMPL
      REAL*8 D_LIV, R_LIV ! Diameter and radius of LIV in nm
      REAL*8 D_BIV ! Diameter and radius of BIV in nm
C!    ----- Local scalars
```

----- FORTRAN source code of program Radial_Weight -----

```
      CHARACTER FILENM*80, HEADER*80
      INTEGER*4 I, IX, IY, IZ, J, K, L, M, N, NRVAL
      REAL*8 DELTAR, POINTS, RSQR, FNORM, X, Y, YY, Z
C!    ----- Local arrays
C!    RSQMAX: Square of outer radius of cylindrical shells
C!    RSQMIN: Square of inner radius of cylindrical shells
C!    RVALUE: Radial coordinate of the inner cylinder shell
C!    SCORE:  Number of data points in intersection of sphere and
C!            cylindrical shells of adjacent multiples of sphere radius
C!    XYOUT:  Auxiliary array for tabulating output
      REAL*8 RSQMAX(MCYLSH), RSQMIN(MCYLSH),
     &       RVALUE(MXRVAL,MCYLSH), SCORE(MXRVAL,MCYLSH), XYOUT(MXYCOL)
C!    ------
      LOGICAL DEBUG
C!<<<<End declarations<<<<<<<<<<<<<<<<<<<<<<<<<<<<<<<<<<<<<<<<<<<<<<<<

      PRINT*, 'Program Radial_Weight Version: '//VDATE

      IF(1.EQ.1) THEN ! DUMMY block for readability: Initialize 1111111
C       Initialize Random number see sectionds
        IX = 12345
        IY = 23498
        IZ = 76345
        POINTS=ZERO
        DO I=1,MXRVAL
          DO J=1,MCYLSH
            RVALUE(I,J)=ZERO
            SCORE(I,J)=ZERO
          END DO
        END DO
        D_BIV=2.0
      END IF          ! DUMMY block for readability: Initialize 1111111

      IF(2.EQ.2) THEN ! DUMMY block for readability: Read input 2222222

        PRINT*,'Diameter of BIV (D_BIV) in nm'
        READ(*,*) D_BIV
        PRINT*,'D_BIV =',D_BIV,'nm '

        PRINT*,'Diameter of LIV (D_LIV) in nm'
        READ(*,*) D_LIV
        PRINT*,'D_LIV =',D_LIV,'nm'

        PRINT*,'Number of radial shells (<=',MCYLSH,')'
        READ(*,*) NCYLSH
        IF(NCYLSH.GT.MCYLSH) NCYLSH = MCYLSH
        PRINT*,'NCYLSH = ',NCYLSH

        PRINT*,'Number of steps per radius of LIV (<=',MRSTEP,')'
        READ(*,*) NRSTEP
        IF(NRSTEP.GT.MRSTEP) NRSTEP = MRSTEP
        PRINT*,'NRSTEP = ',NRSTEP

        PRINT*,'Number of batches (after which a message is printed)'
        READ(*,*) NBATCH
        PRINT*,'NBATCH = ',NBATCH

        PRINT*,'Number of random samples per batch'
        READ(*,*) NSAMPL
        PRINT*,'NSAMPL = ',NSAMPL
```

----- FORTRAN source code of program Radial_Weight -----

```
      END IF          ! DUMMY block for readability: Read input 2222222

      IF(3.EQ.3) THEN ! DUMMY block for readability: Init arrays 333333
        R_LIV=D_LIV/2.
        NRVAL=NRSTEP*3+1

        DELTAR=ONE/FLOAT(NRSTEP)
        DO J=1,NRVAL
          RVALUE(J,1)=FLOAT(J-1)*DELTAR
        END DO !J=1,NRVAL
        DO I=2,NCYLSH
          DO J=1,NRVAL
            RVALUE(J,I)=FLOAT(I-2)+FLOAT(J-1)*DELTAR
          END DO !J=1,NRVAL
        END DO !I=2,NCYLSH

        DO I=1,NCYLSH
            X=FLOAT(I-1)
            RSQMIN(I)=X*X
            RSQMAX(I)=RSQMIN(I)+2*X+ONE
        END DO !I=1,NCYLSH

      END IF          ! DUMMY block for readability: Init arrays 333333

      IF(4.EQ.4) THEN ! DUMMY block for readability: Main loop 44444444

      DO N=1,NBATCH ! Main Loop
      DO I=1,NSAMPL ! Main Loop
C!      Algorithm: Random sample data point in one octant and check
C!                 whether point is in sphere. Then increase # tries by
C!                 one and check for L-th possible radial distance from
C!                 primary trajectory within the K-th annulus whether
C!                 the point shifted by the radial distance along x axis
C!                 falls within the K-th annulus.
C!                 To reduce number of calls to random number generator
C!                 the value -x is also considered.
        Y=RANDY(IX, IY, IZ)
        YY=Y*Y
        X=RANDY(IX, IY, IZ)
        IF(X*X+YY.LE.ONE) THEN ! Inside sphere cross section
          Z=SQRT(ONE-X*X-YY)
          DO J=1,2 ! exploit symmetric points
            POINTS=POINTS+ONE
            DO K=1,NCYLSH
              DO L=1,NRVAL
                RSQR=(X-RVALUE(L,K))*(X-RVALUE(L,K))+YY
                IF(RSQR.GE.RSQMIN(K).AND.RSQR.LT.RSQMAX(K)) THEN
                  SCORE(L,K)=SCORE(L,K)+Z
                END IF
              END DO ! L=1,NRVAL
            END DO ! K=1,NCYLSH
C!          Flip signs
            X=-X
          END DO ! J=1,2
        END IF !(X*X+Y*Y+.LE.ONE)
C!
      END DO !  I=1,NSAMPL ! Main Loop
      PRINT*, NSAMPL*N,'/',NSAMPL*NBATCH
      END DO !  N=1,NBATCH ! Main Loop

      END IF          ! DUMMY block for readability: Main loop 44444444
```

----- FORTRAN source code of program Radial_Weight -----

```
      PRINT*, 'Estimate for Pi:',2.*POINTS/NSAMPL/NBATCH
      PRINT*, 'Ratio with Pi:',2.*POINTS/NSAMPL/NBATCH/PI

      IF(7.EQ.7) THEN ! DUMMY block for readability: Normalize 77777
C!      Notes:
C!      a) Normalization to number of points within sphere cross-section
C!         for y>0 corresponds to division by area of 1/2 of unit
C!         circle, i.e., pi/2.
C!      b) Multiplying by pi the effectively produces factor 2 needed
C!         to compensate that only y>0 was scored.
C!      c) Lead factor 2 as only half the chord length was scored above
        FNORM=2.*PI/POINTS ! 10-MAR-2021
        DO K=1,NCYLSH
          DO L=1,NRVAL
            RVALUE(L,K)=RVALUE(L,K)*R_LIV
            SCORE(L,K)=SCORE(L,K)*FNORM
          END DO ! L=1,NRVAL
        END DO ! K=1,NCYLSH
      END IF          ! DUMMY block for readability: Normalize 77777

      IF(9.EQ.9) THEN ! DUMMY block for readability: Write output 99999

        OPEN(LUN,FILE='RW_VolumeFraction.dat',STATUS='UNKNOWN')
        WRITE(LUN,*) '"Output from program Radial_Weight Version '
     &               //VDATE//'"'
        WRITE(LUN,*) '"* Weighting factors of annuli for unit sphere *"'
        WRITE(LUN,*) 'D_LIV/nm= ',D_LIV,' D_BIV/nm= ',D_BIV,
     &               ' NSAMPL= ',NSAMPL*NBATCH
        WRITE(LUN,'(1000a26)') ('      X            Y      ',J=1,NCYLSH)
        FNORM=3./4./PI
        DO I=1,NRVAL
          DO J=1,NCYLSH
            XYOUT(2*J-1)=RVALUE(I,J)
            XYOUT(2*J)= SCORE(I,J)*FNORM
          END DO
          WRITE(LUN,'(1000(2f13.6))') (XYOUT(J),J=1,2*NCYLSH)
        END DO
        CLOSE(LUN)

        OPEN(LUN,FILE='RW_BIV_TARGETS.dat',STATUS='UNKNOWN')
        WRITE(LUN,*) '"Output from program Radial_Weight Version '
     &               //VDATE//'"'
        WRITE(LUN,*) '"* Annulus weight for LIV&BIV of Schneider2019 *"'
        WRITE(LUN,*) 'D_LIV/nm= ',D_LIV,' D_BIV/nm= ',D_BIV,
     &               ' NSAMPL= ',NSAMPL*NBATCH
        WRITE(LUN,'(1000a26)') ('      X            Y      ',J=1,NCYLSH)
C!      The factor in the following code line is the product of
C!      a) the scaling factor for the sphere's volume (calculation above
C!         was for unit sphere) R_LIV**3
C!      b) the volume density of targets (1 per BIV volume)
        FNORM=6./PI*(R_LIV/D_BIV)**3
        DO I=1,NRVAL
          DO J=1,NCYLSH
            XYOUT(2*J-1)=RVALUE(I,J)
            XYOUT(2*J)= SCORE(I,J)*FNORM
          END DO
          WRITE(LUN,'(1000(2f13.6))') (XYOUT(J),J=1,2*NCYLSH)
        END DO
        CLOSE(LUN)
```

----- FORTRAN source code of program Radial_Weight -----

```
         OPEN(LUN,FILE='RW_SAN_TARGETS.dat',STATUS='UNKNOWN')
         WRITE(LUN,*) '"Output from program Radial_Weight Version '
     &               //VDATE//'"'
         WRITE(LUN,*) '"* Weighting factors of annuli for rescaled LIVs '
     &               //'& BIVs corresponding to Sonwabile''s data  *"'
         WRITE(LUN,'(4(a,f8.2),a,i10)') 'D_LIV/nm= ',D_LIV,' D_BIV/nm= ',
     &               D_BIV, ' D_cyl/nm= ',DCYLNM,' H_cyl/nm= ',HCYLNM,
     &               ' NSAMPL= ',NSAMPL*NBATCH
C!       Modifications as of 12-MAR-2021:
C!       1. Calculate the ratio of a) diameter of a sphere of same volume
C!          as the cylinder and b) diameter of the BIV
         FNORM=EXP(LOG(1.5*DCYLNM*DCYLNM*HCYLNM)/3.)/D_BIV
C!       2. Rescale LIV and BIV and radial distances
         D_LIV=FNORM*D_LIV
         R_LIV=FNORM*R_LIV
         D_BIV=FNORM*D_BIV
         DO I=1,NRVAL
           DO J=1,NCYLSH
             RVALUE(I,J)=FNORM*RVALUE(I,J)
           END DO
         END DO
C
         WRITE(LUN,'(2(2(a,f8.2),a))') ' D_cyl/nm= ',DCYLNM,
     &               ' H_cyl/nm= ',HCYLNM, ' --> rescaled values: ',
     &               ' D_LIV/nm= ',D_LIV,' D_BIV/nm= ',D_BIV
C!      End modifications 12-MAR-2021
         WRITE(LUN,'(1000a26)') ('      X            Y      ',J=1,NCYLSH)
C!       The factor in the following code line is the product of
C!       a) the scaling factor for the sphere's volume (calculation above
C!          was for unit sphere) R_LIV**3
C!       b) the volume density of targets (1 per cylinder volume)
         FNORM=R_LIV**3/(PI/4.*DCYLNM*DCYLNM*HCYLNM)
         DO I=1,NRVAL
           DO J=1,NCYLSH
             XYOUT(2*J-1)=RVALUE(I,J)
             XYOUT(2*J)= SCORE(I,J)*FNORM
           END DO
           WRITE(LUN,'(1000(2f13.6))') (XYOUT(J),J=1,2*NCYLSH)
         END DO
         CLOSE(LUN)

       END IF          ! DUMMY block for readability: Write output 99999


C
       END ! PROGRAM Radial_Weight
C!______________________________________________________________________


       REAL*8 FUNCTION RANDY(IX, IY, IZ)
C      Random Number Generator from
C      Wichmann, B.A. and I.D. Hill, Algorithm AS 183: An Efficient
C      and Portable Pseudo-Random Number Generator,
C      Applied Statistics, 31, 188-190, 1982.
       INTEGER*4 MX, MY, MZ, NX, NY, NZ
       REAL*4 AX, AY, AZ, ONE
       PARAMETER (MX=171, MY=172, MZ=170, NX=30269, NY=30307, NZ=30323,
     &           AX=30269., AY=30307., AZ=30323., ONE=1.0)
       IX = MOD(MX * IX, NX)
       IY = MOD(MY * IY, NY)
       IZ = MOD(MZ * IZ, NZ)
```

```
----- FORTRAN source code of program Radial_Weight -----
      RANDY = AMOD(FLOAT(IX)/AX + FLOAT(IY)/AY + FLOAT(IZ)/AZ, 1.)
      RETURN
      END ! FUNCTION RANDY
```

*Excel VBA source code of routine convol*

This routine is used in an Excel Workbook in which worksheet “Trackdata” includes the results from Braunroth et al. (2020) for the radial dependence of parameter $F_2$, which are multiplied by $2\pi r$ in wotksheet “Tracks”. Worksheet “RR_SAN” contains the output file for the weighting functions for the different annuli calculated with the code listed in Subsection “FORTRAN source code of program Radial_Weight”. The routine calculates the integral on the right-hand side of Eq. (31) in the paper and writes the results into worksheet “Convol”. These results are further processed in additional worksheets to calculate the mean number of ionization clusters (ICs) for a single event or a given value of dose, from which the the probabilities for cluster volumes with a single IC or more than one IC are calculated from Poisson statistics.

```
----- Excel VBA source code of routine convol -----
Sub convol()

Dim Weight As Range, Track As Range, Ziel As Range
Dim EPrim As Integer, Ringe As Integer, Zeilen As Integer, I As Integer,
J As Integer, K As Integer, L As Integer
Dim Count As Double
'
  I = 1
  Set Weight = Worksheets("RR_SAN").Cells(6, 2)
  dx = Weight.Cells(2, 1).Value

  Set Track = Worksheets("Tracks").Cells(3, 1)
  Set Ziel = Worksheets("Convol").Cells(1, 1)
  EPrim = Application.WorksheetFunction.Count(Track.EntireRow) / 2
  Ringe = Application.WorksheetFunction.Count(Weight.EntireRow) / 2
  Zeilen = Application.WorksheetFunction.Count(Weight.EntireColumn)
  Application.ScreenUpdating = False
  Application.Calculation = xlCalculationManual

  For I = 1 To EPrim
    Application.StatusBar = I & "/" & EPrim
    Ziel.Cells(1, I + 1).Value = Track.Offset(-2, 2 * I - 1).Value
    For J = 1 To Ringe
      Ziel.Cells(J + 1, 1).Value = J - 1
      Count = Weight.Cells(1, 2 * J).Value * Track.Cells(1, 2 * I).Value
      For K = 1 To Zeilen
        X = Weight.Cells(K, 2 * J - 1).Value
        L = 2
        While Track.Cells(L, 2 * I - 1).Value < X
          L = L + 1
        Wend
        XL = Track.Cells(L - 1, 2 * I - 1).Value
        XH = Track.Cells(L, 2 * I - 1).Value
        YL = Track.Cells(L - 1, 2 * I).Value
        YH = Track.Cells(L, 2 * I).Value
        Y = (X - XL) * (YH - YL) / (XH - XL) + YL
        Count = Count + Y * Weight.Cells(K, 2 * J).Value
      Next K
      Ziel.Cells(J + 1, I + 1).Value = Count * dx

    Next J
  Next I

  Application.ScreenUpdating = True
  Application.Calculation = xlCalculationAutomatic
```

| ----- Excel VBA source code of routine convol ----- |
|---|
| `Application.StatusBar = False`<br><br>`End Sub` |

*FORTRAN source code of program IC_3D*

This program reads particle tracks from an Ascii file listing the energy transfer points from ionizations in particle tracks and determines ionization clusters and saves the cluster positions (and their complexity) to an output file named IC_*'input_filename'*. It expects an input file in the format (number of event, *z*-position, radial position, azimuth, energy deposited, *x*-position, *y*-position) or (number of event, particle type, process, *x*-position, *y*-position, *z*-position).

Uses subroutines CLUSTR and BLINIT (included from file BLINIT.f, see section "FORTRAN subroutine BLINIT") and expects input of the name of a file listing the debugging options (example see section "Sample debug options file for use with programs IC_3D and ROI_3D").

Notes:

- **The program must be executed in the directory where the data files are located**.
- The program expects input of the name of a file listing the debugging options (example see section "Sample debug options file for use with programs IC_3D and ROI_3D").
- Inputs of parameters and input file names are prompted for unless they are entered via a text file. (Manual input is initiated by entering 0 with the first prompt.) A sample input file is listed in the table below.

----- Sample input file for IC_3D -----

```
210605.16   ! Command file version number ...
IC_3D       ! ... for this program
DEBUG.opt   ! Debug options file name
2           ! DBIV in nm
0.          ! ZROI(1) Start of region of interest
10000.      ! ZROI(2) End of region of interest
TZ%%%XY     ! CODE for input line structure
4           ! NHEADL Number of Header lines
5           ! NFILES
SPRE_p100MeVA_mai2015.dat
SPRE_p50MeVA_mai2015.dat
SPRE_p10MeVA_mai2015.dat
SPRE_p25MeVA_mai2015.dat
SPRE_p3MeVA_mai2015.dat
```

Meaning of the input lines:

(1) Version date and time of the input file structure in format YYMMDD.HHMM
(2) Name of the program
(3) Name of the file listing the debug options (see section "Sample debug options file for use with programs IC_3D and ROI_3D")
(4) Diameter of the target spheres for scoring ionization clusters (basic interaction volume of ( Schneider et al. 2019, 2020))
(5) *z*-coordinate of the start of the region of interest
(6) *z*-coordinate of the end of the region of interest
(7) Encoding of the input file structure. Each letter stand for the meaning of an entry in an input line. T means the number of the primary particle track, X,Y, and Z the x, y, and z coordinates of the transfer point, E the energy deposit (optional). Example: The string 'TZ%%EXY ' indicates that there are seven entries in each line, of which the first is the track number, the second the z coordinate, the fifth the energy, and the sixth and seventh the x and y coordinates.
(8) number of header lines in the input files
(9) number of input files to process
(10) further lines: names of the input files

----- FORTRAN source code of program IC_3D -----

```
      PROGRAM IC_3D
C!>>> Declarations >>>>>>>>>>>>>>>>>>>>>>>>>>>>>>>>>>>>>>>>>>>>>>>>>>>
      IMPLICIT NONE
```

----- FORTRAN source code of program IC_3D -----

```
C!    ----- Parameters: Version Date and number
      CHARACTER VDATE*11
      REAL*8 VINPUT  ! Version number for command files YYMMDD.HHMM
C!                     (date and time  file structure was last changed)
C!    #################################################################
      PARAMETER(VDATE='05-JUN-2021',VINPUT=210605.16d0) ! ############
C!    #################################################################
C!    05-JUN-2021 HR:
C!      - Added timestap to output files
C!      - Reworked identification of input file structure
C!      - Added version check for input files
C!    03-JUN-2021 HR:
C!      - Optimized output
C!    02-JUN-2021 HR:
C!      - Fixed soft bug with output of header line
C!      - Fixed problem with formatted output
C!    28-MAY-2021 HR:
C!      - Fixed bug with length of SRDATE variable
C!    23-MAY-2021 HR:
C!      - Modified call to subroutines to get their version date
C!    23-MAY-2021 HR:
C!      - Modified call to subroutines to get their version date
C!    20-MAY-2021 HR:
C!      - Adapted to potential site sizes >= 10 nm
C!    28-MAR-2021 HR:
C!      - Cleaned up unused variables
C!    20-MAR-2021 HR:
C!      - Created this clone of ROI_3D.f for ionization CLUSTR output
C!    ----- Parameters: General purpose constants
      INTEGER*4 LUN, LUNT
      REAL*8 ONE, ZERO
      PARAMETER(LUN=11, LUNT=12, ONE=1.0, ZERO=0.0)
C!    ----- Parameters for lattice orientation
      INTEGER*4 MAZMTH, MDIV, MTHETA, MDIR
      PARAMETER(MAZMTH=1, MDIV=0, MTHETA=2**MDIV,
     &           MDIR=MAZMTH*(MTHETA*(MTHETA+1))/2)
C!    ----- Parameters for track and radial distance histograms
      INTEGER*4 MIONIZ
      PARAMETER(MIONIZ=100000)
C!    ----- Input parameters for geometry
      REAL*8 DLATC  ! Cell lattice constant
      REAL*8 DSITE  ! Diameter of spherical target
C!    ----- Functions
      CHARACTER TSTAMP*24
C!    ----- Local scalars
      CHARACTER FILENM*80, FILOUT*84, HEADER*80, PREFIX*9
      CHARACTER*11 VDATES(2), SRDATE
      INTEGER*4 I, IFILE, IT, ITA, J, K, NAZMTH, NDIV, NFILES,
     &          NFORMT, NHEADL, NTRACS
      INTEGER*4 NIONIZ
      LOGICAL ASKINP
      REAL*8 DUMMY, PIBY4, VCHECK
      REAL*8 ZROI(2)
C!    ----- Local arrays
      REAL*8 RLINE(8)
C!    ----- Global variables
      INTEGER*4 NDIR
      REAL*8 ADJNTB(3,3,MDIR)
      COMMON /LATTICE/ ADJNTB, NDIR
C!    ------
      REAL*8 XYZ(MIONIZ,3)
```

----- FORTRAN source code of program IC_3D -----

```
      COMMON /TRACKS/ XYZ
C!    ------
      LOGICAL DEBUG(4), SRFLAG
      COMMON /VERBOSE/ DEBUG, SRFLAG
C!    ------
      INTEGER*4 ICSIZE(MIONIZ),NSITES
      COMMON /CLSTRS/ ICSIZE, NSITES
C!    ------
C!     CODE: Character string encoding the meaning of the entries in a
C!           line of the input file as follows:
C!           'T'     - number of the primary particle track
C!           'X','Y','Z' - x, y, and z coordinates of the transfer point
C!           'E'     - energy deposit (if applicable)
C!           'I'     - ionization cluster size (if applicable)
C!           '%'     - additional data that are not used
C!           Example: The string 'TZ%%EXY ' indicates that there are 7
C!                    entries in each line, of which the first is the
C!                    track number, the second the z coordinate, the
C!                    fifth the energy, and the sixth and seventh the x
C!                    and y coordinates
C!     IDT: Array of the columns indices of (1-3) x,y,z coordinates
C!                   (4) energy deposit (if present), (5) track number,
C!                   (6) number of ionizations in cluster (if present)
C!                   (7-8) are there for future use
      CHARACTER CODE*8
      INTEGER*2 IDT(8),NDT
      COMMON /RFORMT/IDT,NDT,CODE
C!<<<<End declarations<<<<<<<<<<<<<<<<<<<<<<<<<<<<<<<<<<<<<<<<<<<<<<<<

      PRINT*, 'Program IC_3D Version: '//VDATE
      SRFLAG=.TRUE. ! Invoke message on Version date from Subroutine
CLUSTR

      IF(1.EQ.1) THEN ! DUMMY block for readability: Read input 1111111
        DEBUG(1)=.FALSE. ! Main program
        DEBUG(2)=.FALSE. ! CLUSTR main sections
        DEBUG(3)=.FALSE. ! CLUSTR IDIR loop
        DEBUG(4)=.FALSE. ! CLUSTR IPOS loop details

        PIBY4=ATAN(ONE)
        PREFIX='IC_0.0nm_'

C!      Check command file version
        PRINT*, 'Enter 0 for manual input or version (YYMMDD.HHMM) '//
     &          'of command file structure '
        READ(*,*) FILENM
        READ(FILENM(1:11),*) VCHECK
        ASKINP=(VCHECK.EQ.ZERO)
        IF(.NOT.ASKINP) THEN
          IF(VCHECK.LT.VINPUT) THEN
            PRINT*, 'Command file structure ',VCHECK,' older than '//
     &              'current version ',VINPUT,'=> STOP.'
            STOP
          END IF
          READ(*,*) FILENM
          IF(FILENM(1:5).NE.'IC_3D') THEN
            PRINT*, 'Command file appears not to be for IC_3D but '//
     &               'for '//FILENM//'=> STOP.'
            STOP
          END IF
        END IF
```

----- FORTRAN source code of program IC_3D -----

```
C!      Read debug options
        IF(ASKINP) PRINT*, 'Enter debug options file name or - for none'
        READ(*,*) FILENM
        OPEN(LUN,FILE=FILENM,STATUS='OLD',ERR=10)
        DO I=1,4
          READ(LUN,*) DEBUG(I)
        END DO ! I=1,4
        CLOSE(LUN)
  10    CONTINUE

C!      Read geometry parameters
        IF(ASKINP) PRINT*, 'Enter site diameter in nm'
        READ(*,*) DSITE
        DLATC=DSITE*SQRT(2.)*EXP(LOG(PIBY4/3)/3.)
                IF(DSITE.LT.10.0d0) THEN
          WRITE(PREFIX(4:6),'(f3.1)') DSITE
        ELSE
          IF(DSITE.LT.100.0d0) THEN
            WRITE(PREFIX(4:6),'(f3.0)') DSITE
          ELSE
            WRITE(PREFIX(4:6),'(i3)') INT(DSITE)
          END IF
        END IF

        IF(ASKINP) PRINT*, 'z position of begin of region of interest'
        READ(*,*) ZROI(1)
        IF(ASKINP) PRINT*, 'z position of end of region of interest'
        READ(*,*) ZROI(2)

        IF(ASKINP) PRINT*, 'Enter 8 character code for data file '//
     &       'structure where ''T'' indicates the track ID,'//
     &       ' ''X'',''Y'' and ''Z'' the respective coordinates, '//
     &       '''E'' the energy deposit (if present) and ''&'' any '//
     &       'other data'
        READ(*,*) CODE
        CALL RFINIT()

        IF(ASKINP) PRINT*, 'Number of header lines'
        READ(*,*) NHEADL

        IF(ASKINP) PRINT*, 'Number of files to process'
        READ(*,*) NFILES
      END IF          ! DUMMY block for readability: Read input 1111111

      IF(2.EQ.2) THEN ! DUMMY block for readability: Initialize 2222222
        NAZMTH=MAZMTH
        NDIV=MDIV
        IF(DEBUG(1)) PRINT*, 'Before CALL BLINIT'
        CALL BLINIT(NAZMTH,NDIV,DLATC,SRDATE) ! Init reziprocal lattice
        VDATES(1)=SRDATE
        IF(DEBUG(1)) PRINT*, 'After CALL BLINIT'
      END IF          ! DUMMY block for readability: Initialize 2222222

      DO IFILE=1,NFILES
        PRINT*, 'Enter file name ',IFILE
        READ(*,*) FILENM

        IF(4.EQ.4) THEN ! DUMMY block for readability: Main Loop  444444
          IF(DEBUG(1)) PRINT*, 'Begin of Block 4'
C!        Init counters
```

----- FORTRAN source code of program IC_3D -----

```
      NIONIZ=1
      NTRACS=1

      IF (NFORMT.EQ.1) THEN
        FILOUT=PREFIX//FILENM(6:80)
      ELSE
        FILOUT=PREFIX//FILENM
      END IF

      OPEN(LUN,FILE=FILENM,STATUS='OLD')
      DO I=1,NHEADL
        READ(LUN,'(a)') HEADER
      END DO
      READ(LUN,*) (RLINE(I),I=1,NDT)
      ITA=NINT(RLINE(IDT(5)))
      DO I=1,3
        XYZ(NIONIZ,I)=RLINE(IDT(I))
      END DO

      NIONIZ=NIONIZ+1

100   CONTINUE ! >>>>>>>>>>>>>>>>>>>>>>>>>>>>>>>>>>>>>>>>>>>>>>>>>>>
      READ(LUN,*,END=200,ERR=200) (RLINE(I),I=1,NDT)
      IT=NINT(RLINE(IDT(5)))
      DO I=1,3
        XYZ(NIONIZ,I)=RLINE(IDT(I))
      END DO

      IF(IT.EQ.ITA) THEN
        NIONIZ=NIONIZ+1
      ELSE ! Entry from a new track was read
        NIONIZ=NIONIZ-1 ! Decrease to number of ionizations in track
        IF(DEBUG(1)) PRINT*, 'Before CALL CLUSTR'
        IF(MOD(NTRACS,10).EQ.0) PRINT*, NTRACS
        CALL CLUSTR(NIONIZ,ZROI,SRDATE)
        VDATES(2)=SRDATE
        IF(NTRACS.EQ.1) THEN
          OPEN(LUNT,FILE=FILOUT,STATUS='UNKNOWN')
          WRITE(LUNT,'(a)') 'Output from IC_3D.f - Version '//
     &         VDATE//' BLINIT:'//VDATES(1)//' CLUSTR:'//VDATES(2)
     &               //' on '//TSTAMP()
          WRITE(LUNT,'(a,f8.3,a)') 'Ionization cluster positions'//
     &        ' in ion tracks for DSITE=',DSITE,' nm '
          WRITE(LUNT,'(a)') 'Processed input data file: '//FILENM
          WRITE(LUNT,'(a)') 'Track x/nm y/nm z/nm ICS'
        END IF
        DO K=1,NSITES-1
          DO I=K+1,NSITES
            IF(XYZ(K,3).GT.XYZ(I,3)) THEN
              DO J=1, 3
                DUMMY=XYZ(K,J)
                XYZ(K,J)=XYZ(I,J)
                XYZ(I,J)=DUMMY
              END DO ! J=1, 3
            END IF
          END DO ! I=K+1,NSITES
        END DO !  K=1,NSITES-1
        DO K=1,NSITES
          IF(ICSIZE(K).GT.1) THEN
            CALL WRLINE(LUNT,IT,XYZ(K,1),XYZ(K,2),XYZ(K,3),
     &                        ICSIZE(K))
```

----- FORTRAN source code of program IC_3D -----

```
            END IF
          END DO ! K=1,NSITES

          IF(DEBUG(1)) PRINT*, 'After CALL CLUSTR'
          DO J=1,3
            XYZ(1,J)=XYZ(NIONIZ+1,J) ! First ionization in new track
          END DO
          ITA=IT ! Remember new track number
          NTRACS=NTRACS+1
          NIONIZ=2 ! There was already one ionization
        END IF
        GOTO 100 ! <<<<<<<<<<<<<<<<<<<<<<<<<<<<<<<<<<<<<<<<<<<<<<<<<<<<<
 200    CONTINUE
          CALL CLUSTR(NIONIZ,ZROI,SRDATE)
          DO K=1,NSITES-1
            DO I=K+1,NSITES
              IF(XYZ(K,3).GT.XYZ(I,3)) THEN
                DO J=1, 3
                  DUMMY=XYZ(K,J)
                  XYZ(K,J)=XYZ(I,J)
                  XYZ(I,J)=DUMMY
                END DO ! J=1, 3
              END IF
            END DO ! I=K+1,NSITES
          END DO !  K=1,NSITES-1
          DO K=1,NSITES
            IF(ICSIZE(K).GT.1) THEN
              CALL WRLINE(LUNT,IT,XYZ(K,1),XYZ(K,2),XYZ(K,3),
     &                             ICSIZE(K))
            END IF
          END DO ! K=1,NSITES
          CLOSE(LUNT)
        CLOSE(LUN)
      END IF          ! DUMMY block for readability: Main Loop  4444444

      END DO ! IFILE=1,NFILES

      END PROGRAM ! IC_3D
C!______________________________________________________________________

      INCLUDE 'BLINIT.f'       ! <<<<<<<<<<<<<<<<<<<<<<<<<<<<<<<<<<<<<<<
      INCLUDE 'RFINIT.f'
      INCLUDE 'TSTAMP.f'


C!*********************************************************************
      SUBROUTINE CLUSTR(NIONIZ,ZROI,SRDATE)
C!    ****************************************************************
C!    Determine targets with ionization clusters
C!    ****************************************************************
C!>>> Declarations >>>>>>>>>>>>>>>>>>>>>>>>>>>>>>>>>>>>>>>>>>>>>>>>>>
      IMPLICIT NONE
C!    ----- Parameters: Version Date
      CHARACTER VDATE*11
      PARAMETER(VDATE='23-MAY-2021')
C!    23-MAY-2021 HR: Added outout of version date to calling program
C!    20-MAY-2021 HR: Fixed severe bug with sorting the hit targets
C!    25-MAR-2021 HR: Removed unneeded variables
C!    20-MAR-2021 HR: Created from ROI_3D_Subs.f
C!    ----- Parameters: General purpose constants
      REAL*8 ONE, ZERO
```

----- FORTRAN source code of program IC_3D -----

```
      PARAMETER(ONE=1.0, ZERO=0.0)
C!    ----- Parameters for lattice orientation
      INTEGER*4 MAZMTH, MDIV, MTHETA, MDIR
      PARAMETER(MAZMTH=1, MDIV=0, MTHETA=2**MDIV,
     &           MDIR=MAZMTH*(MTHETA*(MTHETA+1))/2)
C!    ----- Parameters for track and radial distance histograms
      INTEGER*4 KMAX, MIONIZ, MXRPOS, MXYPOS, MXTARG
      PARAMETER(KMAX=4, MIONIZ=100000, MXRPOS=41, MXTARG=1001,
     &          MXYPOS=4*MXRPOS*(MXRPOS-1)+1)
C!    ----- Local scalars
      INTEGER*4 I, IDIR, IR, J, K
      INTEGER*4 IHOLD, NIONIS
      REAL*8 SAVXYZ, SPROD
C!    ----- Local arrays
      INTEGER*4 ITARGT(MIONIZ,3)
      INTEGER*4 ICSITE(KMAX)
C!           ICSITE is the histogram of absolute frequencies of
C!                  targets with 1, 2, ... >=KMAX ionizations
C!                  occuring for a particular orientation of the
C!                  lattice w.r.t. to the track
C!    ----- Global variables
      INTEGER*4 NDIR
      REAL*8 ADJNTB(3,3,MDIR)
      COMMON /LATTICE/ ADJNTB, NDIR
C!    ------
      REAL*8 XYZ(MIONIZ,3)
      COMMON /TRACKS/ XYZ
C!    ------
      REAL*8 RADIST(MXRPOS),FREQRD(MXTARG,MXRPOS,KMAX)
      COMMON /HISTOG/ RADIST, FREQRD
C!           RADIST is the vector of radial distances
C!           FREQRD initially holds the sum, the sum of squares and the
C!                  sum of variances per track over all tracks.
C!                  In the main program this is converted in the end to
C!                  mean and variance over all analyzed tracks plus the
C!                  the average intra-track variance do to orientation
C!    ------
      LOGICAL DEBUG(4), SRFLAG
      COMMON /VERBOSE/ DEBUG, SRFLAG
C!    ------ Added 14-MAR-2021
      INTEGER*4 ICSIZE(MIONIZ),NSITES
      COMMON /CLSTRS/ ICSIZE, NSITES
C!    ----- Scalars and arrays passed from the calling module
      INTEGER*4 NIONIZ
      REAL*8  ZROI(2)
C!    ----- Scalars and arrays passed to the calling module
      CHARACTER*11 SRDATE
C!<<<<End declarations<<<<<<<<<<<<<<<<<<<<<<<<<<<<<<<<<<<<<<<<<<<<<<<<

      SRDATE=VDATE

      IF(SRFLAG) THEN
        PRINT*, 'Subroutine CLUSTR Version: ',VDATE
        SRFLAG=.FALSE.
      END IF
      IF(DEBUG(2)) PRINT*, 'CLUSTR Input',NIONIZ

      DO K=1,KMAX ! Empty local histograms
        ICSITE(K)=0
      END DO ! I=1,KMAX ! Empty local histograms
```

----- FORTRAN source code of program IC_3D -----

```
      IF(DEBUG(2)) PRINT*, 'CLUSTR vor Main Loop NDIR=',NDIR
      IF(DEBUG(2)) PRINT*, 'CLUSTR vor Main Loop NIONIZ=',NIONIZ
      DO IDIR=1,NDIR ! Loop over all orientations
        IF(IDIR.GT.NDIR) GOTO 100
        IF(DEBUG(3)) PRINT*, 'CLUSTR Begin Loop IDIR',IDIR
C!      # Find target volume for all  ionizations in track
        NIONIS=0
        DO I=1,NIONIZ
          IF(XYZ(I,3).GE.ZROI(1).AND.XYZ(I,3).LE.ZROI(2)) THEN
            NIONIS=NIONIS+1
            DO J=1,3
              SPROD= ADJNTB(1,J,IDIR)*XYZ(I,1)
     &              +ADJNTB(2,J,IDIR)*XYZ(I,2)
     &              +ADJNTB(3,J,IDIR)*(XYZ(I,3))
              ITARGT(NIONIS,J)=NINT(SPROD)
            END DO ! J=1,3
          END IF
        END DO ! I=1,NIONIZ

        IF(DEBUG(3)) PRINT*, 'CLUSTR vor sort volume indices',NIONIS
        IF(IDIR.GT.NDIR) STOP
C!      # Sort target volume indices ascending
        DO I=1,NIONIS
          DO J=I+1,NIONIS
            IF(    (ITARGT(I,1).GT.ITARGT(J,1))
     &         .OR.(ITARGT(I,1).EQ.ITARGT(J,1).AND.
     &              ITARGT(I,2).GT.ITARGT(J,2))
     &         .OR.(ITARGT(I,1).EQ.ITARGT(J,1).AND.
     &              ITARGT(I,2).EQ.ITARGT(J,2).AND.
     &              ITARGT(I,3).LE.ITARGT(J,3))) THEN
              DO IR=1,3
                IHOLD=ITARGT(I,IR)
                ITARGT(I,IR)=ITARGT(J,IR)
                ITARGT(J,IR)=IHOLD
                SAVXYZ=XYZ(I,IR)     ! 14-MAR-2021
                XYZ(I,IR)=XYZ(J,IR)  ! 14-MAR-2021
                XYZ(J,IR)=SAVXYZ     ! 14-MAR-2021
              END DO ! IR=1,3
            END IF
          END DO ! J=1,NIONIS
        END DO ! I=1,NIONIS

        IF(DEBUG(3)) PRINT*, 'CLUSTR vor find unique volumes',NIONIS
C!      # Find unique target volumes and score ionizations
        NSITES=1
        ICSIZE(NSITES)=1
        DO I=2,NIONIS
          IF(    ITARGT(I,1).NE.ITARGT(I-1,1)
     &       .OR.ITARGT(I,2).NE.ITARGT(I-1,2)
     &       .OR.ITARGT(I,3).NE.ITARGT(I-1,3)) THEN
            DO J=1,3 ! 14-MAR-2021
              XYZ(NSITES,J)=XYZ(NSITES,J)/REAL(ICSIZE(NSITES))
            END DO
            NSITES=NSITES+1
            ICSIZE(NSITES)=1
            DO J=1,3
              ITARGT(NSITES,J)=ITARGT(I,J)
              XYZ(NSITES,J)=XYZ(I,J)
            END DO ! J=1,3
          ELSE
            ICSIZE(NSITES)=ICSIZE(NSITES)+1
```

----- FORTRAN source code of program IC_3D -----

```
            DO J=1,3 ! 14-MAR-2021
              XYZ(NSITES,J)=XYZ(NSITES,J)+XYZ(I,J)
            END DO
          END IF
        END DO ! I=2,NIONIS

      END DO ! IDIR=1,NDIR ! Loop over all orientations
 100  CONTINUE

      IF(DEBUG(2)) PRINT*, 'CLUSTR vor EXIT'

      END SUBROUTINE CLUSTR
C!_______________________________________________________________________

      SUBROUTINE WRLINE(LUN,IT,X,Y,Z,ICS)
      INTEGER*4 ICS,IT,LUN,NCH
      REAL*8 X,Y,Z
      CHARACTER FORMTS*80
      COMMON /FRMT/FORMTS,NCH
      FORMTS='('
      NCH=1
      CALL WRINT(IT)
      CALL WRFLT(X,3)
      CALL WRFLT(Y,3)
      CALL WRFLT(Z,3)
      CALL WRINT(ICS)
      WRITE(FORMTS(NCH:NCH),'(a1)') ')'
      WRITE(LUN,FORMTS) IT,X,Y,Z,ICS
      END SUBROUTINE WRLINE

C!_______________________________________________________________________
      SUBROUTINE WRINT(I)
      INTEGER*4 I, LINTEG, NCH, NST
      CHARACTER FORMTS*80, WFORMT*10
      COMMON /FRMT/FORMTS,NCH
      IF(I.EQ.0) THEN
        LINTEG=2
      ELSE
        IF(I.GT.0) THEN
          LINTEG=2+INT(LOG10(REAL(I)))
        ELSE
          LINTEG=3+INT(LOG10(ABS(REAL(I))))
        END IF
      END IF
      NST=NCH+1
      NCH=NCH+3
      WFORMT='(a1,i1,a1)'
      IF(LINTEG.GE.10) THEN
        NCH=NCH+1
        WFORMT='(a1,i2,a1)'
      END IF
      WRITE(FORMTS(NST:NCH),WFORMT) 'i',LINTEG,','
      END

C!_______________________________________________________________________
      SUBROUTINE WRFLT(DF,IDIG)
      INTEGER*4 IDIG,NCH,NST
      REAL*8 DF
      CHARACTER FORMTS*80, WFORMT*16
      COMMON /FRMT/FORMTS,NCH
      IF(DF.EQ.0.0) THEN
```

----- FORTRAN source code of program IC_3D -----

```
          LFLOAT=3
        ELSE
          LFLOAT=4+INT(LOG10(ABS(DF)))
          IF(LFLOAT.LT.4) LFLOAT=4
          IF(DF.GT.0) LFLOAT=LFLOAT-1
        END IF
        LFLOAT=LFLOAT+IDIG
        NST=NCH+1
        NCH=NCH+5
        WFORMT='(a1,i1,a1,i1)'
        IF(LFLOAT.GE.10) THEN
          WRITE(WFORMT(6:6),'(i1)') 2
          NCH=NCH+1
        END IF
        IF(IDIG.GT.10) THEN
          WRITE(WFORMT(12:12),'(i1)') 2
          NCH=NCH+1
        END IF
        WRITE(FORMTS(NST:NCH),WFORMT) 'f',LFLOAT,'.',IDIG,','
        END
```

*FORTRAN source code of program ROI_3D*

This program reads track data (output files of IC_3D or original simulation data) and calculates and outputs the following results:

- Frequency distributions of the number of Wigner-Seitz cells in a (large) spherical region that receive ionization clusters for track at different impact parameters and "true" (infinite radial integral) and conditional (track intersects the spherical target) single event distributions (output file name 3D_*'input_filename'*)
- Bivariate distributions of Wigner Seitz cells containing single or multiple ionization clusters. (Output file name 3B_*'input_filename'*).
- Ratio of bivariate frequency distribution of Wigner Seitz cells containing single or multiple ionization clusters to product of marginal frequencies. (Output file name 3C_*'input_filename'*)

Uses subroutines TARG3D and BLINIT (included from file BLINIT.f, see section "FORTRAN subroutine BLINIT"). Calls subroutine TARG3D after each track has been read and then scores ICs in Wigner Seitz cells.

Notes:

- **The program must be executed in the directory where the data files are located**.
- The program expects input of the name of a file listing the debugging options (example see section "Sample debug options file for use with programs IC_3D and ROI_3D").
- Inputs of parameters and input file names are prompted for unless they are entered via a text file. (Manual input is initiated by entering 0 with the first prompt.) A sample input file is listed in the table below.

----- Sample input file for ROI_3D -----

```
210605.16   ! Command file version number ...
ROI_3D      ! ... for this program
DEBUG.opt   ! Debug options file name
12          ! DLIV in nm
6000.       ! DROI in nm
9900.       ! DBEAM in nm
0    1      ! NTARG(1), NTARG(2) (
5           ! KMAX (if < 2, then maximum allowed is used)
1           ! NZROI Number of regions of interest
5000. 2000. ! ZROIC(1) and DZROI
TXYZI       ! CODE for input line structure (T=track #, I=ICS)
4           ! NHEADL Number of Header lines
1           ! NFILES
IC_2.0nm_p50MeVA_mai2015.dat
```

Meaning of the input lines:
(1) Version date and time of the input file structure in format YYMMDD.HHMM

----- Sample input file for ROI_3D -----

(2) Name of the program
(3) Name of the file listing the debug options (see section "Sample debug options file for use with programs IC_3D and ROI_3D")
(4) Diameter of the target sphere within which the clusters are scored (lethal interaction volume of ( Schneider et al. 2019) or cluster volume of ( Schneider et al. 2020))
(5) Diameter of the spherical region of interest within which the target spheres with clusters are scored (e.g. a cell nucleus or a compartment of it).
(6) Diameter of the primary beam (circular cross section)
(7) Maximum elements of the histograms of frequencies of targets with one cluster or with two and more . If NTARG(1) < 1, then the maximum allowed value is used. If NTARG(2) = 1 then the maximum allowed value is used, if NTARG(2)=0 then correlations are not calculated.
(8) Number of clusters for which individual frequencies are scored. (The last element holds the cumulative frequencies of cluster numbers KMAX or higher).
(9) Number of regions of interest of size DROI to be scored along the direction of the primary particle track.
(10) ZROIC(1) is the *z*-coordinate of the center of the first ROI; DZROI is the increment in ZROI position.
(11) Encoding of the input file structure. Each letter stands for the meaning of an entry in an input line. T means the number of the primary particle track, X,Y, and Z the x, y, and z coordinates of the transfer point, E the energy deposit (optional). Example: The string 'TZ%%EXY ' indicates that there are seven entries in each line, of which the first is the track number, the second the z coordinate, the fifth the energy, and the sixth and seventh the x and y coordinates.
(12) number of header lines in the input files
(13) Number of input files to process
(14) further lines: names of the input files

-----
FORTRAN source code of program ROI_3D -----

```
      PROGRAM ROI_3D
C!>>> Declarations >>>>>>>>>>>>>>>>>>>>>>>>>>>>>>>>>>>>>>>>>>>>>>>>>>>
      IMPLICIT NONE
C!    ----- Parameters: Version Date and number
      CHARACTER VDATE*11
      REAL*8 VINPUT  ! Version number for command files YYMMDD.HHMM
C!                   (date and time  file structure was last changed)
C!    ################################################################
      PARAMETER(VDATE='05-JUN-2021',VINPUT=210605.16d0) ! ############
C!    ################################################################
C!    05-JUN-2021 HR:
C!      - Added timestap to output files
C!      - Reworked identification of input file structure
C!    23-MAY-2021 HR:
C!      - Modified call to subroutines to get their version date
C!    17-APR-2021 HR:
C!      - Increased MXTRG2 owing to data set with larger values
C!    13-APR-2021 HR:
C!      - Fixed soft bug: Results were overwritten due to same filename
C!      - Added reduction of output to non-zero values
C!    02-APR-2021 HR:
C!      - Added output of correlation matrix
C!      - Detected and fixed bug in section with convolution
C!    28-MAR-2021 HR:
C!      - Changed normalization of traversing tracks
C!    20-MAR-2021 HR:
C!      - Added input of options file name
C!      - Added discrimination between site size and lattice constant
C!    14-MAR-2021 HR:
C!      - Fixed bug with treatment of last track
C!      - Added common block CLSTRS for info on cluster positions
C!    16-OCT-2020 HR:
C!      - Added readin of geometry parameters via options file
```

----- FORTRAN source code of program ROI_3D -----

```
C!      - Added readin of multiple file names to process with options
C!      - Added processing of multiple ROIs in track data set
C!    15-MAR-2020 HR:
C!      - Added VDATE and modified COMMON BLOCK VERBOSE
C!      - Added variable NXYPOS to fix bug in TARG3D
C!    ----- Parameters: General purpose constants
      INTEGER*4 LUN
      REAL*8 EPS, ONE, ZERO
      PARAMETER(LUN=11, EPS=1.0e-8, ONE=1.0, ZERO=0.0)
C!    ----- Parameters for lattice orientation
      INTEGER*4 MAZMTH, MDIV, MTHETA, MDIR
      PARAMETER(MAZMTH=1, MDIV=0, MTHETA=2**MDIV,
     &           MDIR=MAZMTH*(MTHETA*(MTHETA+1))/2)
C!    ----- Parameters for track and radial distance histograms
      INTEGER*4 KMAX, MIONIZ, MXRPOS, MXTARG, MXTRG2, MXYPOS, MXZROI
      PARAMETER(KMAX=9, MIONIZ=100000, MXRPOS=101, MXTARG=1025,
     &          MXTRG2=513, MXYPOS=4*MXRPOS*(MXRPOS-1)+1, MXZROI=100)
C!    ----- Input parameters for geometry
      REAL*8 DBEAM  ! Diameter of beam in nm
      REAL*8 DLATC  ! Cell lattice constant
      REAL*8 DROI   ! Diameter of region of interest (cell nucleus size)
      REAL*8 DSITE  ! Diameter of spherical target
      REAL*8 DZROI  ! Increment in position of region of interest
C!    ----- Functions
      CHARACTER TSTAMP*24
C!    ----- Local scalars
      CHARACTER*11 VDATES(2), SRDATE
      CHARACTER FILENM*80, FILOUT*85, HEADER*80, PARAM*2, PREFIX*10
      INTEGER*4 I, IFILE, IT, ITA, J, K, KMX, L, NAZMTH, NDIV, NFILES,
     &          NHEADL, NZROI, NPHI, NTRACS
      INTEGER*4 NIONIZ, NRPOS, NXYPOS
      LOGICAL ASKINP
      REAL*8 COSPHI, DELTAR, DUMMY, FNORM, PIBY4, RROI,
     &       SINPHI, SPOSIN, VCHECK
C!    ----- Local arrays
      INTEGER*4 NTARG(2), NTARGK(KMAX)
      REAL*8 CORRSE(MXTARG,MXTRG2),CONVOL(MXTARG),FREQBV(MXTARG,MXTRG2)
      REAL*8 FREQSE(MXTARG,KMAX),FREQTE(MXTARG,KMAX),ZROIC(MXZROI)
      REAL*8 RLINE(8)
C!    ----- Global variables
      INTEGER*4 NDIR
      REAL*8 ADJNTB(3,3,MDIR)
      COMMON /LATTICE/ ADJNTB, NDIR
C!    ------
      INTEGER*4 IRAD(MXYPOS)
      INTEGER*4 NRROI2 ! Integer of Square of ROI radius
      REAL*8 XROI(MXYPOS), YROI(MXYPOS), ZROI(3)
      COMMON /ROICTR/  XROI, YROI, ZROI, IRAD, NRROI2
C!    ------
      REAL*8 XYZ(MIONIZ,3)
      COMMON /TRACKS/ XYZ
C!    ------
      REAL*8 CORR12(MXTARG,MXTRG2,MXRPOS)
      REAL*8 RADIST(MXRPOS),FREQRD(MXTARG,MXRPOS,KMAX)
C!*      COMMON /HISTOG/ RADIST, FREQRD
      COMMON /HISTOG/ RADIST, FREQRD, CORR12
C!          RADIST is the vector of radial distances
C!          FREQRD initially holds the sum, the sum of squares and the
C!                 sum of variances per track over all tracks.
C!                 In the main program this is converted in the end to
```

-----
FORTRAN source code of program ROI_3D -----

```
C!                    mean and variance over all analyzed tracks plus the
C!                    the average intra-track variance do to orientation
C!    ------
      LOGICAL DEBUG(4), SRFLAG
      COMMON /VERBOSE/ DEBUG, SRFLAG
C!    ------
      INTEGER*4 ICSIZE(MIONIZ),NSITES
      COMMON /CLSTRS/ ICSIZE, NSITES
C!    ------
C!     CODE: Character string encoding the meaning of the entries in a
C!           line of the input file as follows:
C!           'T'     - number of the primary particle track
C!           'X','Y','Z' - x, y, and z coordinates of the transfer point
C!           'E'     - energy deposit (if applicable)
C!           'I'     - ionization cluster size (if applicable)
C!           '%'     - additional data that are not used
C!           Example: The string 'TZ%%EXY ' indicates that there are 7
C!                     entries in each line, of which the first is the
C!                     track number, the second the z coordinate, the
C!                     fifth the energy, and the sixth and seventh the x
C!                     and y coordinates
C!     IDT: Array of the columns indices of (1-3) x,y,z coordinates
C!                   (4) energy deposit (if present), (5) track number,
C!                   (6) number of ionizations in cluster (if present)
C!                   (7-8) are there for future use
      CHARACTER CODE*8
      INTEGER*2 IDT(8),NDT
      COMMON /RFORMT/IDT,NDT,CODE
C!<<<<End declarations<<<<<<<<<<<<<<<<<<<<<<<<<<<<<<<<<<<<<<<<<<<<<<<

      PRINT*, 'Program ROI_3D Version: '//VDATE
      SRFLAG=.TRUE. ! Invoke message on Version date from Subroutine
TARG3D

      IF(1.EQ.1) THEN ! DUMMY block for readability: Read input 1111111
        DEBUG(1)=.FALSE. ! Main program
        DEBUG(2)=.FALSE. ! TARG3D main sections
        DEBUG(3)=.FALSE. ! TARG3D IDIR loop
        DEBUG(4)=.FALSE. ! TARG3D IPOS loop details

        PIBY4=ATAN(ONE)
        PREFIX='3D__0.0nm_'
        DO K=1, KMAX
          NTARGK(K)=0
        END DO

C!      Check command file version
        PRINT*, 'Enter 0 for manual input or version (YYMMDD.HHMM) '//
     &          'of command file structure '
        READ(*,*) VCHECK
        ASKINP=(VCHECK.EQ.ZERO)
        IF(.NOT.ASKINP) THEN
          IF(VCHECK.LT.VINPUT) THEN
            PRINT*, 'Command file structure ',VCHECK,' older than '//
     &              'current version ',VINPUT,'=> STOP.'
            STOP
          END IF
          READ(*,*) FILENM
          IF(FILENM(1:6).NE.'ROI_3D') THEN
            PRINT*, 'Command file appear not to be for ROI_3D but '//
```

-----
FORTRAN source code of program ROI_3D -----

```
     &               'for '//FILENM//'=> STOP.'
            STOP
          END IF
        END IF

C!      Read debug options
        IF(ASKINP) PRINT*, 'Enter debug options file name or - for none'
        READ(*,*) FILENM
        OPEN(LUN,FILE=FILENM,STATUS='OLD',ERR=10)
        DO I=1,4
          READ(LUN,*) DEBUG(I)
        END DO ! I=1,4
        CLOSE(LUN)
  10    CONTINUE

C!      Read geometry parameters
        IF(ASKINP) PRINT*, 'Enter site diameter in nm'
        READ(*,*) DSITE
        DLATC=DSITE*SQRT(2.)*EXP(LOG(PIBY4/3.)/3.)
        IF(DSITE.LT.10.0) THEN
          WRITE(PREFIX(5:7),'(f3.1)') DSITE
        ELSE
          WRITE(PREFIX(4:7),'(f4.1)') DSITE
        END IF

        IF(ASKINP) PRINT*, 'Enter ROI diameter in nm'
        READ(*,*) DROI
        RROI=DROI/2.
        NRROI2=INT(RROI*RROI/DLATC/DLATC) ! Note: This is correct as
DLATC is the unit of length
        DELTAR=DROI/40.

        IF(ASKINP) PRINT*, 'Enter beam diameter in nm <= ',5.*DROI
        READ(*,*) DBEAM
        NRPOS=1+NINT(DBEAM/2./DELTAR)
        IF(NRPOS.GT.MXRPOS) THEN
          NRPOS=MXRPOS
          PRINT*, 'Beam diameter ',DBEAM,' nm is too large. <<<<<<<<<<<'
          DBEAM=2.*DELTAR*REAL(MXRPOS-1)
          PRINT*, '>>> maximum possible value ',DBEAM,' is used.'
        END IF

        IF(ASKINP) PRINT*, 'Enter maximum # of targets in histogram'
        READ(*,*) NTARG(1), NTARG(2)
        IF(NTARG(1).GT.MXTARG.OR.NTARG(1).LT.1) NTARG(1)=MXTARG
        IF(NTARG(2).GT.MXTRG2.OR.NTARG(2).EQ.1) NTARG(2)=MXTRG2

        IF(ASKINP) PRINT*, 'Enter maximum ionization cluster '//
     &                     'complexity (KMAX)'
        READ(*,*) KMX
        IF(KMX.GT.KMAX.OR.KMX.LT.2) KMX=KMAX

        IF(ASKINP) PRINT*, 'Enter # of regions of interest along track'
        READ(*,*) NZROI
        IF (NZROI.EQ.1) THEN
          IF(ASKINP) PRINT*, 'Enter z position of region of interest'
          READ(*,*) ZROIC(1)
          DZROI=ZERO
        ELSE
          IF(ASKINP) PRINT*, 'Enter position of first region of '//
```

-----
FORTRAN source code of program ROI_3D -----

```
     &            'interest (ROI) and increment in ROI position'
          READ(*,*) ZROIC(1), DZROI
          DO I=2,NZROI
            ZROIC(I)=ZROIC(I-1)+DZROI
          END DO
        END IF

        IF(ASKINP) PRINT*, 'Enter 8 character code for data file '//
     &       'structure where ''T'' indicates the track ID,'//
     &       ' ''X'',''Y'' and ''Z'' the respective coordinates, '//
     &       '''E'' the energy deposit (if present) and ''&'' any '//
     &       'other data'
        READ(*,*) CODE
        CALL RFINIT()

        IF(ASKINP) PRINT*, 'Number of header lines'
        READ(*,*) NHEADL

        IF(ASKINP) PRINT*, 'Number of files to process'
        READ(*,*) NFILES

      END IF          ! DUMMY block for readability: Read input 1111111

      IF(2.EQ.2) THEN ! DUMMY block for readability: Initialize 2222222
*         IF(DEBUG(1)) PRINT*, 'Hier v'
        NAZMTH=MAZMTH
        NDIV=MDIV
        IF(DEBUG(1)) PRINT*, 'Hier vor CALL BLINIT'
        CALL BLINIT(NAZMTH,NDIV,DLATC,SRDATE) ! Init reziprocal lattice
        VDATES(1)=SRDATE
        IF(DEBUG(1)) PRINT*, 'Hier nach CALL BLINIT', NRPOS, NTARG

        DO I=1,NRPOS
C!        Define radial offsets of track w.r.t. ROI center
          RADIST(I)=REAL(I-1)*DELTAR
C!        x&y positions of track w.r.t. ROI center (piecake method)
          IF(I.EQ.1) THEN
            NXYPOS=1
            NPHI=0
            IRAD(NXYPOS)=1
            XROI(NXYPOS)=ZERO
            YROI(NXYPOS)=ZERO
          ELSE
            NPHI=NPHI+8
            NXYPOS=NXYPOS+1
            IRAD(NXYPOS)=I
            XROI(NXYPOS)=RADIST(I)
            YROI(NXYPOS)=ZERO
            COSPHI=COS(PIBY4/REAL(I-1))
            SINPHI=SIN(PIBY4/REAL(I-1))
            DO J=2, NPHI
              NXYPOS=NXYPOS+1
              IRAD(NXYPOS)=I
              XROI(NXYPOS)=XROI(NXYPOS-1)*COSPHI-YROI(NXYPOS-1)*SINPHI
              YROI(NXYPOS)=XROI(NXYPOS-1)*SINPHI+YROI(NXYPOS-1)*COSPHI
            END DO
          END IF
        END DO ! DO I=1,NRPOS
      END IF          ! DUMMY block for readability: Initialize 2222222
```

----- FORTRAN source code of program ROI_3D -----

```
      DO IFILE=1,NFILES
        IF(ASKINP) PRINT*, 'Enter file name ',IFILE
        READ(*,*) FILENM
        IF(.NOT.ASKINP) PRINT*, 'Processing file ',FILENM

        IF(4.EQ.4) THEN ! DUMMY block for readability: Main Loop  444444
          IF(DEBUG(1)) PRINT*, 'Hier beginnt Block 4'
          DO J=1,NRPOS
C!          Init counters
            DO I=1,NTARG(1)
              DO K=1,KMAX
                FREQRD(I,J,K)=ZERO
              END DO
            END DO
C!          Begin 02-APR-2021 >>>>>>>>>>>>>>>>>
            DO I=1,NTARG(1)
              DO L=1,NTARG(2)
                CORR12(I,L,J)=ZERO
              END DO
            END DO
C!          END 02-APR-2021 <<<<<<<<<<<<<<<<<<
          END DO ! DO J=1,NRPOS
          NIONIZ=1
          NTRACS=1

          OPEN(LUN,FILE=FILENM,STATUS='OLD')
          DO I=1,NHEADL
            READ(LUN,*) HEADER
          END DO
   90     READ(LUN,*) (RLINE(I),I=1,NDT)
          IF(IDT(6).GT.0) THEN
            IF(NINT(RLINE(IDT(6))).LT.2) GOTO 90 ! No ionization cluster
          END IF
          ITA=NINT(RLINE(IDT(5)))
          DO I=1,3
            XYZ(NIONIZ,I)=RLINE(IDT(I))
          END DO

          NIONIZ=NIONIZ+1
          IF(DEBUG(1)) PRINT*, 'Hier 4'
  100     CONTINUE ! >>>>>>>>>>>>>>>>>>>>>>>>>>>>>>>>>>>>>>>>>>>>>>>>>>
          READ(LUN,*,END=200,ERR=200) (RLINE(I),I=1,NDT)
          IF(IDT(6).GT.0) THEN
            IF(NINT(RLINE(IDT(6))).LT.2) GOTO 100 ! No ionization cluster
          END IF
          IT=NINT(RLINE(IDT(5)))
          DO I=1,3
            XYZ(NIONIZ,I)=RLINE(IDT(I))
          END DO

          IF(IT.EQ.ITA) THEN
            NIONIZ=NIONIZ+1
          ELSE ! Entry from a new track was read
            NIONIZ=NIONIZ-1 ! Decrease to number of ionizations in track
            IF(DEBUG(1)) PRINT*, 'Hier vor CALL TARG3D'
            IF(MOD(NTRACS,10).EQ.0) PRINT*, NTRACS
            DO J=1,NZROI
              ZROI(2)=ZROIC(J)
              ZROI(1)=ZROI(2)-DROI/2.-DLATC/2.
              ZROI(3)=ZROI(2)+DROI/2.+DLATC/2.
```

-----
FORTRAN source code of program ROI_3D -----

```
            CALL TARG3D(NIONIZ,NRPOS,NTARG,NXYPOS,SRDATE)
          END DO ! J=1,NZROI

          IF(DEBUG(1)) PRINT*, 'Hier nach CALL TARG3D'
          DO J=1,3
            XYZ(1,J)=XYZ(NIONIZ+1,J) ! First ionization in new track
          END DO
          ITA=IT ! Remember new track number
          NTRACS=NTRACS+1
          NIONIZ=2 ! There was already one ionization
        END IF
        GOTO 100 ! <<<<<<<<<<<<<<<<<<<<<<<<<<<<<<<<<<<<<<<<<<<<<<<<<<<<
  200   CONTINUE
          DO J=1,NZROI
            ZROI(2)=ZROIC(J)
            ZROI(1)=ZROI(2)-DROI/2.-DLATC/2.
            ZROI(3)=ZROI(2)+DROI/2.+DLATC/2.
            CALL TARG3D(NIONIZ,NRPOS,NTARG,NXYPOS,SRDATE)
          END DO ! J=1,NZROI
          VDATES(2)=SRDATE
        CLOSE(LUN)
      END IF          ! DUMMY block for readability: Main Loop  4444444

      IF(6.EQ.6) THEN ! DUMMY block for readability: Normalize  666666
C!      #Normalization
        FNORM=ONE/REAL(NTRACS)
        FNORM=ONE/REAL(NTRACS)/REAL(NZROI)
*        SUMCNT=ZERO
        DO J=1,NRPOS
          DO I=1,NTARG(1)
            DO K=1,KMAX
              FREQRD(I,J,K)=FNORM*FREQRD(I,J,K)
            END DO ! K=1,KMAX
C!          Start 02-APR-2021 new >>>>>>>>>>>>>>>>>>>>
            DO L=1,NTARG(2)
              CORR12(I,L,J)=FNORM*CORR12(I,L,J)
            END DO ! NTARG(2)
C!          End 02-APR-2021 new <<<<<<<<<<<<<<<<<<<<<<<<<<
          END DO ! I=1,NTARG(1)
        END DO ! J=1,NRPOS
        DO K=KMX-1,2,-1
          DO J=1,NRPOS
            DO I=1,NTARG(1)
              CONVOL(I)=ZERO
            END DO ! I=1,NTARG(1)
            DO I=1,NTARG(1)
              DO L=0, I-1
                CONVOL(I)=CONVOL(I)+FREQRD(I-L,J,K)*FREQRD(L+1,J,K+1)
              END DO ! L=0, I-1
            END DO ! I=1,NTARG(1)
            DO I=1,NTARG(1)
              FREQRD(I,J,K)=CONVOL(I)
            END DO ! I=1,NTARG(1)
          END DO ! J=1,NRPOS
        END DO ! K=KMX,1,-1
      END IF          ! DUMMY block for readability: Normalize  666666

      IF(8.EQ.8) THEN ! DUMMY block for readability: Prepare Output
8888888
C!      Calculate single-event distributions
```

-----
FORTRAN source code of program ROI_3D -----

```
          DO K=1,KMX
            NTARGK(K)=1
            DO I=1,NTARG(1)
              FREQSE(I,K)=FREQRD(I,1,K)
              SPOSIN=ONE
              DUMMY=ZERO
              DO J=2,NRPOS
                DUMMY=DUMMY+8.
                FREQSE(I,K)=FREQSE(I,K)+DUMMY*FREQRD(I,J,K)
                SPOSIN=SPOSIN+DUMMY
                IF (RADIST(J).LE.RROI) THEN
                  FREQTE(I,K)=FREQSE(I,K)/SPOSIN
                END IF
              END DO
              FREQSE(I,K)=FREQSE(I,K)/SPOSIN
              IF(FREQSE(I,K).GE.EPS.OR.FREQTE(I,K).GE.EPS) NTARGK(K)=I
            END DO
          END DO

          IF(NTARG(2).GT.0) THEN
            IF(NTARGK(2).GT.MXTRG2) THEN
              PRINT*, 'Major problem: 2nd array dimension MXTRG2=',
     &                MXTRG2,' < max. number of 2+ clusters NTARGK(2)=',
     &                NTARGK(2)
              STOP
            END IF
            DO I=1,NTARGK(1)
              DO L=1,NTARGK(2)
                FREQBV(I,L)=CORR12(I,L,1)
                SPOSIN=ONE
                DUMMY=ZERO
                DO J=2,NRPOS
                  DUMMY=DUMMY+8.
                  FREQBV(I,L)=FREQBV(I,L)+DUMMY*CORR12(I,L,J)
                  SPOSIN=SPOSIN+DUMMY
                END DO
                FREQBV(I,L)=FREQBV(I,L)/SPOSIN
                IF(FREQSE(I,1)*FREQSE(L,2).GT.ZERO) THEN
                  CORRSE(I,L)=FREQBV(I,L)/(FREQSE(I,1)*FREQSE(L,2))
                ELSE
                  CORRSE(I,L)=FREQBV(I,L)
                END IF
              END DO
            END DO
          END IF
        END IF          ! DUMMY block for readability: Prepare Output
8888888

       IF(9.EQ.9) THEN ! DUMMY block for readability: OUTPUT  9999999999
C!        #Write results to output file
          WRITE(PREFIX(2:2),'(a)') 'D'
          FILOUT=PREFIX//FILENM
          PRINT*, 'Write output to '//FILOUT
          OPEN(LUN,FILE=FILOUT,STATUS='UNKNOWN')
          WRITE(LUN,*) ' *** Output from PROGRAM ROI_3D Version '//
     &              VDATE//' BLINIT:'//VDATES(1)//' TARG3D:'//VDATES(2)
     &                   //' on '//TSTAMP()
          WRITE(LUN,*) ' Filename: '//FILOUT
          WRITE(LUN,'(a10,10i6)') ' NTARG(K)=',(NTARGK(I),I=1,KMAX)
          WRITE(LUN,'(6(a,f8.3))') ' DLATC=',DLATC,' nm    DROI= ',DROI,
```

-----
FORTRAN source code of program ROI_3D -----

```
     &                             ' nm   DSITE=',DSITE,' nm   DBEAM= ',
     &                             DBEAM,' nm'
          PARAM='P '
          DO K=1,KMX
            IF(K.GT.1) PARAM='F '
            WRITE(PARAM(2:2),'(I1)') K
            WRITE(LUN,'(1X,2a8,103a15)') 'Para-', '#Sites', 'Average',
     &                        'Average',('Distance/nm',J=1,NRPOS)
            WRITE(LUN,'(1X,2a8,2a15,101f15.6)') 'meter', '/track',
     &                       'total','inside',(RADIST(J),J=1,NRPOS)
            DO I=1,NTARGK(K)
              WRITE(LUN,'(1X,a8,i8,103f15.8)') PARAM, I-1,
     &                                   FREQSE(I,K),FREQTE(I,K),
     &                                  (FREQRD(I,J,K),J=1,NRPOS)
            END DO
            WRITE(LUN,*) '____________________________________________'
            WRITE(LUN,*) '*********************************************'
          END DO
          CLOSE(LUN)

          IF(NTARG(2).GT.0) THEN ! begin 02-APR-2021 >>>>>>>>>>>>
            WRITE(PREFIX(2:2),'(a)') 'B'
            FILOUT=PREFIX//FILENM
            PRINT*, 'Write output to '//FILOUT
            OPEN(LUN,FILE=FILOUT,STATUS='UNKNOWN')
            WRITE(LUN,*) '*** Output from PROGRAM ROI_3D Version '//
     &              VDATE//' BLINIT:'//VDATES(1)//' TARG3D:'//VDATES(2)
     &                   //' on '//TSTAMP()
            WRITE(LUN,*) 'Filename: '//FILOUT
            WRITE(LUN,'(6(a,f8.3))') ' DLATC=',DLATC,' nm    DROI= ',
     &            DROI,' nm   DSITE=',DSITE,' nm   DBEAM= ',DBEAM,' nm'
            WRITE(LUN,*) 'Correlations P1 and F2'
            WRITE(LUN,*) NTARGK(1),NTARGK(2)
            DO I=1,NTARGK(1)
              WRITE(LUN,'(1X,1000e15.8)') (FREQBV(I,L),L=1,NTARGK(2))
            END DO
            CLOSE(LUN)

            WRITE(PREFIX(2:2),'(a)') 'C'
            FILOUT=PREFIX//FILENM
            PRINT*, 'Write output to '//FILOUT
            OPEN(LUN,FILE=FILOUT,STATUS='UNKNOWN')
            WRITE(LUN,*) '*** Output from PROGRAM ROI_3D Version '//
     &              VDATE//' BLINIT:'//VDATES(1)//' TARG3D:'//VDATES(2)
     &                   //' on '//TSTAMP()
            WRITE(LUN,*) 'Filename: '//FILOUT
            WRITE(LUN,'(6(a,f8.3))') ' DLATC=',DLATC,' nm    DROI= ',
     &            DROI,' nm   DSITE=',DSITE,' nm   DBEAM= ',DBEAM,' nm'
            WRITE(LUN,*) 'Correlations P1 and F2'
            WRITE(LUN,*) NTARGK(1),NTARGK(2)
            DO I=1,NTARGK(1)
              WRITE(LUN,'(1X,1000e15.8)') (CORRSE(I,L),L=1,NTARGK(2))
            END DO
            CLOSE(LUN)
          END IF ! (NTARG(2).GT.0)
        END IF          ! DUMMY block for readability: OUTPUT  999999999

      END DO ! IFILE=1,NFILES

      END PROGRAM ! ROI_3D
```

-----
FORTRAN source code of program ROI_3D -----

```
C!____________________________________________________________________________

      INCLUDE 'BLINIT.f'  ! <<<<<<<<<<<<<<<<<<<<<<<<<<<<<<<<<<<<<<
      INCLUDE 'RFINIT.f'
      INCLUDE 'TSTAMP.f'


C!**********************************************************************
      SUBROUTINE TARG3D(NIONIZ,NRPOS,NTARG,NXYPOS,SRDATE)
C!    ****************************************************************
C!    Determine targets with ionizations
C!    ****************************************************************
C!>>> Declarations >>>>>>>>>>>>>>>>>>>>>>>>>>>>>>>>>>>>>>>>>>>>>>>>>>
      IMPLICIT NONE
C!    ----- Parameters: Version Date
      CHARACTER VDATE*11
      PARAMETER(VDATE='23-MAY-2021')
C!    23-MAY-2021 HR: Added outout of version date to calling program
C!    20-MAY-2020 HR: Fixed severe bug with sorting the hit targets
C!    02-APR-2021 HR: Included scoring of correlations
C!    28-MAR-2021 HR: Cleaned up unused variables
C!    15-MAR-2020 HR:
C!      - Added VDATE and modified COMMON BLOCK VERBOSE and added output
of VDATE on first call
C!      - Added input variable NXYPOS to fix bug in processing off-axis
positions
C!      - Fixed bug: Case that no ionizations occur at all was not
properly handled.
C!    14-MAR-2021 HR:
C!      - Added output of cluster positions for analysis of cluster of
clusters
C!    ----- Parameters: General purpose constants
      REAL*8 ONE, ZERO
      PARAMETER(ONE=1.0, ZERO=0.0)
C!    ----- Parameters for lattice orientation
      INTEGER*4 MAZMTH, MDIV, MTHETA, MDIR
      PARAMETER(MAZMTH=1, MDIV=0, MTHETA=2**MDIV,
     &           MDIR=MAZMTH*(MTHETA*(MTHETA+1))/2)
C!    ----- Parameters for track and radial distance histograms
      INTEGER*4 KMAX, MIONIZ, MXRPOS, MXTARG, MXTRG2, MXYPOS
      PARAMETER(KMAX=9, MIONIZ=100000, MXRPOS=101, MXTARG=1025,
     &          MXTRG2=513, MXYPOS=4*MXRPOS*(MXRPOS-1)+1)
C!    ----- Local scalars
      INTEGER*4 I, IDIR, IDIST, IPOS, IR, J, K
      INTEGER*4 IHOLD, NCOUNT, NIONIS
      REAL*8 WEIGHT, SAVXYZ, SPROD
C!    ----- Local arrays
      INTEGER*4 ICROI(3), ITARGT(MIONIZ,3)
      INTEGER*4 ICSITE(KMAX)
      REAL*8 CORREL(MXTARG,MXTRG2,MXRPOS)
      REAL*8 FTGICS(MXTARG,MXRPOS,KMAX)
C!          ICSITE is the histogram of absolute frequencies of
C!                 targets with 1, 2, ... >=KMAX ionizations
C!                 occuring for a particular orientation of the
C!                 lattice w.r.t. to the track
C!          FTGICS initially is the sum of the absolute frequencies
C!                 of targets holding a certain ICS for an orientation
C!                 in the particular track.
C!                 In the end it is normalized and trasferred to global
C!                 counters
```

-----
FORTRAN source code of program ROI_3D -----

```
C!    ----- Global variables
      INTEGER*4 NDIR
      REAL*8 ADJNTB(3,3,MDIR)
      COMMON /LATTICE/ ADJNTB, NDIR
C!    ------
      INTEGER*4 IRAD(MXYPOS)
      INTEGER*4 NRROI2 ! Integer of Square of ROI radius
      REAL*8 XROI(MXYPOS), YROI(MXYPOS), ZROI(3)
      COMMON /ROICTR/  XROI, YROI, ZROI, IRAD, NRROI2
C!    ------
      REAL*8 XYZ(MIONIZ,3)
      COMMON /TRACKS/ XYZ
C!    ------
      REAL*8 CORR12(MXTARG,MXTRG2,MXRPOS)
      REAL*8 RADIST(MXRPOS),FREQRD(MXTARG,MXRPOS,KMAX)
C!*      COMMON /HISTOG/ RADIST, FREQRD
      COMMON /HISTOG/ RADIST, FREQRD, CORR12
C!           RADIST is the vector of radial distances
C!           FREQRD initially holds the sum, the sum of squares and the
C!                  sum of variances per track over all tracks.
C!                  In the main program this is converted in the end to
C!                  mean and variance over all analyzed tracks plus the
C!                  the average intra-track variance do to orientation
C!    ------
      LOGICAL DEBUG(4), SRFLAG
      COMMON /VERBOSE/ DEBUG, SRFLAG
C!    ------
      INTEGER*4 ICSIZE(MIONIZ),NSITES
      COMMON /CLSTRS/ ICSIZE, NSITES
C!    ----- Scalars and arrays passed from the calling module
      INTEGER*4 NIONIZ, NRPOS, NTARG(2), NXYPOS
C!    ----- Scalars and arrays passed to the calling module
      CHARACTER*11 SRDATE
C!<<<<End declarations<<<<<<<<<<<<<<<<<<<<<<<<<<<<<<<<<<<<<<<<<<<<<<<<

      SRDATE=VDATE

      IF(SRFLAG) THEN
        PRINT*, 'Subroutine TARG3D Version: ',VDATE
        SRFLAG=.FALSE.
      END IF
      IF(DEBUG(2)) PRINT*, 'TARG3D Input',NIONIZ,NRPOS,NTARG,NXYPOS

C!    Init local histograms
      DO K=1,KMAX
        ICSITE(K)=0
        DO J=1,NRPOS
          DO I=1,NTARG(1)
            FTGICS(I,J,K)=ZERO
          END DO ! I=1,NTARG(1)
        END DO ! J=1,NRPOS
      END DO ! I=1,KMAX

C!    Begin 02-APR-2021 >>>>>>>>>
      DO J=1,NRPOS
        DO I=1,NTARG(1)
          DO K=1,NTARG(2)
            CORREL(I,K,J)=ZERO
          END DO ! K=1,NTARG(2)
        END DO ! I=1,NTARG(1)
```

----- FORTRAN source code of program ROI_3D -----

```
      END DO ! J=1,NRPOS
C!    End 02-APR-2021 <<<<<<<<<<
C!    End init local histograms

      IF(DEBUG(2)) PRINT*, 'TARG3D vor Main Loop NDIR=',NDIR
      IF(DEBUG(2)) PRINT*, 'TARG3D vor Main Loop NIONIZ=',NIONIZ
      DO IDIR=1,NDIR ! Loop over all orientations
        IF(IDIR.GT.NDIR) GOTO 100
        IF(DEBUG(3)) PRINT*, 'TARG3D Begin Loop IDIR',IDIR
C!      # Find target volume for all  ionizations in track
        NIONIS=0
        DO I=1,NIONIZ
          IF(XYZ(I,3).GE.ZROI(1).AND.XYZ(I,3).LE.ZROI(3)) THEN
            NIONIS=NIONIS+1
            DO J=1,3
              SPROD= ADJNTB(1,J,IDIR)*XYZ(I,1)
     &                  +ADJNTB(2,J,IDIR)*XYZ(I,2)
     &                  +ADJNTB(3,J,IDIR)*(XYZ(I,3)-ZROI(2))
              ITARGT(NIONIS,J)=NINT(SPROD)
            END DO ! J=1,3
          END IF
        END DO ! I=1,NIONIZ

        IF(DEBUG(3)) PRINT*, 'TARG3D vor sort volume indices',NIONIS
        IF(IDIR.GT.NDIR) STOP
C!      # Sort target volume indices ascending
        DO I=1,NIONIS
          DO J=I+1,NIONIS
            IF(     (ITARGT(I,1).GT.ITARGT(J,1))
     &         .OR.(ITARGT(I,1).EQ.ITARGT(J,1).AND.
     &              ITARGT(I,2).GT.ITARGT(J,2))
     &         .OR.(ITARGT(I,1).EQ.ITARGT(J,1).AND.
     &              ITARGT(I,2).EQ.ITARGT(J,2).AND.
     &              ITARGT(I,3).LE.ITARGT(J,3))) THEN
              DO IR=1,3
                IHOLD=ITARGT(I,IR)
                ITARGT(I,IR)=ITARGT(J,IR)
                ITARGT(J,IR)=IHOLD
                SAVXYZ=XYZ(I,IR)
                XYZ(I,IR)=XYZ(J,IR)
                XYZ(J,IR)=SAVXYZ
              END DO ! IR=1,3
            END IF
          END DO ! J=1,NIONIS
        END DO ! I=1,NIONIS

        IF(DEBUG(3)) PRINT*, 'TARG3D vor find unique volumes',NIONIS
C!      # Find unique target volumes and score ionizations
        NSITES=1
        ICSIZE(NSITES)=1
        DO I=2,NIONIS
          IF(    ITARGT(I,1).NE.ITARGT(I-1,1)
     &       .OR.ITARGT(I,2).NE.ITARGT(I-1,2)
     &       .OR.ITARGT(I,3).NE.ITARGT(I-1,3)) THEN
            DO J=1,3 ! 14-MAR-2021
              XYZ(NSITES,J)=XYZ(NSITES,J)/REAL(ICSIZE(NSITES))
            END DO
            NSITES=NSITES+1
            ICSIZE(NSITES)=1
            DO J=1,3
```

-----
FORTRAN source code of program ROI_3D -----

```
              ITARGT(NSITES,J)=ITARGT(I,J)
            END DO ! J=1,3
          ELSE
            ICSIZE(NSITES)=ICSIZE(NSITES)+1
            DO J=1,3
              XYZ(NSITES,J)=XYZ(NSITES,J)+XYZ(I,J)
            END DO
          END IF
        END DO ! I=2,NIONIS

        IF(DEBUG(3)) PRINT*, 'TARG3D Begin Loop IPOS',NXYPOS
        DO IPOS=1,NXYPOS ! Loop over all track positions ! 15-MAR-2020
C!        Calculate cell indices of ROI center
          DO J=1,3
            SPROD= ADJNTB(1,J,IDIR)*XROI(IPOS)
     &                 +ADJNTB(2,J,IDIR)*YROI(IPOS)
*     &                 +ADJNTB(3,J,IDIR)*ZROI(2)
            ICROI(J)=NINT(SPROD)
          END DO ! J=1,3

          IF(DEBUG(4)) PRINT*, 'TARG3D vor Score hit targets',ICROI,
     &                       IPOS, XROI(IPOS)
C!        # Score hit targets
          DO I=1,KMAX ! Zero local counter
            ICSITE(I)=0
          END DO ! I=1,KMAX ! Zero local counter

          DO I=1,NSITES ! Count hit targets
            IDIST=0
            DO J=1,3
              DO K=J,3
                IDIST=IDIST+(ITARGT(I,J)-ICROI(J))
     &                     *(ITARGT(I,K)-ICROI(K))
              END DO
            END DO
            NCOUNT=0
            IF(ABS(IDIST).LE.NRROI2) THEN ! count if inside ROI
              NCOUNT=ICSIZE(I) !  Ionization cluster size
              IF(NCOUNT.GT.KMAX) NCOUNT=KMAX
              ICSITE(NCOUNT)=ICSITE(NCOUNT)+1
*             IF(DEBUG(5)) PRINT*, 'Count hit targets',IDIST,NCOUNT,IPOS
            END IF

          END DO ! I=1,NSITES ! Count hit targets

          IF(DEBUG(3)) PRINT*, 'TARG3D vor add to sum arrays',NDIR,IPOS
C!      # Add this histogram to sum arrays
          IR=IRAD(IPOS)
          IF(DEBUG(3)) PRINT*, 'TARG3D vor add to sum arrays',IR,ICSITE
          DO K=1,KMAX ! Update global counters
            NCOUNT=ICSITE(K)+1
            IF(NCOUNT.GT.NTARG(1)) NCOUNT=NTARG(1)
            IF(NCOUNT.GT.0) FTGICS(NCOUNT,IR,K)=FTGICS(NCOUNT,IR,K)+ONE
          END DO
C!        Begin 02-APR-2021 >>>>>>>>>
          IF(NTARG(2).GT.0) THEN
            ICSITE(1)=ICSITE(1)+1
            IF(ICSITE(1).GT.NTARG(1)) ICSITE(1)=NTARG(1)
            NCOUNT=1
            DO K=2,KMAX
```

-----
FORTRAN source code of program ROI_3D -----

```
              NCOUNT=NCOUNT+ICSITE(K)
            END DO
            IF(NCOUNT.GT.NTARG(2)) NCOUNT=NTARG(2)
            CORREL(ICSITE(1),NCOUNT,IR)=
     &                      CORREL(ICSITE(1),NCOUNT,IR)+ONE
          END IF
C!        End 02-APR-2021 <<<<<<<<<<

          IF(DEBUG(3)) PRINT*, 'TARG3D nach sum arrays',NDIR,IDIR
        END DO ! IPOS=1,NXYPOS ! Loop over all track positions
      END DO ! IDIR=1,NDIR ! Loop over all orientations
 100  CONTINUE


C!    # Update global counters
      DO I=1,NRPOS
        IF(I.EQ.1) THEN
          WEIGHT=ONE/REAL(NDIR)
        ELSE
          WEIGHT=ONE/REAL(8*I-8)/REAL(NDIR)
        END IF
        DO J=1,NTARG(1)
          DO K=1,KMAX
            FREQRD(J,I,K)=FREQRD(J,I,K)+WEIGHT*FTGICS(J,I,K)
          END DO  ! K=1,KMAX
C!        Begin 02-APR-2021 >>>>>>>>>
          IF(NTARG(2).GT.0) THEN
            DO K=1,NTARG(2)
              CORR12(J,K,I)=CORR12(J,K,I)+WEIGHT*CORREL(J,K,I)
            END DO ! K=1,NTARG(2)
          END IF ! (NTARG(2).GT.0)
C!        End 02-APR-2021 <<<<<<<<<<
        END DO  ! J=1,NTARG(1)
      END DO ! I=1,NRPOS

      IF(DEBUG(2)) PRINT*, 'TARG3D vor EXIT'

      END SUBROUTINE TARG3D
C!____________________________________________________________________
```

*FORTRAN source code of program ME_ROI_3C*

This program reads data from output files 3B_*.dat produced by ROI_3D and convolutes them with Binomial distributions of a given success probability such as to convert IC to DSB distributions. Outputs:

- MEA_*'input_filename'* multi- and single event frequency distributions after convolution with binomial compared to frequency distributions of ionization clusters
- MEB_*'input_filename'* bivariate multi-event frequency distribution of single and multiple ionization clusters
- MEC_ *'input_filename'* ratio of bivariate frequency distribution to product of marginal frequencies
- MED_*'input_filename'* multi- and single event frequency distributions after convolution with binomial only (smaller file size)
- SEB_*'input_filename'* bivariate single-event frequency distribution of single and multiple ionization clusters
- SEC_*'input_filename'* ratio of bivariate frequency distribution to product of marginal frequencies

Notes:

- **The program must be executed in the directory where the data files are located**.
- The program expects input of the name of a file listing the debugging options (example see section “Sample debug options file for use with programs IC_3D and ROI_3D”).
- Inputs of parameters and input file names are prompted for unless they are entered via a text file. (Manual input is initiated by entering 0 with the first prompt.) A sample input file is listed in the table below.

----- Sample input file for ME_ROI_3C -----

```
210605.16      ! Command file version number ...
ME_ROI_3C      ! ... for this program
1       ! Flag DEBUG for priniting messages
0.01    ! Relative target density
2.0     ! Absorbed dose in Gy
0    0   ! NMAXME(1), NMAXME(2) (if <= 1 then max allowed is used)
1      ! Number of input files to process
3B_12.0nm_IC_2.0nm_p3MeVA_mai2015.dat
3.
```

Meaning of the input lines:

(1) Version date and time of the input file structure in format YYMMDD.HHMM
(2) Name of the program
(3) Flag for priniting messages on progress (value = 1)
(4) Probability parameter of the Binomial distribution used to convert ionization cluster distributions to DSB cluster distributions.
(5) Absorbed dose in Gy
(6) Maximum dimensions of the matrix of bivariate frequencies of the number of targets with a single DSB and multiple DSBs.
(7) Number of input files to process
(8) further odd lines: names of the input files
(9) further even lines: proton energy in MeV

----- FORTRAN source code of program ME_ROI_3C -----

```
      PROGRAM ME_ROI_3C
C!>>> Declarations >>>>>>>>>>>>>>>>>>>>>>>>>>>>>>>>>>>>>>>>>>>>>>>>>>>
      IMPLICIT NONE
C!    ----- Parameters: Version Date and number
      CHARACTER VDATE*11
      REAL*8 VINPUT  ! Version number for command files YYMMDD.HHMM
C!                   (date and time  file structure was last changed)
C!    ################################################################
      PARAMETER(VDATE='05-JUN-2021',VINPUT=210605.1600d0) ! ###########
C!    05-JUN-2021 HR:
C!      - Reorganized command input file.
C!      - Added timestamp in output file
C!    17-APR-2021 HR:
C!      - Cleaned up debug messages
C!    16-APR-2021 HR:
C!      - Adapted to changes done in ROI_3D (leaner output).
C!    13-APR-2021 HR:
C!      - Created
C!    ----- Parameters: General purpose constants
      INTEGER*4 LUN
      REAL*8 ONE, ZERO, PI, WMIN, WMIN2
      PARAMETER(LUN=11, ONE=1.0, ZERO=0.0, PI=3.1415926, WMIN=1d-10,
     &          WMIN2=1d-40)
C!    ----- Parameters for track and radial distance histograms
      INTEGER*4 MXTIC1, MXTIC2, MXTRG1, MXTRG2
      PARAMETER(MXTIC1=1025, MXTIC2=513, MXTRG1=129, MXTRG2=65)
C!    ----- Parameters for fluence calculation from power law
C!          regression to PSTAR data between 1 MeV and 100 MeV
      REAL*8 DPERFL  ! Conversion factor from mass stopping power in
                     ! MeV cm²/g to dose per fluence in Gy nm²
      REAL*8 STPWR1  ! Stopping Power at 1 MeV
      REAL*8 STPEXP  ! Exponent of power law for stopping power
      PARAMETER(DPERFL=1.6e4, STPEXP=-0.7899, STPWR1=276.973)
C!    ----- Functions
```

----- FORTRAN source code of program ME_ROI_3C -----

```
      CHARACTER TSTAMP*24
C!    ----- Local scalars
      CHARACTER FILENM*120, FILOUT*120, HEADER*120
      INTEGER*4 I, I1, I2, IFILE, IJ, J, K, L, NFILES
      LOGICAL ASKINP, DEBUG, HIFLNC
      REAL*8 DBEAM   ! Diameter of beam in nm
      REAL*8 DOSE    ! Absorbed dose in Gray
      REAL*8 DROI    ! Diameter of ROI in nm
      REAL*8 ENERGY  ! Proton energy in MeV
      REAL*8 EVENTS  ! number of events
      REAL*8 FLUENC  ! average number of tracks per ROI cross section
      REAL*8 PVALUE  ! Probability that cluster volume is target
      REAL*8 QVALUE  ! Probability that cluster volume is not a target
      REAL*8 STPWRE  ! Stopping Power at energy E
      REAL*8 DLFLNC, DLWGHT, VCHECK, WGHT
C!    ----- Local arrays
      INTEGER*4 NMAXIC(2), NMAXME(2), NMAXSE(2),
     &          NMCMAX(MXTIC1), NTARGT(8)
      REAL*8 ADEN(2), ANUM(2), APOT(2), PTOK(2), WEIGH(2)
      REAL*8 CORRIC(MXTIC1,MXTIC2)
      REAL*8 PMARG(MXTIC1,8)
      REAL*8 CORRSE(MXTRG1,MXTRG2),CVARSE(MXTRG1,MXTRG2)
      REAL*8 CORRME(MXTRG1,MXTRG2),CVARME(MXTRG1,MXTRG2)
      REAL*8 CTEMP1(MXTRG1,MXTRG2),CTEMP2(MXTRG1,MXTRG2)
      REAL*8 SUMS(8),TGMEAN(8),FNE(MXTRG1,2,2)
C!<<<<End declarations<<<<<<<<<<<<<<<<<<<<<<<<<<<<<<<<<<<<<<<<<<<<<<<<

      PRINT*, 'Program ME_ROI_3C Version: '//VDATE

      IF(1.EQ.1) THEN ! DUMMY block for readability: Read input 1111111
C!    1111111111111111111111111111111111111111111111111111111111111111
C!    Check command file version
      PRINT*, 'Enter 0 for manual input or version (YYMMDD.HHMM) '//
     &          'of command file structure '
      READ(*,*) VCHECK
      ASKINP=(VCHECK.EQ.ZERO)
      IF(.NOT.ASKINP) THEN
        IF(VCHECK.LT.VINPUT) THEN
          PRINT*, 'Command file structure ',VCHECK,' older than '//
     &              'current version ',VINPUT,'=> STOP.'
          STOP
        END IF
        READ(*,*) FILENM
        IF(FILENM(1:9).NE.'ME_ROI_3C') THEN
          PRINT*, 'Command file appear not to be for ROI_3D but '//
     &              'for '//FILENM//'=> STOP.'
          STOP
        END IF
      END IF

      IF(ASKINP) PRINT*, 'Run in debug mode? (1/0)'
      READ(*,*) I
      DEBUG=(I.EQ.1)
      ASKINP=ASKINP.OR.DEBUG

      PRINT*, 'Relative target density (0<PVALUE<1)'
      READ(*,*) PVALUE
      QVALUE=ONE-PVALUE
      IF(ASKINP) PRINT*, 'Absorbed dose in Gy'
      READ(*,*) DOSE
```

----- FORTRAN source code of program ME_ROI_3C -----

```
      IF(ASKINP) PRINT*, 'Size of multi-event bivariate distribution'
      READ(*,*) NMAXME(1), NMAXME(2)
      IF(NMAXME(1).GT.MXTRG1.OR.NMAXME(1).LE.1) NMAXME(1)=MXTRG1
      IF(NMAXME(2).GT.MXTRG2.OR.NMAXME(2).LE.1) NMAXME(2)=MXTRG2

      IF(ASKINP) PRINT*, 'Number of files to process'
      READ(*,*) NFILES
C!    1111111111111111111111111111111111111111111111111111111111111111
      END IF          ! DUMMY block for readability: Read input 1111111

      IF(2.EQ.2) THEN ! DUMMY block for readability: Initialize 2222222
C!    2222222222222222222222222222222222222222222222222222222222222222
C!    2222222222222222222222222222222222222222222222222222222222222222
      END IF          ! DUMMY block for readability: Initialize 2222222

      DO IFILE=1,NFILES
        IF(ASKINP) PRINT*, 'Enter file name ',IFILE
        READ(*,*) FILENM
        IF(ASKINP) PRINT*, 'Enter related proton energy '
        READ(*,*) ENERGY
        STPWRE=STPWR1*EXP(STPEXP*LOG(ENERGY))

        IF(3.EQ.3) THEN ! DUMMY block for readability: Get data  333333
C!      3333333333333333333333333333333333333333333333333333333333333333
        IF(DEBUG) PRINT*, 'Hier beginnt Block 3'

        OPEN(LUN,FILE=FILENM,STATUS='OLD')
        DO I=1,2
          READ(LUN,'(A80)') HEADER
        END DO

        READ(LUN,'(A80)') HEADER
        PRINT*, HEADER
        READ(HEADER(28:36),*) DROI
        READ(HEADER(69:77),*) DBEAM
        FLUENC=0.25*DBEAM*DBEAM*PI*DOSE/DPERFL/STPWRE
        PRINT*, DOSE,ENERGY,STPWRE,FLUENC,DROI,DBEAM

        READ(LUN,'(A)') FILOUT
        READ(LUN,*) NMAXIC(1), NMAXIC(2)
        IF(NMAXIC(1).GT.MXTIC1) THEN
          PRINT*, 'Problem number of lines in '//FILENM//' is ',
     &              NMAXIC(1),' > MXTIC1 =',MXTIC1
          STOP
        END IF
        IF(NMAXIC(2).GT.MXTIC2) THEN
          PRINT*, 'Problem number of columns in '//FILENM//' is ',
     &              NMAXIC(2),' > MXTIC2 =',MXTIC2
          STOP
        END IF

        DO J=1,NMAXIC(1)
          READ(LUN,*) (CORRIC(J,L),L=1,NMAXIC(2))
        END DO

        CLOSE(LUN)
C!      3333333333333333333333333333333333333333333333333333333333333333
        END IF          ! DUMMY block for readability: Get data 3333333

        IF(4.EQ.4) THEN ! DUMMY block for readability: zero arrays 4444
C!      4444444444444444444444444444444444444444444444444444444444444444
```

----- FORTRAN source code of program ME_ROI_3C -----

```
        IF(DEBUG) PRINT*, 'Hier beginnt Block 4 '
        DO I=1,NMAXIC(1)
          NMCMAX(I)=0
          DO J=1,NMAXIC(2)
            IF(CORRIC(I,J).GT.WMIN2) NMCMAX(I)=J
          END DO
        END DO

        DO I=1,MXTIC1
          DO K=1,8
            PMARG(I,K)=ZERO
          END DO
        END DO

        DO I=1,8
          NTARGT(I)=0
        END DO

        DO I=1,MXTRG1
          DO J=1,MXTRG2
            CVARSE(I,J)=ZERO
            CORRSE(I,J)=ZERO
            CVARME(I,J)=ZERO
            CORRME(I,J)=ZERO
            CTEMP1(I,J)=ZERO
            CTEMP2(I,J)=ZERO
          END DO
          DO J=1, 2
            DO K=1,2
              FNE(I,J,K)=ZERO
            END DO
          END DO

        END DO ! DO I=1,NMAXIC(1)
C!      444444444444444444444444444444444444444444444444444444444444444
        END IF          ! DUMMY block for readability: zero arrays 4444

        IF(5.EQ.5) THEN ! DUMMY block for readability: Convolve 5555555
C!      555555555555555555555555555555555555555555555555555555555555555
          IF(DEBUG) PRINT*, 'Hier beginnt Block 5 '
          APOT(1)=ZERO
          PTOK(1)=ONE
          I1=0
          DO I=1,NMAXIC(1)
            IF(I1.LT.NMAXME(1)) I1=I1+1
            APOT(2)=ZERO
            PTOK(2)=ONE
            I2=0
            DO J=1,NMAXIC(2)
              IF(I2.LT.NMAXME(2)) I2=I2+1
              ANUM(1)=APOT(1)
              ADEN(1)=ZERO
              WEIGH(1)=PTOK(1)
              DO K=I,NMAXIC(1)
                IF(WEIGH(1).GT.WMIN2) THEN
                  ANUM(2)=APOT(2)
                  ADEN(2)=ZERO
                  WEIGH(2)=PTOK(2)*WEIGH(1)
                  DO L=J,NMCMAX(K)
                    IF(WEIGH(2).GT.WMIN2) THEN
                      CVARSE(I1,I2)=CVARSE(I1,I2)+WEIGH(2)*CORRIC(K,L)
```

----- FORTRAN source code of program ME_ROI_3C -----

```
                    ADEN(2)=ADEN(2)+ONE
                    ANUM(2)=ANUM(2)+ONE
                    WEIGH(2)=WEIGH(2)*QVALUE*ANUM(2)/ADEN(2)
                  END IF
                END DO ! L=J,NMAXIC(2)
                ADEN(1)=ADEN(1)+ONE
                ANUM(1)=ANUM(1)+ONE
                WEIGH(1)=WEIGH(1)*QVALUE*ANUM(1)/ADEN(1)
              END IF
            END DO ! K=I,NMAXIC(1)
            APOT(2)=APOT(2)+ONE
            PTOK(2)=PTOK(2)*PVALUE
          END DO !J=1,NMAXIC(2)
          APOT(1)=APOT(1)+ONE
          PTOK(1)=PTOK(1)*PVALUE
        END DO !I=1,NMAXIC(1)
C!      555555555555555555555555555555555555555555555555555555555555555
        END IF          ! DUMMY block for readability: Convolve 555555

        IF(6.EQ.6) THEN ! DUMMY block for readability: Margins 6666666
C!      66666666666666666666666666666666666666666666666666666666666666
        IF(DEBUG) PRINT*, 'Hier beginnt Block 6'
C!        marginal distributions of targets with one IC or multi ICs
        DO I=1,NMAXIC(1)
          DO J=1,NMAXIC(2)
            PMARG(I,5)=PMARG(I,5)+CORRIC(I,J)
            PMARG(J,6)=PMARG(J,6)+CORRIC(I,J)
          END DO !I=1,NMAXIC(1)
        END DO !J=1,NMAXIC(2)
        NTARGT(5)=NMAXIC(1)
        NTARGT(6)=NMAXIC(2)
C!        marginal distributions of targets with one DSB or multi DSBs
        DO I=1,NMAXME(1)
          DO J=1,NMAXME(2)
            PMARG(I,3)=PMARG(I,3)+CVARSE(I,J)
            PMARG(J,4)=PMARG(J,4)+CVARSE(I,J)
            IF(CVARSE(I,J).GT.WMIN2) THEN
              NMAXSE(1)=I
              NMAXSE(2)=J
            END IF
          END DO !I=1,NMAXSE(1)
        END DO !J=1,NMAXSE(2)
        NTARGT(3)=NMAXSE(1)
        NTARGT(4)=NMAXSE(1)
C!        "Relative correlation matrix" - ratio bivariate frequency to
C!        product of marginal frequencies
        DO I=1,NMAXSE(1)
          DO J=1,NMAXSE(2)
            IF(PMARG(I,3).GT.WMIN2.AND.PMARG(J,4).GT.WMIN2) THEN
              CORRSE(I,J)=CVARSE(I,J)/(PMARG(I,3)*PMARG(J,4))
            ELSE
              CORRSE(I,J)=CVARSE(I,J)
            END IF
          END DO !I=1,NMAXSE(1)
        END DO !J=1,NMAXSE(2)
C!      66666666666666666666666666666666666666666666666666666666666666
        END IF          ! DUMMY block for readability: Margins 6666666


        IF(7.EQ.7) THEN ! DUMMY block for readability: Main loop  7777777
C!      777777777777777777777777777777777777777777777777777777777777777
```

----- FORTRAN source code of program ME_ROI_3C -----

```
      IF(DEBUG) PRINT*, 'Hier beginnt Block 7'
      NTARGT(1)=NMAXSE(1)
      NTARGT(2)=NMAXSE(2)

      PMARG(1,1)=EXP(-FLUENC) ! Kronecker's delta for J=1 (0 targets)
      PMARG(1,2)=EXP(-FLUENC) ! Kronecker's delta for J=1 (0 targets)
      CVARME(1,1)=EXP(-FLUENC)

      EVENTS=ONE
      HIFLNC=(FLUENC.GT.23.4)
      IF(HIFLNC) THEN
        DLFLNC=DLOG(FLUENC)
        DLWGHT=-FLUENC+DLFLNC
        WGHT=EXP(DLWGHT)
      ELSE
        WGHT=EXP(-FLUENC)*FLUENC
      END IF

      DO I=1,NTARGT(1)
        DO J=1,NTARGT(2)
          CTEMP2(I,J)=CVARSE(I,J)
          CVARME(I,J)=CVARME(I,J)+CTEMP2(I,J)*WGHT
        END DO
      END DO

      DO K=1,2
        TGMEAN(K)=ZERO
        DO I=1,NTARGT(K)
          FNE(I,2,K)=PMARG(I,K+2)
          PMARG(I,K)=PMARG(I,K)+FNE(I,2,K)*WGHT
          IF(DEBUG) TGMEAN(K)=TGMEAN(K)+FNE(I,2,K)*(I-1)
        END DO
      END DO

      IF(DEBUG) THEN
        IF(WGHT.GT.WMIN) THEN
          PRINT*, '#',EVENTS,WGHT,TGMEAN(1)/EVENTS,TGMEAN(2)/EVENTS
        ELSE
          PRINT*, '#',EVENTS,WGHT
        END IF
      END IF


10    CONTINUE
      EVENTS=EVENTS+ONE
      IF(HIFLNC) THEN
        DLWGHT=DLWGHT+DLFLNC-DLOG(EVENTS)
        WGHT=EXP(DLWGHT)
      ELSE
         WGHT=WGHT*FLUENC/EVENTS
      END IF

      DO K=1,2
        TGMEAN(K)=ZERO
        DO I=1,NTARGT(K)
          FNE(I,1,K)=FNE(I,2,K)
          FNE(I,2,K)=ZERO
        END DO
        DO I=1,NTARGT(K)
          DO J=1,NMAXSE(K)
            IJ=I+J-1
```

----- FORTRAN source code of program ME_ROI_3C -----

```
              IF (IJ.GT.NMAXME(K)) IJ=NMAXME(K)
              FNE(IJ,2,K)=FNE(IJ,2,K)+FNE(I,1,K)*PMARG(J,K+2)
            END DO
          END DO
          NTARGT(K)=NTARGT(K)+NMAXSE(K)
          IF(NTARGT(K).GT.NMAXME(K)) NTARGT(K)=NMAXME(K)
          DO I=1,NTARGT(K)
            PMARG(I,K)=PMARG(I,K)+FNE(I,2,K)*WGHT
            IF(DEBUG) TGMEAN(K)=TGMEAN(K)+FNE(I,2,K)*(I-1)
          END DO
        END DO

        DO I=1,NTARGT(1)
          DO J=1,NTARGT(2)
            CTEMP1(I,J)=CTEMP2(I,J)
            CTEMP2(I,J)=ZERO
          END DO
        END DO

        DO I=1,NTARGT(1)
          DO J=1,NTARGT(2)
            DO K=1,I
              DO L=1,J
                CTEMP2(I,J)=CTEMP2(I,J)+
     &                         CTEMP1(K,L)*CVARSE(I+1-K,J+1-L)
              END DO
            END DO
            CVARME(I,J)=CVARME(I,J)+CTEMP2(I,J)*WGHT
          END DO
        END DO

        IF(DEBUG) THEN
          IF(WGHT.GT.WMIN) THEN
            PRINT*, '#',EVENTS,WGHT,TGMEAN(1),TGMEAN(2)
          ELSE
            PRINT*, '#',EVENTS,WGHT
          END IF
        END IF

        IF(EVENTS.LT.FLUENC.OR.WGHT.GT.WMIN) GOTO 10
C!      777777777777777777777777777777777777777777777777777777777777777
        END IF          ! DUMMY block for readability: Main Loop  777777

        IF(8.EQ.8) THEN ! DUMMY block for readability: Margins 8888888
C!      88888888888888888888888888888888888888888888888888888888888888
        IF(DEBUG) PRINT*, 'Hier beginnt Block 8'
        DO I=1,NMAXME(1)
          DO J=1,NMAXME(2)
            PMARG(I,7)=PMARG(I,7)+CVARME(I,J)
            PMARG(J,8)=PMARG(J,8)+CVARME(I,J)
            IF(CVARME(I,J).GT.WMIN2) THEN
              NTARGT(1)=I
              NTARGT(2)=J
            END IF
          END DO !J=1,NMAXME(2)
        END DO !I=1,NMAXME(1)

        DO I=1,NTARGT(1)
          DO J=1,NTARGT(2)
            IF(PMARG(I,7).GT.WMIN2.AND.PMARG(J,8).GT.WMIN2) THEN
              CORRME(I,J)=CVARME(I,J)/(PMARG(I,7)*PMARG(J,8))
```

----- FORTRAN source code of program ME_ROI_3C -----

```
            ELSE
              CORRME(I,J)=CVARME(I,J)
            END IF
          END DO !I=1,NTARGT(1)
        END DO !J=1,NTARGT(2)

        DO K=1,8
          TGMEAN(K)=ZERO
          SUMS(K)=ZERO
          IF(K.GT.6) NTARGT(K)=NTARGT(K-6)
          DO J=1,NTARGT(K)
            TGMEAN(K)=TGMEAN(K)+PMARG(J,K)*(J-1)
            SUMS(K)=SUMS(K)+PMARG(J,K)
          END DO
        END DO ! DO K=1,8
        PRINT*, 'Distributions averages (#hit targets)'
        WRITE(*,'(10f10.4)') (SUMS(J),J=1,8)
        WRITE(*,'(10f10.4)') (TGMEAN(J),J=1,8)
        WRITE(*,*) TGMEAN(3)/TGMEAN(7),TGMEAN(4)/TGMEAN(8)
C!      8888888888888888888888888888888888888888888888888888888888888
        END IF          ! DUMMY block for readability: Margins 8888888

        IF(9.EQ.9) THEN ! DUMMY block for readability: OUTPUT  9999999
C!      9999999999999999999999999999999999999999999999999999999999999
C!        #Write results to output file
          FILOUT='SEB_'//FILENM
          PRINT*, 'Write output to '//FILOUT
          OPEN(LUN,FILE=FILOUT,STATUS='UNKNOWN')
          WRITE(LUN,*) ' *** Output from PROGRAM ME_ROI_3C Version '
     &                 //VDATE//' on '//TSTAMP()
          WRITE(LUN,*) ' Filename: '//FILOUT
          WRITE(LUN,*) HEADER
          WRITE(LUN,*) ' Correlations P1 and F2 for PVALUE= ',PVALUE
          WRITE(LUN,*) NMAXSE(1), NMAXSE(2)
          DO I=1,NMAXSE(1)
            WRITE(LUN,'(1000D15.6)') (CVARSE(I,J),J=1,NMAXSE(2))
          END DO
          CLOSE(LUN)

          FILOUT='SEC_'//FILENM
          PRINT*, 'Write output to '//FILOUT
          OPEN(LUN,FILE=FILOUT,STATUS='UNKNOWN')
          WRITE(LUN,*) ' *** Output from PROGRAM SE_ROI_3C Version '
     &                 //VDATE//' on '//TSTAMP()
          WRITE(LUN,*) ' Filename: '//FILOUT
          WRITE(LUN,*) HEADER
          WRITE(LUN,*) ' Correlations P1 and F2 for PVALUE= ',PVALUE
          WRITE(LUN,*) NMAXSE(1), NMAXSE(2)
          DO I=1,NMAXSE(1)
            WRITE(LUN,'(1000D15.6)') (CORRSE(I,J),J=1,NMAXSE(2))
          END DO
          CLOSE(LUN)

          FILOUT='MEA_'//FILENM
          PRINT*, 'Write output to '//FILOUT
          OPEN(LUN,FILE=FILOUT,STATUS='UNKNOWN')
          WRITE(LUN,*) ' *** Output from PROGRAM ME_ROI_3C Version '
     &                 //VDATE//' on '//TSTAMP()
          WRITE(LUN,*) ' Filename: '//FILOUT
          WRITE(LUN,*) HEADER
          WRITE(LUN,*) ' Multi and single event distribution for '//
```

----- FORTRAN source code of program ME_ROI_3C -----

```
     &                'PVALUE= ',PVALUE,'  DOSE= ',DOSE,' Gy'
          WRITE(LUN,*) ' ***  Compared with distributions for ICs ***'

          WRITE(LUN,'(A9,12a15)') '#CVs','P1','F2','P1','F2','P1','F2',
     &                           'P1','F2'

          WRITE(LUN,'(1X,a8,50a15)') 'Data:',('ME',K=1,2),
     &               ('SE',K=1,2),('IC',K=1,2),('ME',K=1,2)

          DO I=1,NMAXIC(1)
            WRITE(LUN,'(1X,i8,10F15.10)') I-1,(PMARG(I,J),J=1,8)
          END DO
          CLOSE(LUN)

          FILOUT='MEB_'//FILENM
          PRINT*, ' Write output to '//FILOUT
          OPEN(LUN,FILE=FILOUT,STATUS='UNKNOWN')
          WRITE(LUN,*) ' *** Output from PROGRAM ME_ROI_3C Version '
     &                //VDATE//' on '//TSTAMP()
          WRITE(LUN,*) ' Filename: '//FILOUT
          WRITE(LUN,*) HEADER
          WRITE(LUN,*) ' Bivariate freq. P1 and F2 for PVALUE= ',PVALUE
          WRITE(LUN,*) NMAXME(1), NMAXME(2)
          DO I=1,NMAXME(1)
            WRITE(LUN,'(1000D15.6)') (CVARME(I,J),J=1,NMAXME(2))
          END DO
          CLOSE(LUN)

          FILOUT='MEC_'//FILENM
          PRINT*, 'Write output to '//FILOUT
          OPEN(LUN,FILE=FILOUT,STATUS='UNKNOWN')
          WRITE(LUN,*) ' *** Output from PROGRAM ME_ROI_3C Version '
     &                //VDATE//' on '//TSTAMP()
          WRITE(LUN,*) ' Filename: '//FILOUT
          WRITE(LUN,*) HEADER
          WRITE(LUN,*) ' Correlations P1 and F2 for PVALUE= ',PVALUE
          WRITE(LUN,*) NMAXME(1), NMAXME(2)
          DO I=1,NMAXME(1)
            WRITE(LUN,'(1000D15.6)') (CORRME(I,J),J=1,NMAXME(2))
          END DO
          CLOSE(LUN)

          FILOUT='MED_'//FILENM
          PRINT*, 'Write output to '//FILOUT
          OPEN(LUN,FILE=FILOUT,STATUS='UNKNOWN')
          WRITE(LUN,*) ' *** Output from PROGRAM ME_ROI_3C Version '
     &                //VDATE//' on '//TSTAMP()
          WRITE(LUN,*) ' Filename: '//FILOUT
          WRITE(LUN,*) HEADER
          WRITE(LUN,*) ' Multi and single event distribution for '//
     &                'PVALUE= ',PVALUE,'  DOSE= ',DOSE,' Gy'

          WRITE(LUN,'(A9,12a15)') '#CVs','P1','F2','P1','F2'
          WRITE(LUN,'(1X,a8,50a15)') 'Data:',('ME',K=1,2),
     &               ('SE',K=1,2)

          DO I=1,NTARGT(1)
            WRITE(LUN,'(1X,i8,10F15.10)') I-1,(PMARG(I,J),J=1,4)
          END DO
          CLOSE(LUN)
C!      999999999999999999999999999999999999999999999999999999999999999
```

----- FORTRAN source code of program ME_ROI_3C -----

```
         END IF          ! DUMMY block for readability: OUTPUT     9999999
      END DO ! IFILE=1,NFILES

      END PROGRAM ! ME_ROI_3C
C!______________________________________________________________________

        INCLUDE 'TSTAMP.f'
```

*FORTRAN source code of program ME_ROI_3D*

This is the precursor of ME_ROI_3C where only the marginal distributions are considered and there is no convolution with Binomial distributions so that the results apply to IC distributions. Reads output files 3D_*.dat from ROI_3D, calculates multi event distributions for targets with ionization clusters and produces several output files for further use of the results, named by adding a prefix to the input files:

- ME_*'input_filename'* multi event frequency distributions of targets with one or more than one ionization clusters.
- SE_*'input_filename'* single event frequency distributions of targets with one or more than one ionization clusters.
- TE_*'input_filename'* conditional single event frequency distributions of targets with one or more than one ionization clusters for traversal of the ROI by the primary particle trajectory.
- All_*'input_filename'* all of the three above plus single tracks at certain distances plus tracks passing through annuli around region of interest cross section.

Notes:

- **The program must be executed in the directory where the data files are located**.
- The program expects input of the name of a file listing the debugging options (example see section “Sample debug options file for use with programs IC_3D and ROI_3D”).
- Inputs of parameters and input file names are prompted for unless they are entered via a text file. (Manual input is initiated by entering 0 with the first prompt.) A sample input file is listed in the table below.

----- Sample input file for ME_ROI_3D -----

```
210605.17       ! Command file version number ...
ME_ROI_3D       ! ... for this program
1               ! Flag IDEBUG for priniting messages
2.0             ! Absorbed dose in Gy
1               ! Number of input files to process
3B_12.0nm_IC_2.0nm_p3MeVA_mai2015.dat
3.
```

Meaning of the input lines:
(1) Version date and time of the input file structure in format YYMMDD.HHMM
(2) Name of the program
(3) Flag for printing messages on progress (value = 1)
(4) Absorbed dose in Gy
(5) Maximum dimensions of the matrix of bivariate frequencies of the number of targets with a single DSB and multiple DSBs.
(6) Number of input files to process
(7) further odd lines: names of the input files
(8) further even lines: proton energy in MeV

----- FORTRAN source code of program ME_ROI_3D -----

```
      PROGRAM ME_ROI_3D
C!>>> Declarations >>>>>>>>>>>>>>>>>>>>>>>>>>>>>>>>>>>>>>>>>>>>>>>>>
      IMPLICIT NONE
C!    ----- Parameters: Version Date and number
      CHARACTER VDATE*11
      REAL*8 VINPUT  ! Version number for command files YYMMDD.HHMM
C!                    (date and time  file structure was last changed)
```

----- FORTRAN source code of program ME_ROI_3D -----

```
C!    ################################################################
      PARAMETER(VDATE='05-JUN-2021',VINPUT=210605.1700d0) ! ###########
C!    05-JUN-2021 HR:
C!      - Reorganized initialization of counters.
C!      - Added timestamp in output file
C!    14-APR-2021 HR:
C!      - Adapted to changes done in ROI_3D (leaner output).
C!    31-MAR-2021 HR:
C!      - Fixed bug in initialization of convolution
C!    28-MAR-2021 HR:
C!      - Added one-in-all output for easier transfer to Origin etc.
C!    26-MAR-2021 HR:
C!      - Added option for calculating fluence from dose (based on
C!        new input info of proton energy)
C!    15-MAR-2020 HR:
C!      - Added VDATE and modified COMMON BLOCK VERBOSE
C!    ----- Parameters: General purpose constants
      INTEGER*4 LUN
      REAL*8 ONE, ZERO, PI, WMIN
      PARAMETER(LUN=11, ONE=1.0, ZERO=0.0, PI=3.1415926, WMIN=1d-12)
C!    ----- Parameters for track and radial distance histograms
      INTEGER*4 KMAX, MXRPOS, MXTARG
      PARAMETER(KMAX=9, MXRPOS=101, MXTARG=65537)
C!    ----- Parameters for fluence calculation from power law
C!          regression to PSTAR data between 1 MeV and 100 MeV
      REAL*8 DPERFL  ! Conversion factor from mass stopping power in
                     ! MeV cm²/g to dose per fluence in Gy nm²
      REAL*8 STPWR1  ! Stopping Power at 1 MeV
      REAL*8 STPEXP  ! Exponent of power law for stopping power
      PARAMETER(DPERFL=1.6e4, STPEXP=-0.7899, STPWR1=276.973)
C!    ----- Functions
      CHARACTER TSTAMP*24
C!    ----- Local scalars
      CHARACTER FILENM*120, FILOUT*120, HEADER*1200, PARAM*2,
     &          RINFO(5,2)*11
      INTEGER*4 I, IFILE, IJ, ITARG, J, K, KMX, L, NFILES, NRPOS
      INTEGER*4 NTARG
      LOGICAL ASKINP, DEBUG, HIFLNC
      REAL*8 DBEAM   ! Diameter of beam in nm
      REAL*8 DOSE    ! Absorbed dose in Gray
      REAL*8 DROI    ! Diameter of ROI in nm
      REAL*8 ENERGY  ! Proton energy in MeV
      REAL*8 EVENTS  ! number of events
      REAL*8 FLUENC  ! average number of tracks per ROI cross section
      REAL*8 STPWRE  ! Stopping Power at energy E
      REAL*8 DLFLNC, DLWGHT, SPOSIN, SUMPOS, VCHECK, WGHT, TGMEAN(KMAX)
C!    ----- Local arrays
      INTEGER*4 JMAX(2),NTARGK(KMAX),NTARGT(KMAX)
      REAL*8 FREQSE(MXTARG,KMAX),FREQTE(MXTARG,KMAX),
     &       FREQME(MXTARG,KMAX),FNE(MXTARG,2,KMAX),
     &       FREQST(MXRPOS), FREQTR(MXTARG,KMAX,5),
     &       FREQAR(MXTARG,KMAX,5), RELFLU(5)
C!<<<<End declarations<<<<<<<<<<<<<<<<<<<<<<<<<<<<<<<<<<<<<<<<<<<<<<<<<

      PRINT*, 'Program ME_ROI_3D Version: '//VDATE

      IF(1.EQ.1) THEN ! DUMMY block for readability: Read input 1111111

C!      Check command file version
        PRINT*, 'Enter 0 for manual input or version (YYMMDD.HHMM) '//
     &          'of command file structure '
```

----- FORTRAN source code of program ME_ROI_3D -----

```
        READ(*,*) VCHECK
        ASKINP=(VCHECK.EQ.ZERO)
        IF(.NOT.ASKINP) THEN
          IF(VCHECK.LT.VINPUT) THEN
            PRINT*, 'Command file structure ',VCHECK,' older than '//
     &              'current version ',VINPUT,'=> STOP.'
            STOP
          END IF
          READ(*,*) FILENM
          IF(FILENM(1:9).NE.'ME_ROI_3D') THEN
            PRINT*, 'Command file appear not to be for ROI_3D but '//
     &              'for '//FILENM//'=> STOP.'
            STOP
          END IF
        END IF

        IF(ASKINP) PRINT*, 'Run in debug mode? (1/0)'
        READ(*,*) I
        DEBUG=(I.EQ.1)

        IF(ASKINP) PRINT*, 'Absorbed dose in Gy'
        READ(*,*) DOSE

        IF(ASKINP) PRINT*, 'Number of files to process'
        READ(*,*) NFILES

      END IF          ! DUMMY block for readability: Read input 1111111

      IF(2.EQ.2) THEN ! DUMMY block for readability: Initialize 2222222
        RINFO(1,1)='        R=0'
        RINFO(2,1)='R=0.75R_ROI'
        RINFO(3,1)='    R=R_ROI'
        RINFO(4,1)=' R=1.5R_ROI'
        RINFO(5,1)='   R=2R_ROI'
        RINFO(1,2)='    R<R_ROI'
        RINFO(2,2)='1<R/R_ROI<2'
        RINFO(3,2)='2<R/R_ROI<3'
        RINFO(4,2)='3<R/R_ROI<4'
        RINFO(5,2)='4<R/R_ROI<5'
        NTARG=MXTARG
      END IF          ! DUMMY block for readability: Initialize 2222222

      DO IFILE=1,NFILES
        IF(ASKINP) PRINT*, 'Enter file name ',IFILE
        READ(*,*) FILENM
        IF(.NOT.ASKINP) PRINT*, 'Processing file '//FILENM
        IF(ASKINP) PRINT*, 'Enter related proton energy '
        READ(*,*) ENERGY
        STPWRE=STPWR1*EXP(STPEXP*LOG(ENERGY))

        IF(3.EQ.3) THEN ! DUMMY block for readability: Init counters 3333
          DO K=1,KMAX
            DO I=1,MXTARG
              FREQSE(I,K)=ZERO
              FREQTE(I,K)=ZERO
              FREQME(I,K)=ZERO
              DO J=1,2
                FNE(I,J,K)=ZERO
              END DO
              DO J=1,5
                FREQTR(I,K,J)=ZERO
```

----- FORTRAN source code of program ME_ROI_3D -----

```
              FREQAR(I,K,J)=ZERO
            END DO
          END DO !I=1,MXTARG
          TGMEAN(K)=ZERO
        END DO ! DO K=1,KMAX
      END IF          ! DUMMY block for readability: Init counters 3333

      IF(4.EQ.4) THEN ! DUMMY block for readability: Get data  444444
        IF(DEBUG) PRINT*, 'Begin of Block 4'

        OPEN(LUN,FILE=FILENM,STATUS='OLD')
        DO I=1,2
          READ(LUN,'(A80)') HEADER
        END DO
        READ(LUN,'(10X,10i6)') (NTARGK(I),I=1,KMAX)
        DO I=1,KMAX
          IF(NTARGK(I).GT.0) KMX=I
        END DO

        READ(LUN,'(A80)') HEADER
        READ(HEADER(28:36),*) DROI
        READ(HEADER(69:77),*) DBEAM
        NRPOS=1+NINT(DBEAM/DROI*20.)
        IF(NRPOS.GE.41) THEN
          JMAX(1)=5
        ELSE
          JMAX(1)=4
          IF(NRPOS.LT.16) JMAX(1)=1
          IF(NRPOS.LT.21) JMAX(1)=2
          IF(NRPOS.LT.31) JMAX(1)=3
        END IF
        IF(NRPOS.EQ.101) THEN
          JMAX(2)=5
        ELSE
          JMAX(2)=4
          IF(NRPOS.LT.41) JMAX(2)=1
          IF(NRPOS.LT.61) JMAX(2)=2
          IF(NRPOS.LT.81) JMAX(2)=3
        END IF
C!        FLUENC=DROI*DROI*PI/4.*DOSE/DPERFL/STPWRE
C!      Single-event distribution uses 20*RROI as max. impact parameter
        FLUENC=0.25*DBEAM*DBEAM*PI*DOSE/DPERFL/STPWRE
        RELFLU(1)=DROI*DROI/(DBEAM*DBEAM)
        SPOSIN=ONE
        DO I=2,5
          SPOSIN=SPOSIN+2.
          RELFLU(I)=RELFLU(1)*SPOSIN
        END DO
        PRINT*, DOSE,ENERGY,STPWRE,FLUENC,DROI,DBEAM,NRPOS

        DO K=1,KMX
          DO J=1,2
            READ(LUN,*) PARAM
          END DO
          DO J=1,NTARGK(K)
            READ(LUN,*) PARAM,ITARG,FREQSE(J,K), FREQTE(J,K),
     &                      (FREQST(L),L=1,NRPOS)
            FREQTR(J,K,1)=FREQST(1)
            IF(NRPOS.GE.16) FREQTR(J,K,2)=FREQST(16)
            IF(NRPOS.GE.21) FREQTR(J,K,3)=FREQST(21)
            IF(NRPOS.GE.31) FREQTR(J,K,4)=FREQST(31)
```

----- FORTRAN source code of program ME_ROI_3D -----

```
              IF(NRPOS.GE.41) FREQTR(J,K,5)=FREQST(41)
              L=1
              SPOSIN=ZERO
              SUMPOS=ONE
              FREQAR(J,K,L)=FREQST(1)
              DO IJ=20*L-18,20*L+1
                SPOSIN=SPOSIN+8.
                FREQAR(J,K,L)=FREQAR(J,K,L)+SPOSIN*FREQST(IJ)
                SUMPOS=SUMPOS+SPOSIN
              END DO
              FREQAR(J,K,L)=FREQAR(J,K,L)*RELFLU(L)/SUMPOS
              DO L=2,JMAX(2)
                FREQAR(J,K,L)=ZERO
                SUMPOS=ZERO
                DO IJ=20*L-18,20*L+1
                  SPOSIN=SPOSIN+8.
                  FREQAR(J,K,L)=FREQAR(J,K,L)+SPOSIN*FREQST(IJ)
                  SUMPOS=SUMPOS+SPOSIN
                END DO
                FREQAR(J,K,L)=FREQAR(J,K,L)*RELFLU(L)/SUMPOS
              END DO
            END DO
            DO J=1,2
              READ(LUN,*) PARAM
            END DO
          END DO
          CLOSE(LUN)
        END IF          ! DUMMY block for readability: Get data  4444444

        IF(5.EQ.5) THEN ! DUMMY block for readability: Main loop  5555555
          IF(DEBUG) PRINT*, 'Begin of Block 5'

C!        Init counters
          DO K=1,KMX
            FREQME(1,K)=EXP(-FLUENC) ! Kronecker's delta for J=1 (0
targets)
          END DO

          EVENTS=ONE
          HIFLNC=(FLUENC.GT.23.4)
          IF(HIFLNC) THEN
            DLFLNC=DLOG(FLUENC)
            DLWGHT=-FLUENC+DLFLNC
            WGHT=EXP(DLWGHT)
          ELSE
             WGHT=EXP(-FLUENC)*FLUENC
          END IF

          DO K=1,KMX
            DO J=1,NTARGK(K)
              FNE(J,2,K)=FREQSE(J,K)
              IF(DEBUG) TGMEAN(K)=TGMEAN(K)+FNE(J,2,K)*(J-1)
              FREQME(J,K)=FREQME(J,K)+FNE(J,2,K)*WGHT
            END DO
            NTARGT(K)=NTARGK(K)
          END DO ! DO K=1,KMX

          IF(DEBUG) THEN
            IF(WGHT.GT.WMIN) THEN
              PRINT*, '#',EVENTS,WGHT,TGMEAN(1)/EVENTS,TGMEAN(2)/EVENTS
            ELSE
```

----- FORTRAN source code of program ME_ROI_3D -----

```
              PRINT*, '#',EVENTS,WGHT
            END IF
          END IF

   10     CONTINUE
          EVENTS=EVENTS+ONE
          IF(HIFLNC) THEN
            DLWGHT=DLWGHT+DLFLNC-DLOG(EVENTS)
            WGHT=EXP(DLWGHT)
          ELSE
             WGHT=WGHT*FLUENC/EVENTS
          END IF

          DO K=1,KMX
            TGMEAN(K)=ZERO
            DO J=1,NTARGT(K)
              FNE(J,1,K)=FNE(J,2,K)
              FNE(J,2,K)=ZERO
            END DO
            DO J=1,NTARGT(K)
              DO I=1,NTARGK(K)
                IJ=I+J-1
                IF (IJ.GT.MXTARG) IJ=MXTARG
                FNE(IJ,2,K)=FNE(IJ,2,K)+FNE(J,1,K)*FREQSE(I,K)
              END DO
            END DO
            NTARGT(K)=NTARGT(K)+NTARGK(K)
            IF(NTARGT(K).GT.MXTARG) NTARGT(K)=MXTARG
            DO J=1,NTARGT(K)
              FREQME(J,K)=FREQME(J,K)+FNE(J,2,K)*WGHT
              IF(DEBUG) TGMEAN(K)=TGMEAN(K)+FNE(J,2,K)*(J-1)
            END DO
          END DO ! DO K=1,KMX

          IF(DEBUG) THEN
            IF(WGHT.GT.WMIN) THEN
              PRINT*, '#',EVENTS,WGHT,TGMEAN(1),TGMEAN(2)
            ELSE
              PRINT*, '#',EVENTS,WGHT
            END IF
          END IF

          IF(EVENTS.LT.FLUENC.OR.WGHT.GT.WMIN) GOTO 10   ! >>>>>>>>>>
C!        End of inner loop

          DO K=1,KMX
            TGMEAN(K)=ZERO
            DO J=1,NTARGK(K)
              TGMEAN(K)=TGMEAN(K)+FREQSE(J,K)*(J-1)
            END DO
          END DO ! DO K=1,KMX
          PRINT*, 'SE distributions averages (#hit targets) K=1,',KMX
          PRINT*, (TGMEAN(K),K=1,KMX)
          PRINT*, (TGMEAN(K)/EVENTS,K=1,KMX)

          DO K=1,KMX
            TGMEAN(K)=ZERO
            DO J=1,NTARGK(K)
              TGMEAN(K)=TGMEAN(K)+FREQTE(J,K)*(J-1)
            END DO
          END DO ! DO K=1,KMX
```

----- FORTRAN source code of program ME_ROI_3D -----

```
          PRINT*, 'TE distributions averages (#hit targets) K=1,',KMX
          PRINT*, (TGMEAN(K),K=1,KMX)
          PRINT*, (TGMEAN(K)/EVENTS,K=1,KMX)

          DO K=1,KMX
            TGMEAN(K)=ZERO
            DO J=1,NTARGT(K)
              TGMEAN(K)=TGMEAN(K)+FREQME(J,K)*(J-1)
            END DO
          END DO ! DO K=1,KMX
          PRINT*, 'ME distributions averages (#hit targets) K=1,',KMX
          PRINT*, (TGMEAN(K),K=1,KMX)
          PRINT*, (TGMEAN(K)/EVENTS,K=1,KMX)

        END IF          ! DUMMY block for readability: Main Loop  555555

        IF(9.EQ.9) THEN ! DUMMY block for readability: OUTPUT  9999999
C!        #Write results to output file
          FILOUT='SE_'//FILENM
          PRINT*, 'Write output to '//FILOUT
          OPEN(LUN,FILE=FILOUT,STATUS='UNKNOWN')
          WRITE(LUN,*) '*** Output from PROGRAM ME_ROI_3D Version '
     &                //VDATE//' on '//TSTAMP()
          WRITE(LUN,*) 'Filename: '//FILOUT
          WRITE(LUN,*) HEADER
          WRITE(LUN,'(1X,2a8,9(a14,i1))') '#Sites','P1',('F',J,J=2,KMX)
          DO I=1,NTARGK(1)
            SUMPOS=ZERO
            DO K=1,KMX
              IF(SUMPOS.LT.FREQSE(I,K)) SUMPOS=FREQSE(I,K)
            END DO
            IF(SUMPOS.GT.WMIN) NTARG=I
          END DO
          DO I=1,NTARG
            WRITE(LUN,'(1X,i8,10f15.8)') I-1,(FREQSE(I,K),K=1,KMX)
          END DO
          CLOSE(LUN)

          FILOUT='TE_'//FILENM
          PRINT*, 'Write output to '//FILOUT
          OPEN(LUN,FILE=FILOUT,STATUS='UNKNOWN')
          WRITE(LUN,*) '*** Output from PROGRAM ME_ROI_3D Version '
     &                //VDATE//' on '//TSTAMP()
          WRITE(LUN,*) 'Filename: '//FILOUT
          WRITE(LUN,*) HEADER
          WRITE(LUN,'(1X,2a8,9(a14,i1))') '#Sites','P1',('F',J,J=2,KMX)
          DO I=1,NTARGK(1)
            SUMPOS=ZERO
            DO K=1,KMX
              IF(SUMPOS.LT.FREQTE(I,K)) SUMPOS=FREQTE(I,K)
            END DO
            IF(SUMPOS.GT.WMIN) NTARG=I
          END DO
          DO I=1,NTARG
            WRITE(LUN,'(1X,i8,10f15.6)') I-1,(FREQTE(I,K),K=1,KMX)
          END DO
          CLOSE(LUN)

          FILOUT='ME_'//FILENM
          PRINT*, 'Write output to '//FILOUT
          OPEN(LUN,FILE=FILOUT,STATUS='UNKNOWN')
```

----- FORTRAN source code of program ME_ROI_3D -----

```
      WRITE(LUN,*) '*** Output from PROGRAM ME_ROI_3D Version '
     &             //VDATE//' on '//TSTAMP()
      WRITE(LUN,*) 'Filename: '//FILOUT
      WRITE(LUN,*) HEADER
      WRITE(LUN,*) 'DOSE= ',DOSE,' Gy'
      WRITE(LUN,'(1X,2a8,9(a14,i1))') '#Sites','P1',('F',J,J=2,KMX)
      DO I=1,NTARGT(1)
        SUMPOS=ZERO
        DO K=1,KMX
          IF(SUMPOS.LT.FREQME(I,K)) SUMPOS=FREQME(I,K)
        END DO
        IF(SUMPOS.GT.WMIN) NTARG=I
      END DO
      DO I=1,NTARG
        WRITE(LUN,'(1X,i8,10f15.6)') I-1,(FREQME(I,K),K=1,KMX)
      END DO
      CLOSE(LUN)

      FILOUT='ALL_'//FILENM
      KMX=2
      PRINT*, 'Write output to '//FILOUT
      OPEN(LUN,FILE=FILOUT,STATUS='UNKNOWN')
      WRITE(LUN,*) '*** Output from PROGRAM ME_ROI_3D Version '
     &             //VDATE//' on '//TSTAMP()
      WRITE(LUN,*) 'Filename: '//FILOUT
      WRITE(LUN,*) HEADER
      WRITE(LUN,*) 'DOSE= ',DOSE,' Gy'

      WRITE(LUN,'(1X,a8,50(a14,i1))') '#Sites',
     &               'P',1,('F',K,K=2,KMX),'P',1,('F',K,K=2,KMX),
     &               'P',1,('F',K,K=2,KMX),'P',1,('F',K,K=2,KMX),
     &               'P',1,('F',K,K=2,KMX),'P',1,('F',K,K=2,KMX),
     &               'P',1,('F',K,K=2,KMX),'P',1,('F',K,K=2,KMX),
     &               'P',1,('F',K,K=2,KMX),'P',1,('F',K,K=2,KMX),
     &               'P',1,('F',K,K=2,KMX),'P',1,('F',K,K=2,KMX),
     &               'P',1,('F',K,K=2,KMX)

      WRITE(LUN,'(1X,a8,50a15)') 'Data:',('ME',K=1,KMX),
     &            ('SE',K=1,KMX),('TE',K=1,KMX),
     &            ((RINFO(J,2),K=1,KMX),J=1,JMAX(2)),
     &            ((RINFO(J,1),K=1,KMX),J=1,JMAX(1))

      DO I=1,NTARGT(1)
        SUMPOS=ZERO
        DO K=1,KMX
          IF(SUMPOS.LT.FREQSE(I,K)) SUMPOS=FREQSE(I,K)
          IF(SUMPOS.LT.FREQTE(I,K)) SUMPOS=FREQTE(I,K)
          IF(SUMPOS.LT.FREQME(I,K)) SUMPOS=FREQME(I,K)
          DO J=1,JMAX(1)
            IF(SUMPOS.LT.FREQTR(I,K,J)) SUMPOS=FREQTR(I,K,J)
          END DO
          DO J=1,JMAX(2)
            IF(SUMPOS.LT.FREQAR(I,K,J)) SUMPOS=FREQAR(I,K,J)
          END DO
        END DO
        IF(SUMPOS.GT.WMIN) NTARG=I
      END DO
      DO I=1,NTARG
        WRITE(LUN,'(1X,i8,50e15.6)') I-1,(FREQME(I,K),K=1,KMX),
     &            (FREQSE(I,K),K=1,KMX),(FREQTE(I,K),K=1,KMX),
     &            ((FREQAR(I,K,J),K=1,KMX),J=1,JMAX(2))     ,
```

----- FORTRAN source code of program ME_ROI_3D -----

```
     &                  ((FREQTR(I,K,J),K=1,KMX),J=1,JMAX(1))
            END DO
            CLOSE(LUN)

          END IF          ! DUMMY block for readability: OUTPUT     9999999
        END DO ! IFILE=1,NFILES

        END PROGRAM ! ME_ROI_3D
C!______________________________________________________________________

        INCLUDE 'TSTAMP.f'
```

*FORTRAN source code of program ME_ROI_3P*

This is a lean and much faster variant of ME_ROI_3C where only the marginal distributions are considered. The program uses the bivariate distributions output from ROI_3D (3B_*.dat) and convolutes them with Binomial distributions of a given success probability such as to convert IC to DSB distributions.

Output:

- MEP_*'input_filename'* multi- and single event frequency distributions after convolution with binomial compared to frequency distributions of ionization clusters.

Notes:

- **The program must be executed in the directory where the data files are located**.
- The program expects input of the name of a file listing the debugging options (example see section "Sample debug options file for use with programs IC_3D and ROI_3D").
- Inputs of parameters and input file names are prompted for unless they are entered via a text file. (Manual input is initiated by entering 0 with the first prompt.) A sample input file is listed in the table below.

----- Sample input file for ME_ROI_3P -----

```
210605.16      ! Command file version number ...
ME_ROI_3P      ! ... for this program
1   100        ! Flags IDEBUG, IMESSG
0.01           ! Relative target density
6              ! Number of dose values
0.5 1.0 2.0 4.0 8.0 16.0 ! Absorbed doses in Gy
3B_12.0nm_IC_2.0nm_p3MeVA_mai2015.dat
3.
```

Meaning of the input lines:

(1) Version date and time of the input file structure in format YYMMDD.HHMM
(2) Name of the program
(3) Flag for printing messages on progress (value = 1) and number of iterations between successive outputs
(4) Relative target density (success probability of the binomial)
(5) Number of dose values in the next line
(6) Absorbed doses in Gy
(7) further odd lines: names of the input files
(8) further even lines: proton energy in MeV

----- FORTRAN source code of program ME_ROI_3P -----

```
      PROGRAM ME_ROI_3P
C!>>> Declarations >>>>>>>>>>>>>>>>>>>>>>>>>>>>>>>>>>>>>>>>>>>>>>>>>>>>>
      IMPLICIT NONE
C!    ----- Parameters: Version Date and number
      CHARACTER VDATE*11
      REAL*8 VINPUT  ! Version number for command files YYMMDD.HHMM
C!                     (date and time  file structure was last changed)
C!    ################################################################
      PARAMETER(VDATE='05-JUN-2021',VINPUT=210605.1600d0) ! ###########
C!    05-JUN-2021 HR:
```

----- FORTRAN source code of program ME_ROI_3P -----

```
C!        - Reorganized command input file.
C!        - Added timestamp in output file
C!      02-MAY-2021 HR:
C!        - Introduced PREFIX to output file name for inclusion of dose.
C!      15-APR-2021 HR:
C!        - Adapted to changes done in ROI_3D (leaner output).
C!      13-APR-2021 HR:
C!        - Created from ME_ROI_3D
C!      ----- Purpose
C!      Calculate multi-ebent distribution of targets with single and
C!      and multiple ionization clusters after filtering with Bernoulli
C!      process.
C!      ----- Parameters: General purpose constants
        INTEGER*4 LUN
        REAL*8 ONE, ZERO, PI, WMIN, WMIN2
        PARAMETER(LUN=11, ONE=1.0, ZERO=0.0, PI=3.1415926, WMIN=1d-10,
     &           WMIN2=1d-40)
C!      ----- Parameters for track and radial distance histograms
        INTEGER*4 MXTARG, MXTRG1, MXTRG2, MXNDOS, MXNCOL
        PARAMETER(MXTARG=65636, MXTRG1=1025, MXTRG2=513, MXNDOS=8,
     &           MXNCOL=2*MXNDOS+4)
C!      ----- Parameters for fluence calculation from power law
C!            regression to PSTAR data between 1 MeV and 100 MeV
        REAL*8 DPERFL  ! Conversion factor from mass stopping power in
                       ! MeV cm²/g to dose per fluence in Gy nm²
        REAL*8 STPWR1  ! Stopping Power at 1 MeV
        REAL*8 STPEXP  ! Exponent of power law for stopping power
        PARAMETER(DPERFL=1.6e4, STPEXP=-0.7899, STPWR1=276.973)
C!      ----- Functions
        CHARACTER TSTAMP*24
C!      ----- Local scalars
        CHARACTER FILENM*120, FILOUT*120, HEADER*120, PREFIX*4
        INTEGER*4 I, IDOSE, IFILE, IJ, IMESSG(2), J, K, L, NCOLS, NDOSES,
     &           NFILES
        LOGICAL ASKINP, DEBUG(2), HIFLNC
        REAL*8 DBEAM   ! Diameter of beam in nm
        REAL*8 DROI    ! Diameter of ROI in nm
        REAL*8 ENERGY  ! Proton energy in MeV
        REAL*8 EVENTS  ! number of events
        REAL*8 FLUENC  ! average number of tracks per ROI cross section
        REAL*8 PVALUE  ! Probability that cluster volume is target
        REAL*8 QVALUE  ! Probability that cluster volume is not a target
        REAL*8 STPWRE  ! Stopping Power at energy E
        REAL*8 DLFLNC, DLWGHT, VCHECK, WGHT
C!      ----- Local arrays
        CHARACTER CDOSES(MXNDOS)*7
        INTEGER*4 ICOL(2), ICOLIC(2), ICOLSE(2)
        INTEGER*4 NMAX(2), NMAXME(2), NMAXSE(2), NMCMAX(MXTRG1), NTARGT(8)
        REAL*8 ADEN(2), ANUM(2), APOT(2), PTOK(2), WEIGH(2)
        REAL*8 DOSE(MXNDOS)     ! Absorbed dose in Gray
        REAL*8 BVARIC(MXTRG1,MXTRG2)
        REAL*8 PMARG(MXTARG,MXNCOL)
        REAL*8 CORRSE(MXTRG1,MXTRG2),BVARSE(MXTRG1,MXTRG2)
        REAL*8 SUMS(MXNCOL),TGMEAN(MXNCOL),FNE(MXTARG,2,2)
C!<<<<End declarations<<<<<<<<<<<<<<<<<<<<<<<<<<<<<<<<<<<<<<<<<<<<<<<<<

        PRINT*, 'Program ME_ROI_3P Version: '//VDATE

        IF(1.EQ.1) THEN ! DUMMY block for readability: Read input 1111111
C!        Check command file version
          PRINT*, 'Enter 0 for manual input or version (YYMMDD.HHMM) '//
```

----- FORTRAN source code of program ME_ROI_3P -----

```
     &          'of command file structure '
        READ(*,*) VCHECK
        ASKINP=(VCHECK.EQ.ZERO)
        IF(.NOT.ASKINP) THEN
          IF(VCHECK.LT.VINPUT) THEN
            PRINT*, 'Command file structure ',VCHECK,' older than '//
     &              'current version ',VINPUT,'=> STOP.'
            STOP
          END IF
          READ(*,*) FILENM
          IF(FILENM(1:9).NE.'ME_ROI_3P') THEN
            PRINT*, 'Command file appear not to be for ROI_3D but '//
     &              'for '//FILENM//'=> STOP.'
            STOP
          END IF
        END IF

        IF(ASKINP) THEN
          PRINT*, 'Run in debug mode? (1/0) '
          READ(*,*) IMESSG(1)
          PRINT*, 'Interval between printing intermediate results '//
     &            '(0=suppress this output) '
          READ(*,*) IMESSG(2)
        ELSE
          READ(*,*) (IMESSG(I),I=1,2)
        END IF
        DEBUG(1)=(IMESSG(1).EQ.1)
        DEBUG(2)=(IMESSG(2).GE.1)
        ASKINP=ASKINP.OR.DEBUG(1)

        PRINT*, 'Relative target density (0<PVALUE<1)'
        READ(*,*) PVALUE
        QVALUE=ONE-PVALUE
        IF(ASKINP) PRINT*, 'Number of dose values (<=',MXNDOS,')'
        READ(*,*) NDOSES
        IF(NDOSES.GT.MXNDOS) NDOSES=MXNDOS
        IF(ASKINP) PRINT*, 'Absorbed doses in Gy'
        READ(*,*) (DOSE(I),I=1,NDOSES)

        IF(ASKINP) PRINT*, 'Number of files to process'
        READ(*,*) NFILES
      END IF          ! DUMMY block for readability: Read input 1111111

      IF(2.EQ.2) THEN ! DUMMY block for readability: Initialize 2222222
        DO I=1,2
          ICOLSE(I)=2*NDOSES+I
          ICOLIC(I)=ICOLSE(I)+2
        END DO
        NMAXME(1)=MXTARG
        NMAXME(2)=MXTARG
      END IF          ! DUMMY block for readability: Initialize 2222222

      DO IFILE=1,NFILES
        IF(ASKINP) PRINT*, 'Enter file name ',IFILE
        READ(*,*) FILENM
        IF(ASKINP) PRINT*, 'Enter related proton energy '
        READ(*,*) ENERGY
        STPWRE=STPWR1*EXP(STPEXP*LOG(ENERGY))

        IF(3.EQ.3) THEN ! DUMMY block for readability: Init counters 3333
          DO K=1,2
```

----- FORTRAN source code of program ME_ROI_3P -----

```
          DO I=1,MXTARG
            DO J=1,2
              FNE(I,J,K)=ZERO
            END DO
          END DO !I=1,MXTARG
          TGMEAN(K)=ZERO
        END DO ! DO K=1,2
      END IF          ! DUMMY block for readability: Init counters 3333

      IF(4.EQ.4) THEN ! DUMMY block for readability: Get data  444444
        IF(DEBUG(1)) PRINT*, 'Begin of Block 4'

        OPEN(LUN,FILE=FILENM,STATUS='OLD')
        DO I=1,2
          READ(LUN,'(A80)') HEADER
        END DO

        READ(LUN,'(A80)') HEADER
        PRINT*, FILENM
        PRINT*, HEADER
        READ(HEADER(28:36),*) DROI
        READ(HEADER(69:77),*) DBEAM
        PRINT*, (DOSE(J),J=1,NDOSES),ENERGY,STPWRE,DROI,DBEAM

        READ(LUN,'(A)') FILOUT
        READ(LUN,*) NMAX(1), NMAX(2)

        DO J=1,NMAX(1)
          READ(LUN,*) (BVARIC(J,L),L=1,NMAX(2))
        END DO

        CLOSE(LUN)

        DO J=1,NMAX(1)
          NMCMAX(J)=0
        END DO

        DO J=1,NMAX(1)
          DO L=1,NMAX(2)
            BVARSE(J,L)=ZERO
            CORRSE(J,L)=ZERO
            IF(BVARIC(J,L).GT.WMIN2) THEN
              NMCMAX(J)=L
            END IF
          END DO
        END DO

      END IF          ! DUMMY block for readability: Get data 4444444

      IF(5.EQ.5) THEN ! DUMMY block for readability: Convolve 5555555
        IF(DEBUG(1)) PRINT*, 'Begin of Block 5 '
        APOT(1)=ZERO
        PTOK(1)=ONE
        DO I=1,NMAX(1)
          APOT(2)=ZERO
          PTOK(2)=ONE
          DO J=1,NMAX(2)
            ANUM(1)=APOT(1)
            ADEN(1)=ZERO
            WEIGH(1)=PTOK(1)
            DO K=I,NMAX(1)
```

----- FORTRAN source code of program ME_ROI_3P -----

```
              IF(WEIGH(1).GT.WMIN2) THEN
                ANUM(2)=APOT(2)
                ADEN(2)=ZERO
                WEIGH(2)=PTOK(2)*WEIGH(1)
                DO L=J,NMCMAX(K)
                  IF(WEIGH(2).GT.WMIN2) THEN
                    BVARSE(I,J)=BVARSE(I,J)+WEIGH(2)*BVARIC(K,L)
                    ADEN(2)=ADEN(2)+ONE
                    ANUM(2)=ANUM(2)+ONE
                    WEIGH(2)=WEIGH(2)*QVALUE*ANUM(2)/ADEN(2)
                  END IF
                END DO ! L=J,NMAX(2)
                ADEN(1)=ADEN(1)+ONE
                ANUM(1)=ANUM(1)+ONE
                WEIGH(1)=WEIGH(1)*QVALUE*ANUM(1)/ADEN(1)
              END IF
            END DO ! K=I,NMAX(1)
            APOT(2)=APOT(2)+ONE
            PTOK(2)=PTOK(2)*PVALUE
          END DO !J=1,NMAX(2)
          APOT(1)=APOT(1)+ONE
          PTOK(1)=PTOK(1)*PVALUE
        END DO !I=1,NMAX(1)
      END IF          ! DUMMY block for readability: Convolve 555555

      IF(5.EQ.5) THEN ! DUMMY block for readability: Margins 6666666
        IF(DEBUG(1)) PRINT*, 'Begin of Block 6'
        DO I=1,MXTARG
          DO K=1,MXNCOL
            PMARG(I,K)=ZERO
          END DO
        END DO

        DO I=1,NMAX(1)
          DO J=1,NMAX(2)
            PMARG(I,ICOLIC(1))=PMARG(I,ICOLIC(1))+BVARIC(I,J)
            PMARG(J,ICOLIC(2))=PMARG(J,ICOLIC(2))+BVARIC(I,J)
            IF(BVARSE(I,J).GT.WMIN2) THEN
              NMAXSE(1)=I
              NMAXSE(2)=J
            END IF
          END DO !I=1,NMAX(1)
        END DO !J=1,NMAX(2)

        DO I=1,NMAXSE(1)
          DO J=1,NMAXSE(2)
            PMARG(I,ICOLSE(1))=PMARG(I,ICOLSE(1))+BVARSE(I,J)
            PMARG(J,ICOLSE(2))=PMARG(J,ICOLSE(2))+BVARSE(I,J)
          END DO !I=1,NMAXSE(1)
        END DO !J=1,NMAXSE(2)

        DO I=1,NMAXSE(1)
          DO J=1,NMAXSE(2)
            IF(PMARG(I,ICOLSE(1)).GT.WMIN2.AND.
   &           PMARG(J,ICOLSE(2)).GT.WMIN2) THEN
              CORRSE(I,J)=BVARSE(I,J)
   &                        /(PMARG(I,ICOLSE(1))*PMARG(J,ICOLSE(2)))
            ELSE
              CORRSE(I,J)=BVARSE(I,J)
            END IF
          END DO !I=1,NMAXSE(1)
```

----- FORTRAN source code of program ME_ROI_3P -----

```
          END DO !J=1,NMAXSE(2)

        END IF          ! DUMMY block for readability: Margins 6666666

        IF(7.EQ.7) THEN ! DUMMY block for readability: Main loop  777777
          DO IDOSE=1,NDOSES
            IF(DEBUG(1)) PRINT*, 'Begin of Block 6, dose =',DOSE(IDOSE)

C!        Init counters

            FLUENC=0.25*DBEAM*DBEAM*PI*DOSE(IDOSE)/DPERFL/STPWRE
            DO K=1,2
              NTARGT(K)=NMAXSE(K)
              ICOL(K)=2*(IDOSE-1)+K
              PMARG(1,ICOL(K))=EXP(-FLUENC) ! Kronecker's delta for J=1
(0 targets)
            END DO

            EVENTS=ONE
            HIFLNC=(FLUENC.GT.23.4)
            IF(HIFLNC) THEN
              DLFLNC=DLOG(FLUENC)
              DLWGHT=-FLUENC+DLFLNC
              WGHT=EXP(DLWGHT)
            ELSE
              WGHT=EXP(-FLUENC)*FLUENC
            END IF

             DO K=1,2
              TGMEAN(ICOL(K))=ZERO
              DO I=1,NTARGT(K)
                FNE(I,2,K)=PMARG(I,ICOLSE(K))
                PMARG(I,ICOL(K))=PMARG(I,ICOL(K))+FNE(I,2,K)*WGHT
                IF(DEBUG(2)) TGMEAN(ICOL(K))=TGMEAN(ICOL(K))
     &                                        +FNE(J,2,K)*(J-1)
              END DO
            END DO

            IF(DEBUG(1)) THEN
              IF(WGHT.GT.WMIN) THEN
                PRINT*, '#',EVENTS,WGHT,TGMEAN(ICOL(1))/EVENTS,
     &                  TGMEAN(ICOL(2))/EVENTS
              ELSE
                PRINT*, '#',EVENTS,WGHT
              END IF
            END IF

  10        CONTINUE
            EVENTS=EVENTS+ONE
            IF(HIFLNC) THEN
              DLWGHT=DLWGHT+DLFLNC-DLOG(EVENTS)
              WGHT=EXP(DLWGHT)
            ELSE
               WGHT=WGHT*FLUENC/EVENTS
            END IF

            DO K=1,2
              DO I=1,NTARGT(K)
                FNE(I,1,K)=FNE(I,2,K)
                FNE(I,2,K)=ZERO
              END DO
```

----- FORTRAN source code of program ME_ROI_3P -----

```
            DO I=1,NTARGT(K)
              DO J=1,NMAXSE(K)
                IJ=I+J-1
                IF (IJ.GT.NMAXME(K)) IJ=NMAXME(K)
                FNE(IJ,2,K)=FNE(IJ,2,K)+FNE(I,1,K)*PMARG(J,ICOLSE(K))
              END DO
            END DO
            NTARGT(K)=NTARGT(K)+NMAXSE(K)
            IF(NTARGT(K).GT.NMAXME(K)) NTARGT(K)=NMAXME(K)
            DO I=1,NTARGT(K)
              PMARG(I,ICOL(K))=PMARG(I,ICOL(K))+FNE(I,2,K)*WGHT
            END DO
          END DO

          IF(DEBUG(2).AND.MOD(NINT(EVENTS),IMESSG(2)).EQ.0) THEN
            IF(WGHT.GT.WMIN) THEN
              DO K=1,2
                TGMEAN(ICOL(K))=ZERO
                DO J=1,NTARGT(K)
                  TGMEAN(ICOL(K))=TGMEAN(ICOL(K))+FNE(J,2,K)*(J-1)
                END DO
              END DO
              PRINT*, '#',EVENTS,WGHT,TGMEAN(ICOL(1)),TGMEAN(ICOL(2))
            ELSE
              PRINT*, '#',EVENTS,WGHT
            END IF
          END IF

          IF(EVENTS.LT.FLUENC.OR.WGHT.GT.WMIN) GOTO 10
          PRINT*, DOSE

        END DO ! IDOSE=1,NDOSES
      END IF          ! DUMMY block for readability: Main Loop  777777

      IF(8.EQ.8) THEN ! DUMMY block for readability: Margins 8888888
        IF(DEBUG(1)) PRINT*, 'Begin of Block 8'

        NCOLS=2*NDOSES+4
        NMAX(1)=0
        DO K=1,NCOLS
          TGMEAN(K)=ZERO
          SUMS(K)=ZERO
          IF(K.GT.2) NTARGT(K)=NTARGT(K-2)
          DO J=1,NTARGT(K)
            TGMEAN(K)=TGMEAN(K)+PMARG(J,K)*(J-1)
            SUMS(K)=SUMS(K)+PMARG(J,K)
            IF(PMARG(J,K).GT.WMIN.AND.J.GT.NMAX(1)) NMAX(1)=J
          END DO
        END DO ! DO K=1,NCOLS
        PRINT*, 'Distributions averages (#hit targets)'
        WRITE(*,'(20f10.4)') (SUMS(J),J=1,NCOLS)
        WRITE(*,'(20f10.4)') (TGMEAN(J),J=1,NCOLS)
        WRITE(*,*) TGMEAN(ICOLSE(1))/TGMEAN(ICOLIC(1)),
     &             TGMEAN(ICOLSE(2))/TGMEAN(ICOLIC(2))
        DO K=1,NCOLS-4
          IF(MOD(K,2).EQ.1) THEN
            TGMEAN(K)=TGMEAN(K)/TGMEAN(ICOLSE(1))
          ELSE
            TGMEAN(K)=TGMEAN(K)/TGMEAN(ICOLSE(2))
          END IF
        END DO
```

----- FORTRAN source code of program ME_ROI_3P -----

```
          WRITE(*,'(20f10.4)') (TGMEAN(J),J=1,NCOLS-4)

        END IF          ! DUMMY block for readability: Margins 8888888

        IF(9.EQ.9) THEN ! DUMMY block for readability: OUTPUT  9999999
C!        #Write results to output file
          FILOUT='SEB_'//FILENM
          PRINT*, 'Write output to '//FILOUT
          OPEN(LUN,FILE=FILOUT,STATUS='UNKNOWN')
          WRITE(LUN,*) '*** Output from PROGRAM ME_ROI_3P Version '
     &             //VDATE//' on '//TSTAMP()
          WRITE(LUN,*) 'Filename: '//FILOUT
          WRITE(LUN,*) HEADER
          WRITE(LUN,*) 'Corre1ations P1 and F2 for PVALUE= ',PVALUE
          WRITE(LUN,*) NMAXSE(1), NMAXSE(2)
          DO I=1,NMAXSE(1)
            WRITE(LUN,'(1000F15.10)') (BVARSE(I,J),J=1,NMAXSE(2))
          END DO
          CLOSE(LUN)

          FILOUT='SEC_'//FILENM
          PRINT*, 'Write output to '//FILOUT
          OPEN(LUN,FILE=FILOUT,STATUS='UNKNOWN')
          WRITE(LUN,*) '*** Output from PROGRAM SE_ROI_3C Version '
     &             //VDATE//' on '//TSTAMP()
          WRITE(LUN,*) 'Filename: '//FILOUT
          WRITE(LUN,*) HEADER
          WRITE(LUN,*) 'Corre1ations P1 and F2 for PVALUE= ',PVALUE
          WRITE(LUN,*) NMAXSE(1), NMAXSE(2)
          DO I=1,NMAXSE(1)
            WRITE(LUN,'(1000F15.10)') (CORRSE(I,J),J=1,NMAXSE(2))
          END DO
          CLOSE(LUN)

          PREFIX='MEP_'
          FILOUT=PREFIX//FILENM
          PRINT*, 'Write output to '//FILOUT
          OPEN(LUN,FILE=FILOUT,STATUS='UNKNOWN')
          WRITE(LUN,*) '*** Output from PROGRAM ME_ROI_3P Version '
     &             //VDATE//' on '//TSTAMP()
          WRITE(LUN,*) 'Filename: '//FILOUT
          WRITE(LUN,*) HEADER
          WRITE(LUN,*) 'Single event distribution for PVALUE= ',PVALUE

          CDOSES(1)='      P1'
          CDOSES(2)='     P2+'
          WRITE(LUN,'(A9,50a15)') '#',((CDOSES(K),K=1,2),J=1,NDOSES),
     &                         'P1','P2+','P1','P2+'
          PRINT*, DOSE
          DO I=1,NDOSES
            WRITE(CDOSES(I),'(f5.1,a2)') DOSE(I),'Gy'
          END DO

          WRITE(LUN,'(1X,a8,50a15)') 'targets',
     &             ((CDOSES(J),K=1,2),J=1,NDOSES),
     &             ('SE',K=1,2),('IC',K=1,2)
          DO I=1,NMAX(1)
            WRITE(LUN,'(1X,i8,50F15.10)') I-1,(PMARG(I,J),J=1,NCOLS)
          END DO
          CLOSE(LUN)
```

| ----- FORTRAN source code of program ME_ROI_3P ----- |
|---|
| <pre>        END IF          ! DUMMY block for readability: OUTPUT     9999999<br>      END DO ! IFILE=1,NFILES<br><br>      END PROGRAM ! ME_ROI_3P<br>C!______________________________________________________________________<br><br>        INCLUDE 'TSTAMP.f'</pre> |

*FORTRAN subroutine BLINIT*

This subroutine initializes the Bravais lattice used for scoring clusters and is called from the programs IC_3D and ROI_3D. The respective source codes have an INCLUDE statement that expects the code in the second row of the table below in an Ascii-File named BLINIT.f.

----- FORTRAN subroutine BLINIT -----

```
C!*********************************************************************
      SUBROUTINE BLINIT(NAZMTH,NDIV,DLATC,SRDATE)
C!    ***************************************************************
C!    Initialization of the base vectors of the reziprocal lattices
C!    ***************************************************************
C!    This routine calculates the elements of the array ADJNTB in the
C!    common block LATTICE, i.e. the set of base vectors of the
C!    reciprocal lattice of a cubic fcc Bravais lattice for different
C!    orientations of the primary particle's trajectory with respect to
C!    the Bravais lattice of target volumes.
C!    25.02.2020: The algorithm for deterministic sampling has been
C!    implemented such that for NDIV>0 the spherical triangle defined by
C!    the north pole and the azimuths +/- 2/3*pi on the equator is
C!    divided into four congruent triangles, then these smaller
C!    triangles are again divided into four congruent triangles etc.
C!    until finally we have 4^NDIV triangles. The centers of gravity of
C!    these triangles define directions of the primary particle's
C!    trajectory that are homogenously distributed over the section of
C!    the unit sphere that is sufficient to consider owing to the
C!    symmetry of the cubic fcc Bravais lattice of target volumes.
C!    In addition, also a rotation of the track in itself is considered
C!    to improve statistics.
C!    Declarations ---------------------------------------------------
      IMPLICIT NONE
C!    ----- Parameters: Version Date <<<<<<<<<<<<<<<<<<<<<<<<<<<<<<<<<<<
      CHARACTER VDATE*11
      PARAMETER(VDATE='23-MAY-2021') !<<<<<<<<<<<<<<<<<<<<<<<<<<<<<<<<<
C!    23-MAY-2021 HR: Added outout of version date to calling program
C!    23-OCT-2020 HR: Fixed bug in initialization of reciprocal basis
C!    15-MAR-2020 HR: Added VDATE and modified COMMON BLOCK VERBOSE
C!    ----- Parameters: General purpose constants
      REAL*8 ONE, TWO, THREE, FOUR, ZERO
      PARAMETER(ONE=1.0, TWO=2.0, THREE=3.0, FOUR=4.0, ZERO=0.0)
C!    ----- Parameters for lattice orientation
      INTEGER*4 MAZMTH, MDIV, MTHETA, MDIR
      PARAMETER(MAZMTH=1, MDIV=0, MTHETA=2**MDIV,
     &           MDIR=MAZMTH*(MTHETA*(MTHETA+1))/2)
C!    ----- Local scalars
      REAL*8 PIBY2, PIBY3, ROOT2, ROOT3 ! Constants
      REAL*8 COSTHM, COSTHP, DCOSTH, DPHI2, DPHI1, PHI1, PHI2, THETA
      INTEGER*4 IAZ, IDIR, IR, IS, J, K, L, NTHETA
C!    ----- Local arrays
      REAL*8 ADJBAS(3,3), REULER(3,3)
C!    ----- Global variables
      INTEGER*4 NDIR
      REAL*8 ADJNTB(3,3,MDIR)
```

----- FORTRAN subroutine BLINIT -----

```
      COMMON /LATTICE/ ADJNTB, NDIR
C!    ----- Scalars and arrays passed from the calling module
      INTEGER*4 NAZMTH, NDIV
      REAL*8 DLATC
C!    ----- Scalars and arrays passed to the calling module
      CHARACTER*11 SRDATE
C!    ------
      LOGICAL DEBUG(4), SRFLAG
      COMMON /VERBOSE/ DEBUG, SRFLAG
C!<<<<End declarations<<<<<<<<<<<<<<<<<<<<<<<<<<<<<<<<<<<<<<<<<<<<<<<<<<

      IF(1.EQ.1) THEN ! Dummy block: Checks for consistency
        PRINT*, 'SUBROUTINE BLINIT Version: '//VDATE
        IF(DEBUG(1)) PRINT*, 'BLINIT 1'
        IF(NAZMTH.GT.MAZMTH) THEN
          PRINT*, 'ERROR in BLINIT: NAZMTH > MAZMTH ',NAZMTH,MAZMTH
          STOP
        END IF
        IF(NDIV.GT.MDIV) THEN
          PRINT*, 'ERROR in BLINIT: NDIV > MDIV ',NDIV,MDIV
          STOP
        END IF
        SRDATE=VDATE
      END IF          ! Dummy block: Checks for consistency

      IF(2.EQ.2) THEN ! Dummy block for readability: INITIALIZATION
        IF(DEBUG(1)) PRINT*, 'BLINIT 2'
        ROOT2=SQRT(2.0)
        ROOT3=SQRT(3.0)
        ADJBAS(1,1)=ONE/DLATC/ROOT3
        ADJBAS(2,1)=-ONE/DLATC
        ADJBAS(3,1)=-ONE/DLATC/ROOT2/ROOT3
        ADJBAS(1,2)=ONE/DLATC/ROOT3
        ADJBAS(2,2)=ONE/DLATC
        ADJBAS(3,2)=-ONE/DLATC/ROOT2/ROOT3
        ADJBAS(1,3)=ZERO
        ADJBAS(2,3)=ZERO
        ADJBAS(3,3)=ONE/DLATC*ROOT3/ROOT2

        PIBY2=ATAN(ONE)*TWO
        PIBY3=PIBY2*TWO/THREE
      END IF          ! Dummy block for readability: INITIALIZATION

      PRINT*, ADJBAS
      IF(3.EQ.3) THEN ! Dummy block for readability: Get track direction
                      ! and adjust lattice orientation accordingly
        IF(DEBUG(1)) PRINT*, 'BLINIT 3'
        NTHETA=2**NDIV                 ! number of polar angles
        NDIR=NAZMTH*NTHETA*NTHETA      ! number of orientations
        DCOSTH=ONE/REAL(NTHETA)        ! step size in cos theta
        COSTHP=ONE-TWO/THREE*DCOSTH    ! cos theta plus (see section 
paper)
        COSTHM=ONE-FOUR/THREE*DCOSTH   ! cos theta minus (see section 
paper)
        DPHI1=FOUR*PIBY2/NAZMTH        ! step for rotation of track

        PHI1=ZERO
        IDIR=0
        DO IAZ=1,NAZMTH
          DO K=1,NTHETA ! Centers of upward pointing triangles
            DPHI2=TWO*PIBY3/REAL(K) ! step of trajectory azimuth
```

----- FORTRAN subroutine BLINIT -----

```
          PHI2=(ONE/REAL(K)-ONE)*PIBY3 ! first azimuth value
          THETA=ACOS(COSTHP)
          DO J=1,K
            IDIR=IDIR+1
        IF(DEBUG(1)) PRINT*, 'BLINIT vor CALL EULERM'
            CALL EULERM(REULER,THETA,PHI1,PHI2)
        IF(DEBUG(1)) PRINT*, 'BLINIT nach CALL EULERM',REULER
            DO IR=1,3
               DO IS=1,3
                 ADJNTB(IR,IS,IDIR)=ZERO
                 DO L=1,3
                   ADJNTB(IR,IS,IDIR)=ADJNTB(IR,IS,IDIR)
     &                                   +REULER(IR,L)*ADJBAS(L,IS)
                 END DO ! IS=1,3
               END DO ! IR=1,3
            END DO
            PHI2=PHI2+DPHI2  ! increment trajectory azimuth
          END DO ! J=1,K
          COSTHP=COSTHP+DCOSTH  ! increment cosine of polar angle
        END DO ! K=1,NTHETA

      IF(DEBUG(1)) PRINT*, 'BLINIT 3.2 IAZ', IAZ
        DO K=1,NTHETA-1 ! Centers of downward pointing triangles
          DPHI2=TWO*PIBY3/REAL(K) ! step of trajectory azimuth
          PHI2=(ONE/REAL(K)-ONE)*PIBY3 ! first azimuth value
          THETA=ACOS(COSTHM)
          DO J=1,K
            IDIR=IDIR+1
            CALL EULERM(REULER,THETA,PHI1,PHI2)
            DO IR=1,3
               DO IS=1,3
                 ADJNTB(IR,IS,IDIR)=ZERO
                 DO L=1,3
                   ADJNTB(IR,IS,IDIR)=ADJNTB(IR,IS,IDIR)
     &                                  +REULER(IR,L)*ADJBAS(L,IS)
                 END DO ! IS=1,3
               END DO ! IR=1,3
            END DO
            PHI2=PHI2+DPHI2  ! increment trajectory azimuth
          END DO ! J=1,K
          COSTHM=COSTHM+DCOSTH  ! increment cosine of polar angle
        END DO ! K=1,NTHETA
        PHI1=PHI1+DPHI1 ! increment azimuth for track rotation
      END DO ! IAZ=1,NAZMTH
    END IF          ! Dummy block: Adjust lattice orientation

    END SUBROUTINE BLINIT
C!________________________________________________________________________

C!************************************************************************
      SUBROUTINE EULERM(REULER,THETA,PHI1,PHI2)
C!    *******************************************************************
C!    Calculates combination of three inverse Euler rotation matrices
C!    *******************************************************************
C!    Declarations --------------------------------------------------
      IMPLICIT NONE
C!    ----- Scalars and arrays passed from the calling module
      REAL*8 PHI1,PHI2,THETA
      REAL*8 REULER(3,3)
C!    ----- Local scalars
      INTEGER*4 I,J
```

----- FORTRAN subroutine BLINIT -----

```
      REAL*8 RCOS, RSIN
C!    ----- Local arrays
      REAL*8 ROT1(3,3), ROT2(3,3), ROT3(3,3), RAUXIL(3,3)
C!<<<<End declarations<<<<<<<<<<<<<<<<<<<<<<<<<<<<<<<<<<<<<<<<<<<<<<<
      DO I=1,3
        DO J=1,3
          IF(I.EQ.J) THEN
            REULER(I,J)=1.0
            ROT1(I,J)=1.0
            ROT2(I,J)=1.0
            ROT3(I,J)=1.0
          ELSE
            REULER(I,J)=0.0
            ROT1(I,J)=0.0
            ROT2(I,J)=0.0
            ROT3(I,J)=0.0
          ENDIF
          RAUXIL(I,J)=0.0
        END DO
      END DO

      RCOS=COS(PHI2)
      RSIN=SIN(PHI2)
      ROT1(1,1)=RCOS
      ROT1(2,2)=RCOS
      ROT1(1,2)=RSIN
      ROT1(2,1)=-RSIN

      RCOS=COS(THETA)
      RSIN=SIN(THETA)
      ROT2(1,1)=RCOS
      ROT2(3,3)=RCOS
      ROT2(1,3)=RSIN
      ROT2(3,1)=-RSIN

      RCOS=COS(PHI1)
      RSIN=SIN(PHI1)
      ROT3(1,1)=RCOS
      ROT3(2,2)=RCOS
      ROT3(1,2)=RSIN
      ROT3(2,1)=-RSIN

      CALL MMULT(ROT2,ROT1,RAUXIL)
      CALL MMULT(ROT3,RAUXIL,REULER)
      END SUBROUTINE EULERM
C!______________________________________________________________________

C!*********************************************************************
      SUBROUTINE MMULT(A,B,C)
C!    *************************************************************
C!    Matrix multiplication C=A*B
C!    *************************************************************
      REAL*8 A(3,3), B(3,3), C(3,3)
      INTEGER*4 I, J, K
      DO I=1,3
        DO K=1,3
          C(I,K)=0.0
          DO J=1,3
            C(I,K)=C(I,K)+A(I,J)*B(J,K)
          END DO
        END DO
```

----- FORTRAN subroutine BLINIT -----

```
      END DO
      END SUBROUTINE MMULT
C!______________________________________________________________________________

C!**********************************************************************
      SUBROUTINE MVMULT(A,B,C)
C!    *************************************************************
C!    Multiplication of matrix A and vector B: C=A*B
C!    *************************************************************
      REAL*8 A(3,3), B(3), C(3)
      INTEGER*4 I, J
      DO I=1,3
        C(I)=0.0
        DO J=1,3
          C(I)=C(I)+A(I,J)*B(J)
        END DO
      END DO
      END SUBROUTINE MVMULT
C!______________________________________________________________________________
```

*FORTRAN subroutine RFINIT*

This subroutine interpretes the character string encoding the data structure of the lines of the track simulation output files read by IC_3D and ROI_3D.

----- FORTRAN subroutine RFINIT -----

```
      SUBROUTINE RFINIT()
C!    Sets the information for the columns of the input file
C!
C!    ------ Global variables
C!    ------
C!     CODE: Character string encoding the meaning of the entries in a
C!           line of the input file as follows:
C!           'T'     - number of the primary particle track
C!           'X','Y','Z' - x, y, and z coordinates of the transfer point
C!           'E'     - energy deposit (if applicable)
C!           'I'     - ionization cluster size (if applicable)
C!           '%'     - additional data that are not used
C!           Example: The string 'TZ%%EXY ' indicates that there are 7
C!                 entries in each line, of which the first is the
C!                 track number, the second the z coordinate, the
C!                 fifth the energy, and the sixth and seventh the x
C!                 and y coordinates
C!     IDT: Array of the columns indices of (1-3) x,y,z coordinates
C!               (4) energy deposit (if present), (5) track number,
C!               (6) number of ionizations in cluster (if present)
C!               (7-8) are there for future use
      CHARACTER CODE*8
      INTEGER*2 IDT(8),NDT
      COMMON /RFORMT/IDT,NDT,CODE
C!    Local variables
      INTEGER*1 J

      DO J=1,8
       IDT(J)=0
      END DO

      DO J=1,8
```

```
----- FORTRAN subroutine RFINIT -----
       IF(CODE(J:J).EQ.'X'.AND.IDT(1).EQ.0) IDT(1)=J
       IF(CODE(J:J).EQ.'Y'.AND.IDT(2).EQ.0) IDT(2)=J
       IF(CODE(J:J).EQ.'Z'.AND.IDT(3).EQ.0) IDT(3)=J
       IF(CODE(J:J).EQ.'E'.AND.IDT(4).EQ.0) IDT(4)=J
       IF(CODE(J:J).EQ.'T'.AND.IDT(5).EQ.0) IDT(5)=J
       IF(CODE(J:J).EQ.'I'.AND.IDT(6).EQ.0) IDT(6)=J
       IF(CODE(J:J).NE.' ') NDT=J
      END DO

      IF(IDT(1).EQ.0) PRINT*, 'No data column for X'
      IF(IDT(2).EQ.0) PRINT*, 'No data column for Y'
      IF(IDT(3).EQ.0) PRINT*, 'No data column for Z'
      IF(IDT(5).EQ.0) PRINT*, 'No data column for track number'
      IF(IDT(1)*IDT(2)*IDT(3)*IDT(5).EQ.0) STOP

      END
```

*FORTRAN function TSTAMP*

This function gets the system date and time and returns a 24 character text string in the format YYYY-MM-DD HH:MM ±HH:MM, where the latter is the time difference to UTC.

```
----- FORTRAN subroutine TSTAMP -----
      CHARACTER*24 FUNCTION TSTAMP()
      CHARACTER DATE*8, TIME*10, ZONE*5, TIMEST*24
      INTEGER*4 IDT(8)
      CALL DATE_AND_TIME(DATE,TIME,ZONE,IDT)
C!              123456789 123456789 1234
      TIMEST = 'DD-MMM-YYYY HH:MM +HH:MM'
      TSTAMP=DATE(7:8)//'-'//DATE(5:6)//'-'//DATE(1:4)//' '//
     &       TIME(1:2)//':'//TIME(3:4)//' '//ZONE(1:3)//':'//ZONE(4:5)
      RETURN
      END
```

*Sample debug options file for use with programs IC_3D and ROI_3D*

This is simply an text files with four lines containing either .TRUE. or .FALSE.

```
----- Sample debug options file for use with programs IC_3D and ROI_3D -----
.FALSE.      ! DEBUG(1) --> Main program
.FALSE.      ! DEBUG(2) --> TARGTS main sections
.FALSE.      ! DEBUG(3) --> TARGTS IDIR loop
.FALSE.      ! DEBUG(4) --> TARGTS IPOS loop details
```